%% file: Nf4_6_StepScaling.tex
\documentclass[pdftex, aps, prd, letterpaper, superscriptaddress, nofootinbib, twocolumn,
               notitlepage, preprintnumbers, floatfix, longbibliography]{revtex4-1}
\pdfoutput=1
\usepackage{amsmath,amssymb,amsbsy}
\usepackage[utf8]{inputenc}
\usepackage{graphicx,longtable,tabularx,capt-of}
\usepackage{soul,color,topcapt}
\usepackage[colorlinks=true, linkcolor=blue, filecolor=blue, urlcolor=blue, citecolor=blue, pdftex, plainpages=false]{hyperref}

\definecolor{orange}{rgb}{1.0, 0.5, 0}

\newcommand{\vev}[1]{\ensuremath{\left\langle #1 \right\rangle} }

\begin{document}
\title{Gradient flow step-scaling function for SU(3) with \texorpdfstring{$N_f=6$ or $4$}{Nf=6 or 4} fundamental flavors}
\author{Anna Hasenfratz}
\email{anna.hasenfratz@colorado.edu}
\affiliation{Department of Physics, University of Colorado, Boulder, Colorado 80309, USA}
\author{Claudio Rebbi}
\affiliation{Department of Physics and Center for Computational Science, Boston University, Boston, Massachusetts 02215, USA}
\author{Oliver Witzel}
\email{oliver.witzel@uni-siegen.de}
\affiliation{Center for Particle Physics Siegen, Theoretische Physik 1, Naturwissenschaftlich-Technische Fakult\"at,
  Universit\"at Siegen, 57068 Siegen, Germany}

\preprint{FERMILAB-PUB-22-708-V,~~SI-HEP-2022-29}

\date{\today}
% ------------------------------------------------------------------

%%%%%%%%%%%%%%%%%%%%%%%%%%%%%% ABSTRACT %%%%%%%%%%%%%%%%%%%%%%%%%%%%%%%
\begin{abstract}
  Nonperturbative determinations of the renormalization group (RG) $\beta$ function are crucial to understand properties of gauge-fermion systems at strong coupling and connect lattice simulations and the perturbative ultraviolet regime. Choosing well-understood, QCD-like systems with SU(3) gauge group and either six or four fundamental flavors, we investigate their step-scaling $\beta$ function.  In both cases we push the simulations to the boundary of chiral symmetry breaking and study the regime $g^2_{GF} \lesssim 8.2$ with six, and  $g^2_{GF} \lesssim 6.6$ with four flavors. We carefully consider the lattice discretization errors by comparing  three different gradient flows (GF), and for each flow three operators to estimate the renormalized finite volume coupling. We also consider the tree level improvement of the coupling.  Noteworthy outcome is that nonperturbatively determined $\beta$ functions run much slower than perturbatively predicted.
\end{abstract}
\maketitle

%=================================================
\section{Introduction}
%=================================================
Nonperturbative lattice calculations are essential to account for nonperturbative effects when comparing experimental and theoretical predictions in search of new, beyond standard model (SM)  physics effects \cite{USQCD:2022mmc,Davoudi:2022bnl}.  An important part of the lattice program is to connect the energy range accessible  in the lattice simulation to the UV scale where reliable connection to perturbation theory is possible. The nonperturbative renormalization group (RG) $\beta$ function  provides this connection for the renormalized coupling.

Several lattice approaches exist to  predict the RG $\beta$ function. Many of the recent calculations use the gradient flow (GF) renormalized coupling $g_{GF}^2$~\cite{Narayanan:2006rf,Luscher:2009eq,Luscher:2010iy}, and predict the finite volume step-scaling function or the infinite volume continuous $\beta$ function \cite{Fodor:2012td,Luscher:2014kea,Hasenfratz:2019hpg,Fodor:2017die}. In the former  the flow time (or energy scale) is set by the lattice size  $t = (c L)^2/8$ and the step-scaling function $\beta_{c,s}$ is calculated  by comparing the GF coupling on lattice sizes $L$ and $s L$ \cite{Fodor:2012td}.  The finite volume step-scaling method requires that the lattice size provides the only dimensional scale, i.e.~it is not applicable in the confining, chirally broken regime. Thus, when investigating QCD-like systems it is important to keep the lattice volume small so the simulations are performed in the deconfined regime. This condition limits both the  bare coupling and the accessible renormalized coupling range.   
The continuous $\beta(g^2)$ function is defined analogous to the continuum definition and requires that the lattice data are extrapolated to infinite volume \cite{Hasenfratz:2019hpg,Hasenfratz:2019puu,Fodor:2017die}. This method is applicable even when another energy scale emerges, e.g.~in the chirally broken/confining regime. 

Both the step-scaling and continuous $\beta$ functions are defined in the chiral limit. While simulations with zero fermion mass are possible in the finite volume deconfined phase, the chirally broken/confining regime is accessible only with finite mass simulations. Consequently the analysis requires  to also take the $a m_f \to 0$ chiral limit. To date only preliminary results predicting the RG $\beta$ function in the confined phase have been reported for the $N_f=0$ pure gauge system in Ref.~\cite{Peterson:2021lvb} and for $N_f=2$ in Ref.~\cite{Hasenfratz:2022wll}.

 Both methods  predict that QCD-like systems with $N_f=2$ or 3 flavors exhibit a nonperturbative $\beta$ function which runs slower than the universal 2-loop perturbative prediction in the  $0.2 - 4$ GeV energy range and is  close to the perturbative 1-loop result \cite{DallaBrida:2016kgh,Hasenfratz:2019hpg}. $N_f=0$ shows similar behavior  \cite{Peterson:2021lvb}.
Unfortunately the 3-loop GF scheme $\beta$ function shows very poor convergence, deviating from both $\overline{\textrm{MS}}$ and nonperturbative predictions at $g^2_{GF} \gtrsim 1.0- 2.0$  \cite{Harlander:2016vzb}. In  systems with $N_f=10$ and 12 flavors, nonperturbative lattice calculations suggest conformality and the existence of an infrared fixed point (IRFP), 
although there is no consensus between different lattice groups on either system \cite{Hasenfratz:2016dou,Hasenfratz:2017qyr,Hasenfratz:2019dpr,Hasenfratz:2019puu,Hasenfratz:2020ess,Chiu:2016uui,Chiu:2018edw,Fodor:2016zil,Fodor:2017gtj,Fodor:2017nlp,Fodor:2018tdg}.  In any case, if the IRFP exists, it is at rather strong coupling where perturbative predictions are not reliable \cite{Ryttov:2010iz,Ryttov:2016hal,Ryttov:2016ner,Ryttov:2017lkz}.  

The above observations prompted us to initiate a systematic study of the RG $\beta$ function with  $N_f=4$, 6, and 8 flavors to complement our existing $N_f=2$,  10, 12 flavor results and the ongoing work in the pure gauge ($N_f=0$) system \cite{Peterson:2021lvb}. We use chirally symmetric M\"obius domain wall fermions (MDWF) and Symanzik improved gauge action in all cases and investigate lattice artifacts by comparing different gradient flows and operators. 

Both $N_f=4$ and 6 flavors are QCD-like, chirally broken and confining at zero temperature. In this paper we present our findings on the finite volume step-scaling function of these systems. The $N_f=8$ system is likely very close to the conformal sill, possibly even corresponding to the opening of the conformal window \cite{Hasenfratz:2022qan}. However, establishing the nature of $N_f=8$ is very challenging \cite{Appelquist:2018yqe}.  We intend to report on our $N_f=8$ study with MDWF in a future publication.

\begin{figure}[tb]
  \includegraphics[width=\columnwidth]{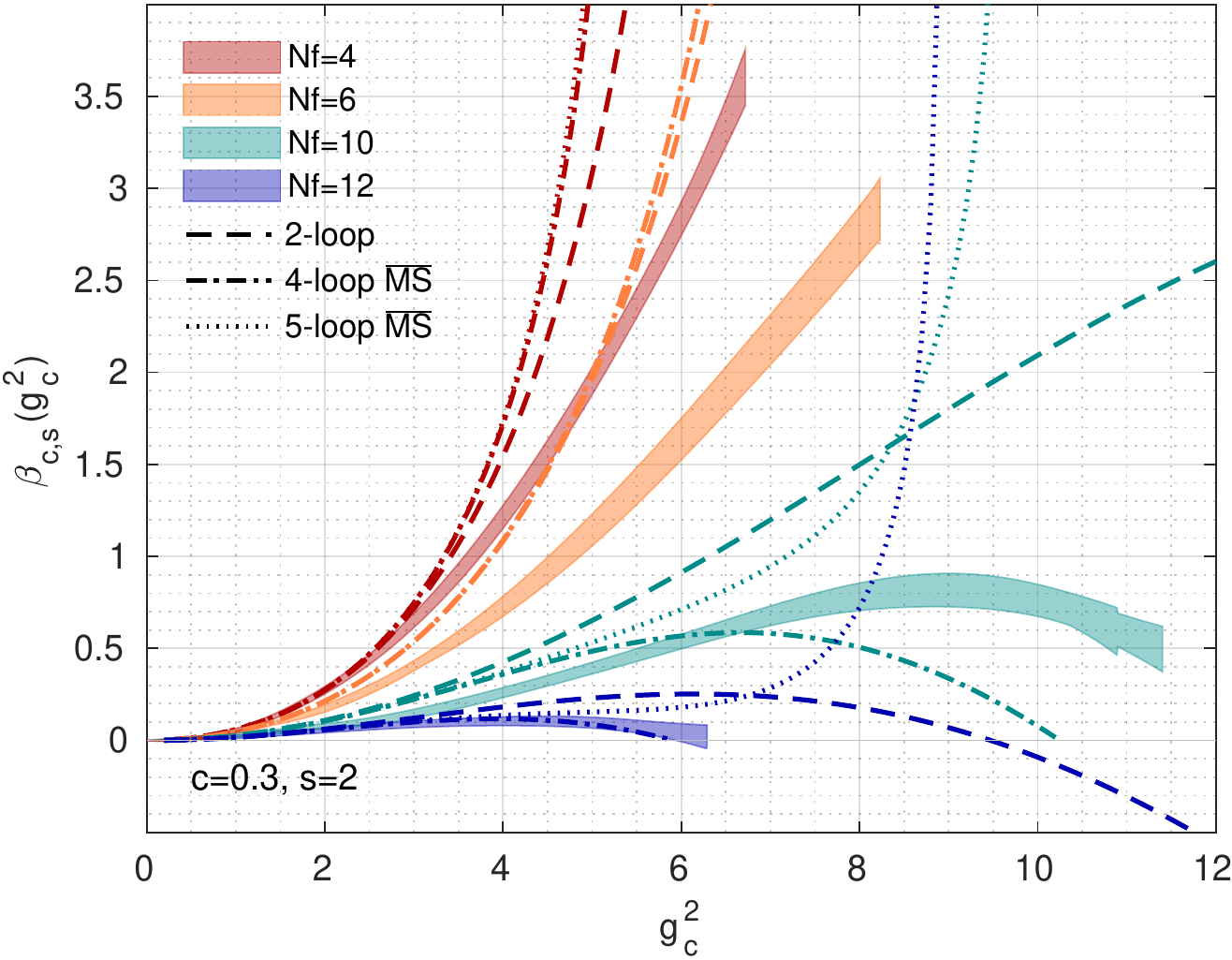}
  \caption{Comparison of the step-scaling functions for $N_f=4$, 6, 10 and 12 using the $c=0.300$ scheme with scale change $s=2$. In addition we show the perturbative predictions for the same scale change at 2-loop (universal) as well as 4- and 5-loop in the $\overline{\textrm{MS}}$ scheme \cite{Baikov:2016tgj}.}
  \label{Fig.All}
\end{figure}
We summarize our findings in Fig.~\ref{Fig.All} where we present the nonperturbative GF step-scaling functions with \mbox{$N_f=4$} (red), 6 (orange), 10 (green) and 12 (blue) flavors in comparison to perturbative predictions.\footnote{We have analyzed the $N_f=2$ system only with the continuous $\beta$ function method and do not present its step scaling function here.}  In the case of the QCD-like systems with four or six flavors, our nonpertubative results indicate that the $\beta$ function runs substantially slower than predicted by perturbation theory. In the case of the conformal $N_f=12$ system, perturbative results at 3- and 4-loop in the $\overline{\textrm{MS}}$-scheme show a qualitatively and quantitatively  similar result as our nonperturbative prediction. However, the 5-loop \mbox{$\overline{\textrm{MS}}$-result} is not a ``small correction'' to the 4-loop $\overline{\textrm{MS}}$-result. The 5-loop result  does not exhibit an IRFP but the poor convergence of the perturbative series makes any prediction for $g^2  \gtrsim 4.0$  questionable. The (near)conformal $N_f=10$ is qualitatively similar to the 3- and 4-loop $\overline{\textrm{MS}}$-predictions but quantitatively the predicted values for the IRFP are rather different. Again the 5-loop result is off, showing a very rapid increase for $g_c^2>8.5$.

This paper is organized as follows: In Section \ref{Sec.Details} we describe the details of our calculations starting with our lattice simulations and the gradient flow measurements. Moreover the definition of the step-scaling $\beta$-function is summarized and the steps of our analysis procedure are given. Our numerical results are presented for SU(3) with six fundamental flavors in Sec.~\ref{Sec.Nf6} and for SU(3) with four fundamental flavors in Sec.~\ref{Sec.Nf4}. We discuss our findings in Sec.~\ref{Sec.Discussion} where we also compare our results to perturbative predictions, as well as nonperturbative results by the Lattice Higgs collaboration for $N_f=4$ resolving the effect of different choices for the scale change $s$.   Subsequently we close with a brief summary. 

%=================================================
\section{Details of our Calculation} \label{Sec.Details}
%=================================================

%=================================================
\subsection{Lattice Simulations}
%=================================================
 As in our previous studies of SU(3) gauge systems with $N_f=2$ \cite{Hasenfratz:2019hpg}, 10 \cite{Hasenfratz:2017qyr,Hasenfratz:2020ess} or 12 \cite{Hasenfratz:2017qyr,Hasenfratz:2019dpr} fundamental fermions, we choose the tree-level improved Symanzik (L\"uscher-Weisz) gauge action \cite{Luscher:1984xn,Luscher:1985zq} and M\"obius domain wall fermions (MDWF) \cite{Brower:2012vk} (domain wall height $M_5=1.0$, M\"obius parameters $b_5=1.5$, $c_5=0.5$) with three levels of stout-smearing \cite{Morningstar:2003gk} ($\varrho=0.1$) for the fermion action.  Dynamical gauge field configurations were generated using the hybrid Monte Carlo (HMC) \cite{Duane:1987de} updating algorithm as implemented  in \texttt{GRID}\footnote{\url{https://github.com/paboyle/Grid}} \cite{Boyle:2015tjk}. We set the fermion  mass $am_f=0.0$ and choose anti-periodic boundary conditions (BC) for the fermions in all four space-time directions but periodic BC for the gauge field. After thermalization, gauge field configurations are saved every five trajectories and each trajectory has a length of $\tau=2$ molecular dynamic time units (MDTU). Our simulations are performed on $(L/a)^4$ hypercubic volumes. 

We choose $L/a=$ 8, 10, 12, 16, 20, 24, 32, 40, and calculate the step-scaling function with scale change $s=2$, i.e.~we consider volume pairs $(8\to16)$, $(10\to20)$, ... $(20\to40)$.  We perform simulations using bare gauge couplings $\beta\equiv 6/g_0^2 \in\{$8.50, 7.00, 6.00, 5.20, 4.80, 4.50, 4.30, 4.20, 4.15, 4.10, 4.05$\}$ for $N_f=6$ and $\beta\in\{$8.50, 7.00, 6.00, 5.20, 4.80, 4.50, 4.30, 4.25, 4.20$\}$ for $N_f=4$. We note however that the strongest couplings are not simulated for all volumes. The goal is to cover (approximately) the same range in the finite volume  renormalized couplings on each volume pair, while keeping the system in the small volume deconfined regime. At the same bare coupling, larger volumes reach larger values of the renormalized coupling and might even transit to the confining regime.  We monitor the emergence of confinement by  computing Polyakov loops.   The bare coupling values listed above for the $N_f=6$ system correspond to the deconfined regime on all volumes. However, according to the Polyakov loop data, the $N_f=4$ system transitions to the confined regime on the  largest volumes.  On  $L/a=40$ volume  the strongest coupling we  use is  $\beta=4.30$ for $N_f=6$ and $\beta=4.60$ for $N_f=4$.  Details of the generated gauge field ensembles including the number of thermalized measurements are shown in Table \ref{Tab.Nf6_nZS_ZS} for $N_f=6$ and in Table \ref{Tab.Nf4_nZS_ZS} for $N_f=4$ in Appendix \ref{Sec.RenCouplings}. On the small volumes we typically accumulate 1200 MDTU, but we use lower statistics on the larger volumes. The largest and numerically most expensive $40^4$ ensembles have about 200 thermalized MDTU each.  Simulations with $\beta > 4.20$ are performed using $L_s=12$ for the extent of the fifth dimension of domain wall fermions, while $L_s=16$ is chosen for $\beta\le 4.20$. As demonstrated in our previous work \cite{Hasenfratz:2017qyr,Hasenfratz:2020ess,Hasenfratz:2019dpr}, this choice ensures that the residual chiral symmetry breaking present for DWF expressed as the residual mass $am_\text{res}$ remains sufficiently small, less than $10^{-5}$. The good chiral properties of MDWF protect our zero mass simulations from effects due to nonzero topological charges. Further the simulated gauge fields are sufficiently smooth and we do not observe any topological artifacts like those that contaminated our $N_f=10$ simulations \cite{Hasenfratz:2020vta}.

%=================================================
\subsection{Gradient Flow Measurements}
%=================================================

Gradient flow measurements are performed on configurations separated by 10 MDTU on all available gauge field configurations. These measurements are carried out using \texttt{Qlua}\footnote{\url{https://usqcd.lns.mit.edu/w/index.php/QLUA}} \cite{Pochinsky:2008zz}. In total we perform three sets of gradient flow measurements choosing different actions for  the kernel. Specifically we obtain data for Wilson (W), Symanzik (S), and Zeuthen (Z) \cite{Ramos:2014kka,Ramos:2015baa} flow,  determining  three operators, Wilson plaquette (W), Symanzik (S) and clover (C) to estimate the energy density $\langle E(t)\rangle$ as a function of the gradient flow time $t$.

We  use the standard definition of the finite volume gradient flow coupling $g_{GF}^2(t;L,\beta)$ \cite{Fodor:2012td},
\begin{align}
 \label{Eq.g2}
g^2_{GF} (t;L,\beta) = \frac{128\pi^2}{3(N^2 - 1)} \frac{1}{C(t,L/a)} \vev{t^2 E(t)},
\end{align}
where the constants in front have been chosen to match the perturbative 1-loop result in the $\overline{\textrm{MS}}$ scheme \cite{Luscher:2010iy} ($N=3$ for the SU(3) gauge group).  $C(t,L/a)$ is a perturbatively computed tree-level improvement term\footnote{Table III in the Appendix of Ref.~\cite{Hasenfratz:2019dpr} lists numerical values for $C(t,L/a)$ for $L/a\le 32$. The values for $L/a=\{40,\,48\}$ are listed in Tab.~\ref{Tab.tlnL40} in Appendix \ref{Sec.tree-level}.} \cite{Fodor:2014cpa}. When analyzing data without tree-level improvement, we replace $C(c,L/a)$ by $1/(1+\delta(t/L^2))$ to compensate for zero modes of the gauge field in periodic volumes \cite{Fodor:2012td}. 

%=================================================
\subsection{Step-scaling \texorpdfstring{$\beta$}{beta} Function}
%=================================================
When defining the finite volume step-scaling function the flow time $t$ is connected to the lattice size $L$ as 
\begin{align}
  t=(c L)^2/8.  
  \label{Eq.RenCon}
\end{align}
The parameter $c$ defines the specific finite volume renormalization scheme. For a scale change $s$, the gradient flow step-scaling $\beta$ function \cite{Fodor:2012td} is given by
\begin{align}
  \beta_{c,s}(g^2_c;L,\beta) = \frac{g^2_c(sL; \beta)- g^2_c(L; \beta)}{\log\;s^2} \,,
  \label{Eq.beta_cs}
\end{align}
where $g_c^2(L,\beta) =g^2_{GF}(t=(c L)^2/8; L,\beta)$. Since the re\-normalized coupling $g^2_c$ is defined at a bare coupling $\beta$, it is contaminated by cutoff effects. Hence an extrapolation to the infinite cutoff continuum limit is required. In the case of the step-scaling function this corresponds to taking $t/a^2 \to \infty$, or equivalently $L/a \to \infty$. At a fixed value of $g^2_c$ we thus tune the bare coupling toward the Gaussian fixed point i.e.~$g_0^2= 6/ \beta \to 0$ as $L/a$ increases. Practically simulations are performed on a limited set of lattice volumes. We compensate for that by simulating at many different values of the bare coupling $\beta$. Combining simulations at different bare coupling, we cover the investigated range of the renormalized coupling and take the $L/a\to \infty$ continuum limit of the step-scaling $\beta_{c,s}(g^2_c;L)$ function at fixed $g_c^2$. As result we obtain the continuum step-scaling $\beta$-function $\beta_{c,s}(g_c^2)$ in the renormalization scheme $c$. 

The specific steps of our analysis are as follows:
\begin{enumerate}
\item We start by calculating discrete $\beta_{c,s}(g_c^2;L)$ functions as defined in Eq.~(\ref{Eq.beta_cs}) for all volume pairs with a scale change of $s=2$. 
\item For each volume pair we next interpolate these discrete $\beta_{c,s}(g_c^2;L)$ functions using a polynomial ansatz motivated by the perturbative expansion
  \begin{align}
    \beta_{c,s}(g_c^2;L) = \sum_{i=0}^{n} b_i g_c^{2i}.
    \label{Eq.fit_form}
  \end{align}
  In practice we observe that $n=3$ is sufficient for a good description of our data over the full range in $g_c^2$ covered by our simulations.  When using the tree-level normalization (tln), discretization effects at weak coupling are sufficiently small, hence we constrain the intercept $b_0=0$. We fit $b_0$ however when analyzing data without tln. After the interpolation, we have finite volume discrete step-scaling functions $\beta_{c,s}(g_c^2; L)$ at continuous values of $g_c^2$.
\item We take the infinite volume continuum limit by extrapolating the interpolated $\beta_{c,s}(g_c^2; L)$ functions at fixed $g^2_c$ values. To check for consistency, we explore different choices for the fit ansatz. In particular we perform a quadratic fit to all volume pairs and a linear fit to the largest three volume pairs. 
\item The continuum result $\beta_{c,s}(g_c^2)$ should be free of discretization effects but may depend on the renormalization scheme $c$ and the choice of the scale change $s$. Further we need to check for possible systematic effects due to the choice of gradient flow/operators or the use of tln.
\end{enumerate}

%=================================================
\subsection{Data Analysis}
%=================================================

To distinguish different flow and operator combinations in our analysis, we introduce the shorthand notation [flow][operator] (indicated by the first capital letters) and prefix ``n'' when including the  tree-level improvement term $C(c,L/a)$ in our analysis. As we will detail below, our preferred analysis uses $O(a^2)$ improved combination of Zeuthen flow with Symanzik operator to which we refer as ``ZS'' without tree-level improvement and ``nZS'' with tree-level improvement. The statistical data analysis presented in the following sections is performed using the $\Gamma$-method \cite{Wolff:2003sm} which estimates and accounts for autocorrelations. Table \ref{Tab.Nf6_nZS_ZS} and \ref{Tab.Nf4_nZS_ZS} list the renormalized couplings at our three chosen renormalization schemes $c=0.300$, 0.275 and 0.250 together with the estimated autocorrelations times for our preferred analysis for $N_f=6$ and 4, respectively. In the following, we present our analysis for $c=0.300$ but for completeness show plots corresponding to $c=0.275$ and $0.250$ in the Appendix \ref{Sec.c0275_c0250}.

\begin{figure*}[t]
  \begin{minipage}{0.49\textwidth}
   \flushright 
   \includegraphics[width=0.96\textwidth]{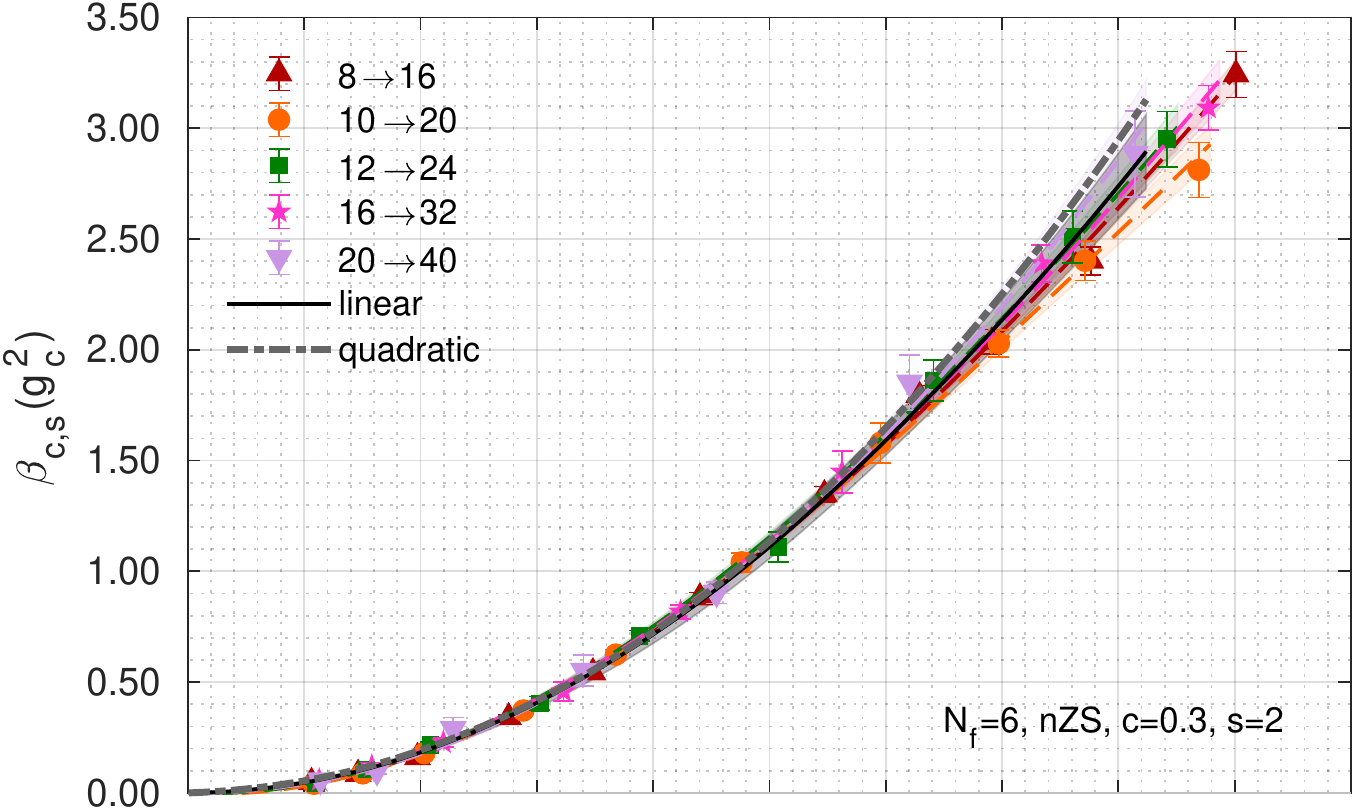}\\
   \includegraphics[width=0.937\textwidth]{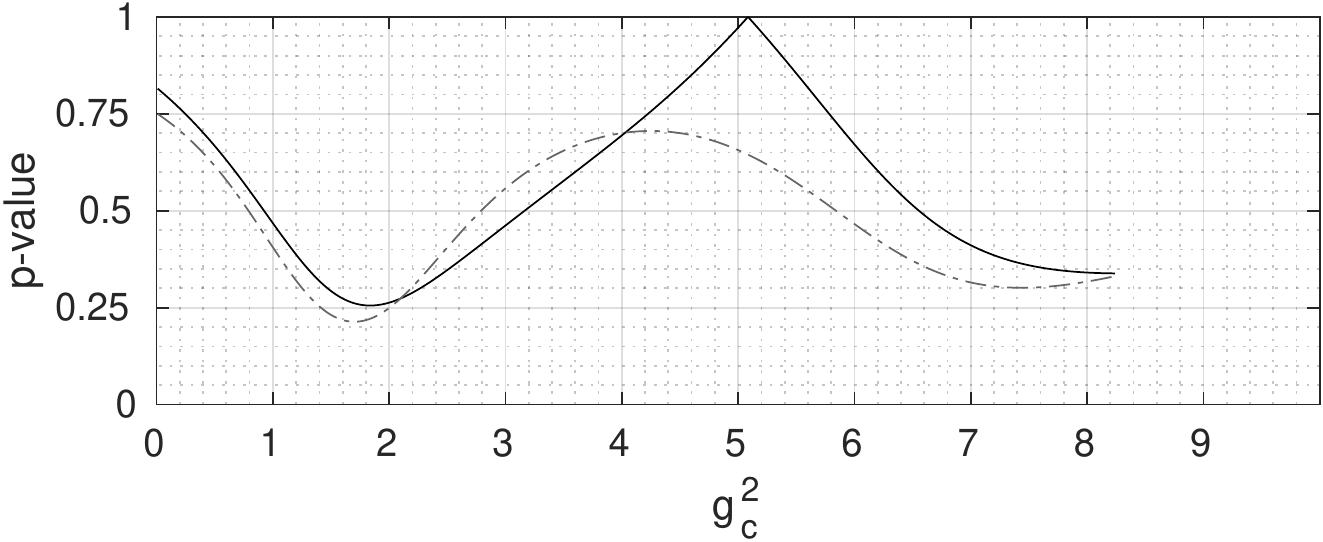} %\\[3mm]
   \includegraphics[width=0.96\textwidth]{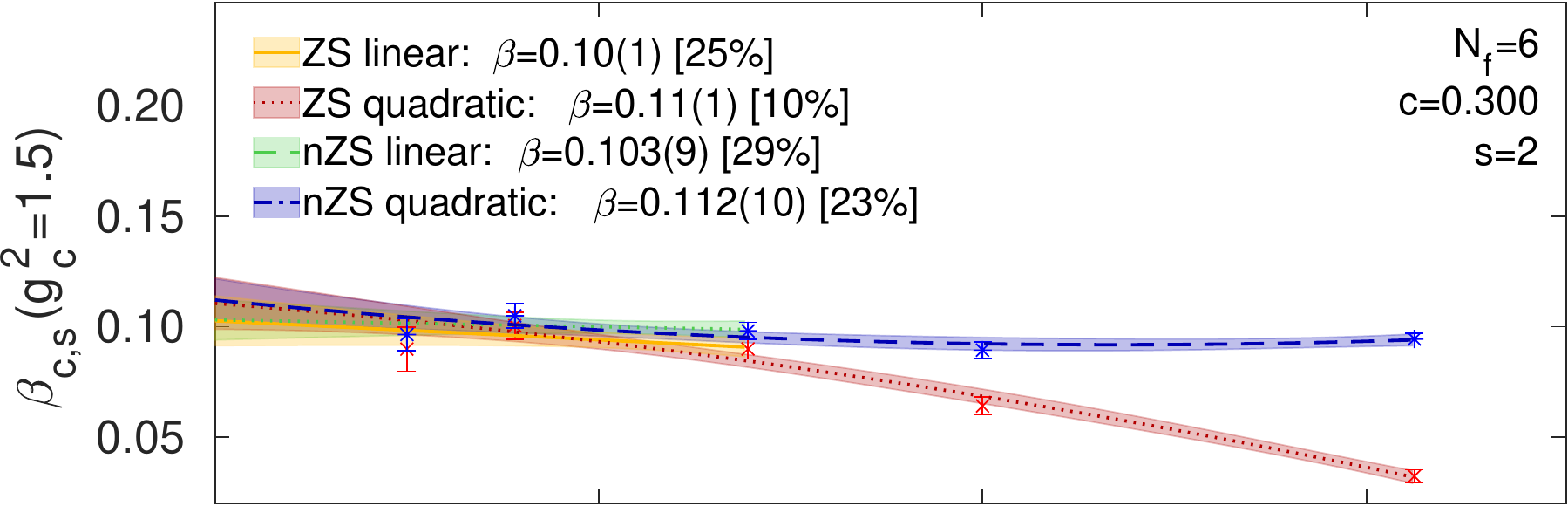}\\
   \includegraphics[width=0.96\textwidth]{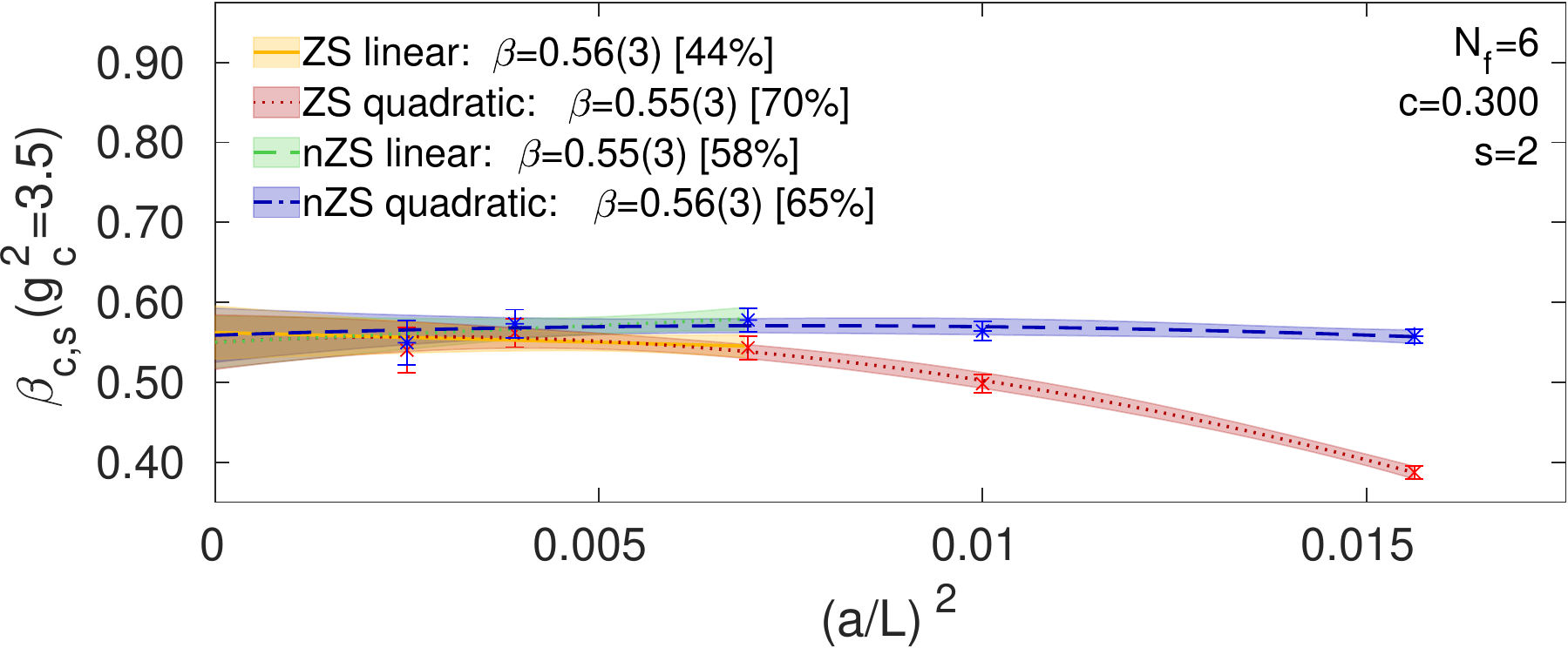}
  \end{minipage}
  \begin{minipage}{0.49\textwidth}
    \flushright
    \includegraphics[width=0.96\textwidth]{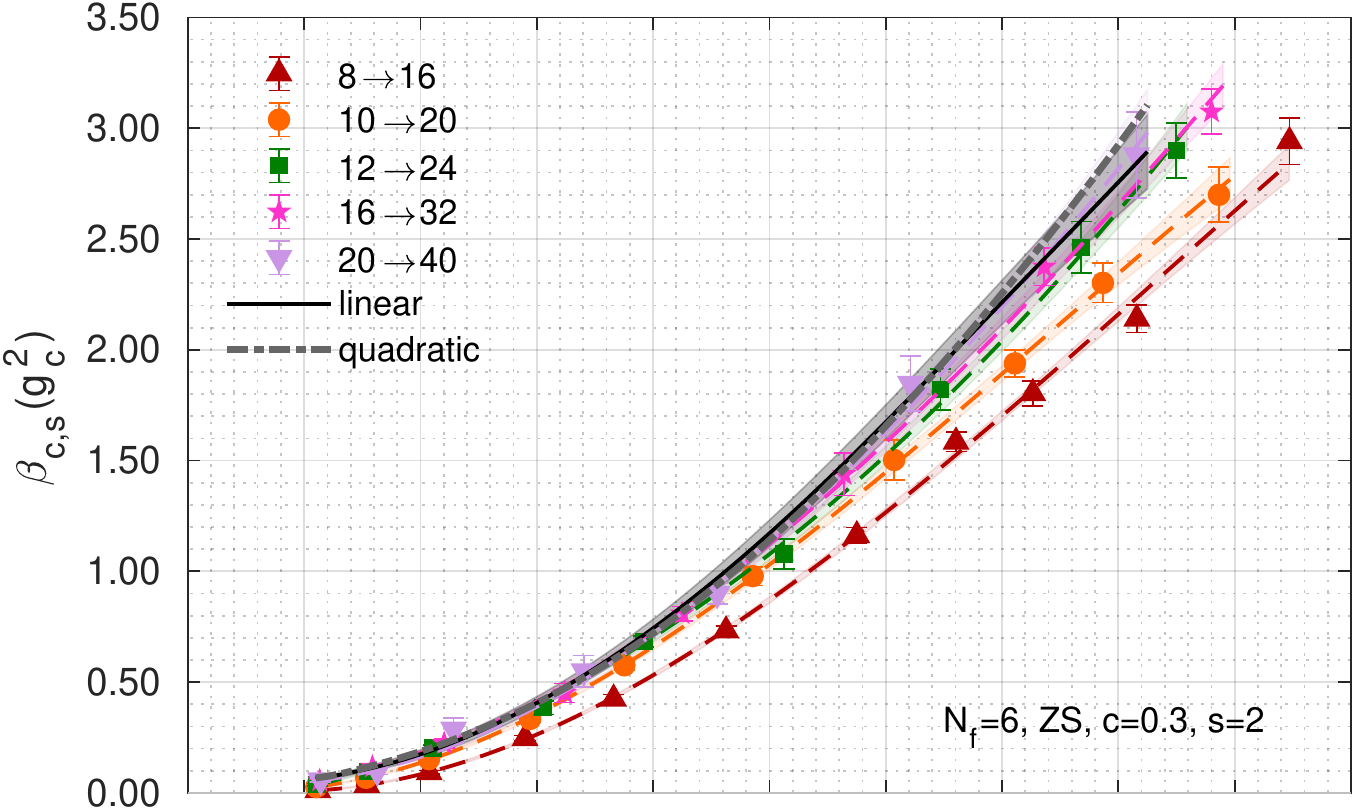}\\    
    \includegraphics[width=0.937\textwidth]{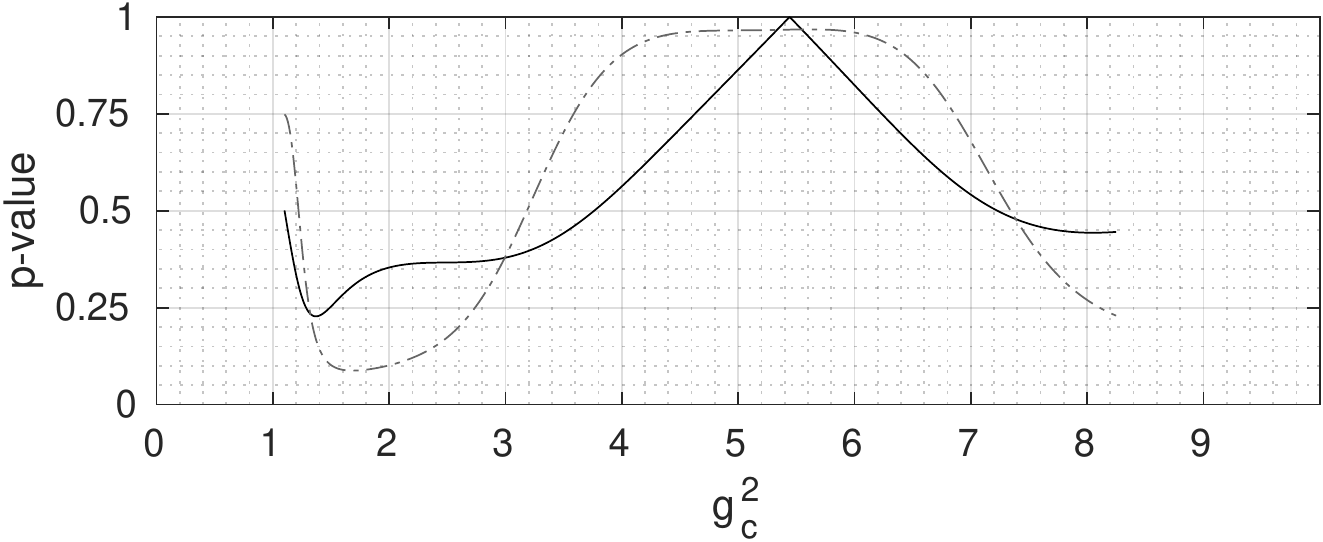} %\\[3mm]
    \includegraphics[width=0.96\textwidth]{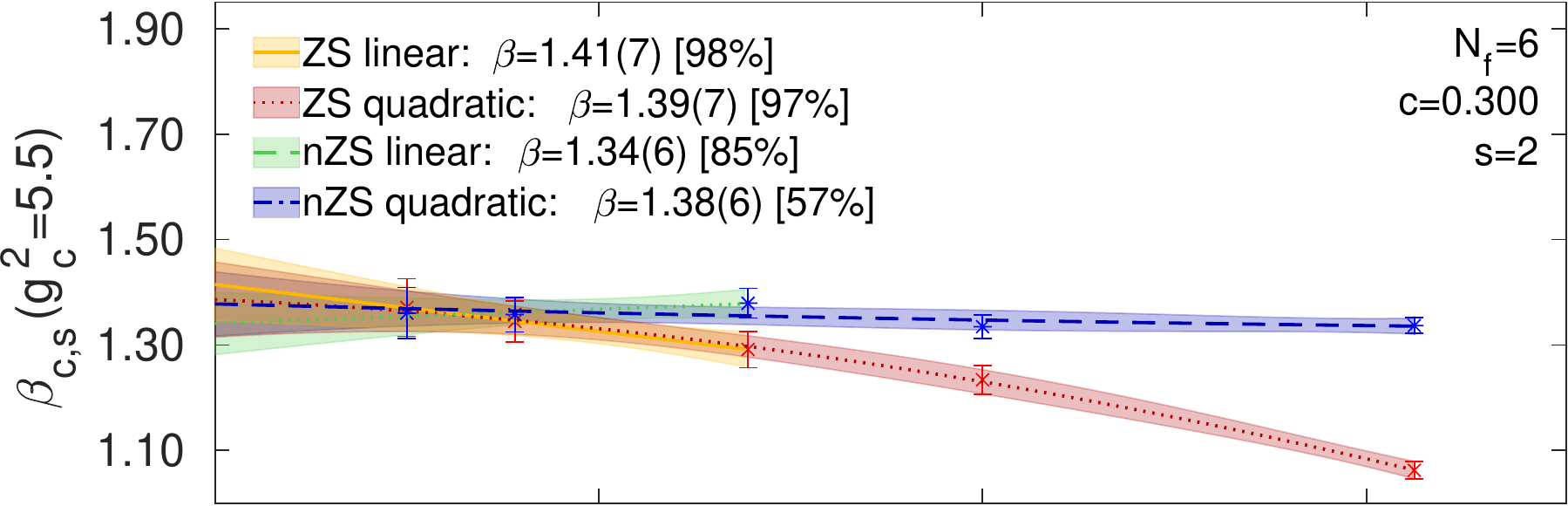}\\
    \includegraphics[width=0.96\textwidth]{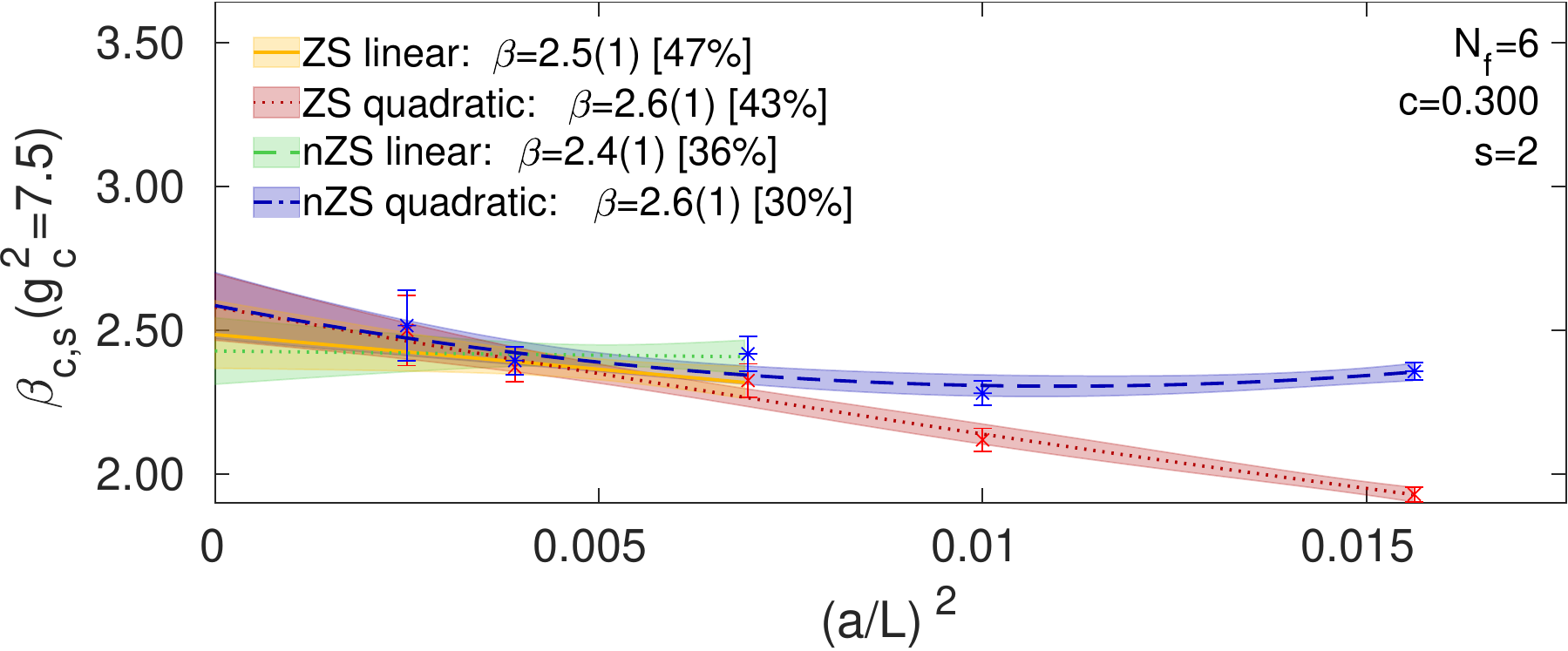}
  \end{minipage}
  \caption{Discrete step-scaling $\beta$-function for $N_f=6$ in the $c=0.300$ gradient flow scheme for our preferred nZS (left) and ZS (right) data sets. The symbols in the top row show our results for the finite volume discrete $\beta$ function with scale change $s=2$. The dashed lines with shaded error bands in the same color of the data points show the interpolating fits. We consider two continuum limits: a linear fit (black line with gray error band) in $a^2/L^2$ to the three largest volume pairs and a quadratic fit to all volume pairs (black dash-dotted line). The $p$-values of the continuum extrapolation fits are shown in the plots in the second row. Further details of the continuum extrapolation at selected $g_c^2$ values are presented in the small panels at the bottom where the legend lists the extrapolated values in the continuum limit with $p$-values in brackets. Only statistical errors are shown.  }  
  \label{Fig.Nf6_beta_c300}
\end{figure*}

%=================================================
\section{SU(3) with six flavors} \label{Sec.Nf6}
%=================================================

Starting from the renormalized couplings shown in Table \ref{Tab.Nf6_nZS_ZS} we calculate the discrete step-scaling function, $\beta_{c,s}(g_c^2; L,\beta)$, (Eq.~(\ref{Eq.beta_cs})) for our five volume pairs with scale change $s=2$ and renormalization scheme $c=0.3$: $8\to 16$, $10\to 20$, $12\to 24$, $16\to 32$, and $20\to 40$. These quantities for Zeuthen flow and Symanzik operator (ZS) are shown in the top row panels of Fig.~\ref{Fig.Nf6_beta_c300} as colored symbols. Plots on the left show our analysis for the tree-level improved combination, whereas plots on the right show the same flow-operator combination without tree-level improvement. 
Deviations between the various volume pairs indicate cutoff effects. Even for our $O(a^2)$ improved ZS combination, the tree-level improvement further reduces discretization effects resulting in data for different volume pairs to sit on top of each other. Quite remarkably, within our statistical errors, volume dependence is observable only for  $g^2_{GF}\gtrsim 6$ when considering the nZS data set. Similar analysis for schemes $c=0.275$ and 0.250 is included in Appendix \ref{Sec.c0275_c0250}.

Next we perform a polynomial interpolation of these data points for any given volume pair to obtain the discrete step-scaling function for continuous values of the renormalized coupling $g_c^2$. As can be seen by the results for the interpolating fits presented in Tab.~\ref{Tab.interpolationsNf6}, a polynomial of degree three is sufficient to describe our data well. As mentioned above, we constrain the intercept to be zero in case of tree-level improved operators because discretization effects are sufficiently small. The corresponding interpolation curve is shown in the top row plots of Fig.~\ref{Fig.Nf6_beta_c300} by the shaded band in the same color as the data points.

The final step is the continuum limit extrapolation. Taking the results of the previous interpolation, we have data at continuous values of $g_c^2$ for five different volume pairs. Here we consider two ans\"atze for the fit:
\begin{itemize}
\item  ``linear'' refers to the extrapolation of the three largest volume pairs $12\to 24$, $16\to 32$, and $20\to 40$ using a linear ansatz in $(a/L)^2$. The resulting continuum limit values are shown by the solid black line with gray error band in the top row plots of Fig.~\ref{Fig.Nf6_beta_c300} and the corresponding $p$-values as a function of $g_c^2$  are presented in the second row plots below.
\item ``quadratic'' uses all five volume pairs and the fit is performed with a quadratic ansatz in $(a/L)^2$. The central value of the resulting continuum limit as well as the corresponding $p$-values are shown by a black dash-dotted line.
\end{itemize}
The excellent agreement between linear and quadratic extrapolation fits for both nZS and ZS analyses for all three renormalization schemes $c$ is apparent. The predicted continuum limits fall into the combined 1-sigma error bands. Further details on the continuum limit extrapolation are demonstrated in the four lowest panels of Fig.~\ref{Fig.Nf6_beta_c300} where we show the detailed continuum extrapolation at four selected values of $g_c^2$ across the range where we have performed simulations. We would like to emphasize that these extrapolations for our $N_f=6$ data set have good $p$-values over the entire range in $g_c^2$ for both analyses and all three $c$-values. At all three $c$-values we also observe a clear improvement due to using the tree-level normalization. 

\begin{figure*}[p]
  \includegraphics[height=0.298\textheight]{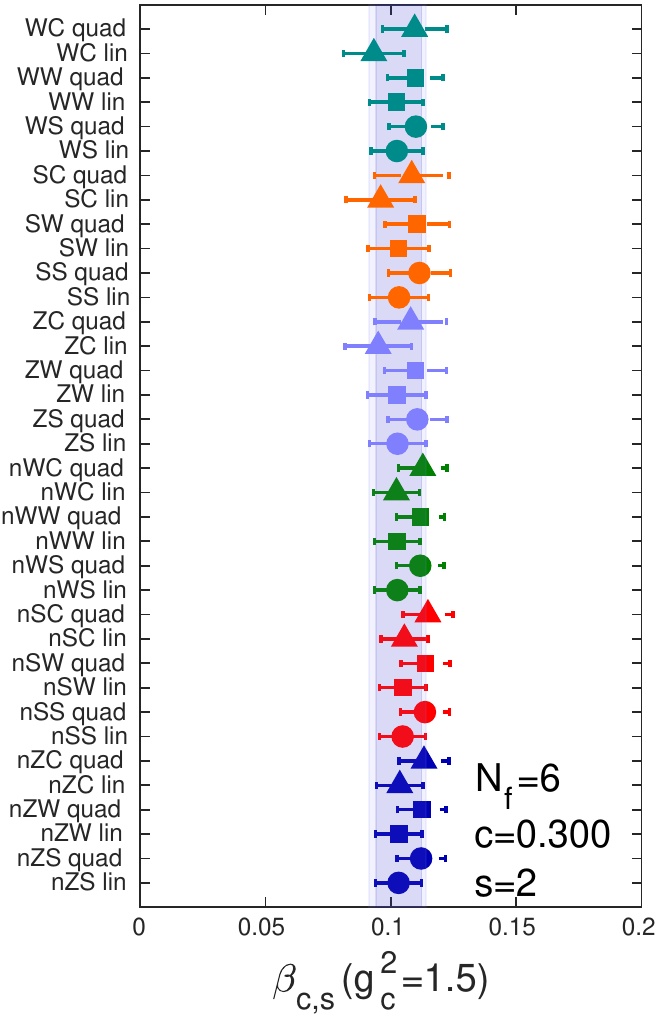}
  \includegraphics[height=0.298\textheight]{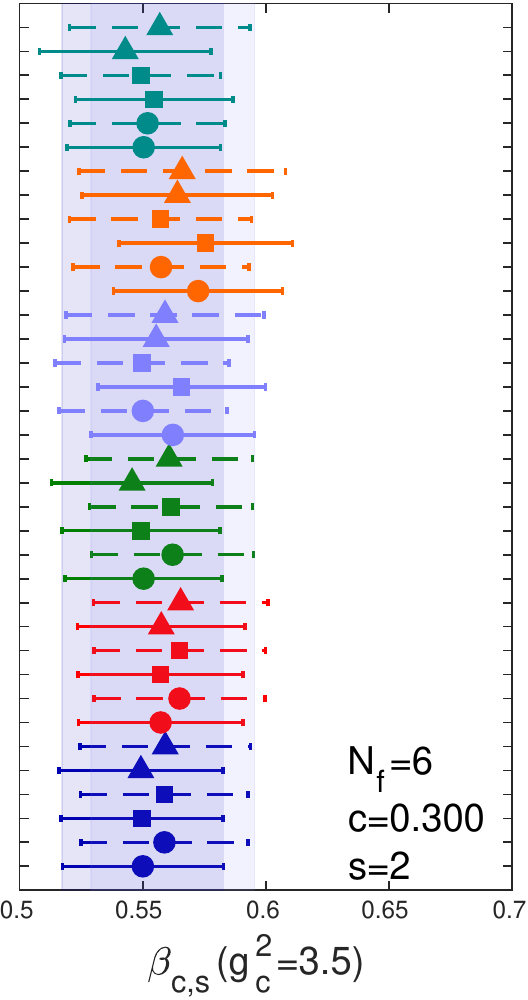}
  \includegraphics[height=0.298\textheight]{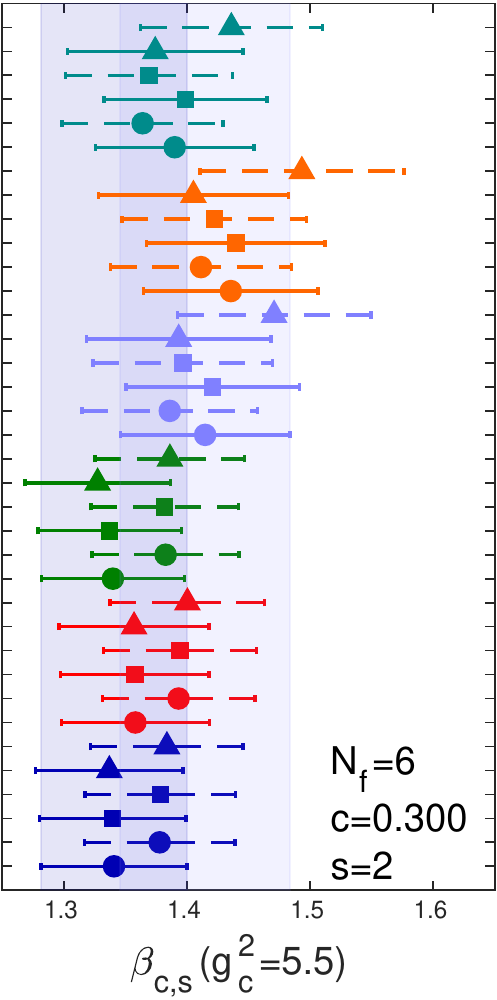}
  \includegraphics[height=0.298\textheight]{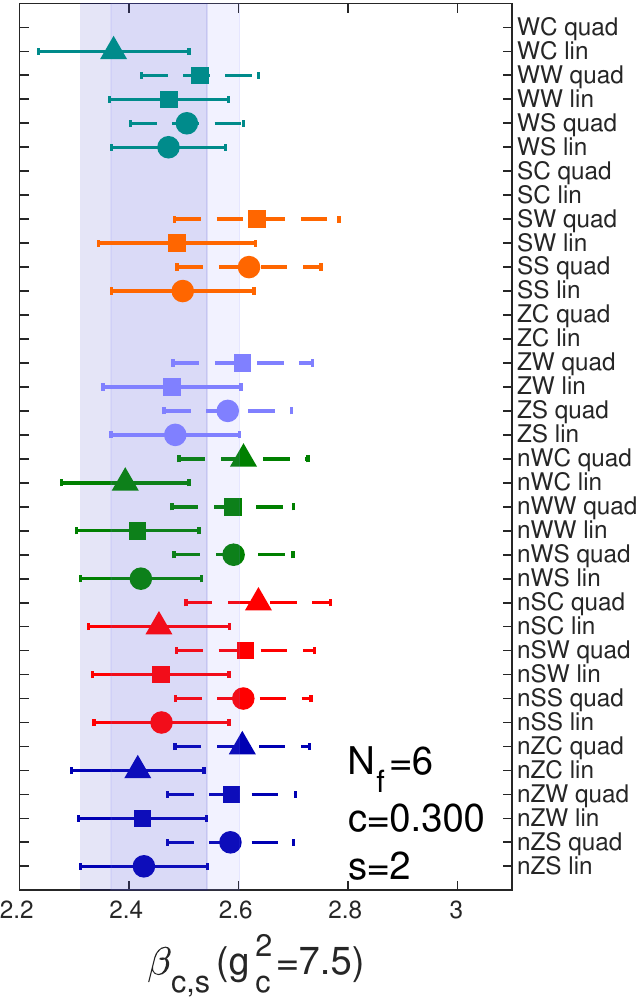}\\  
  \includegraphics[height=0.298\textheight]{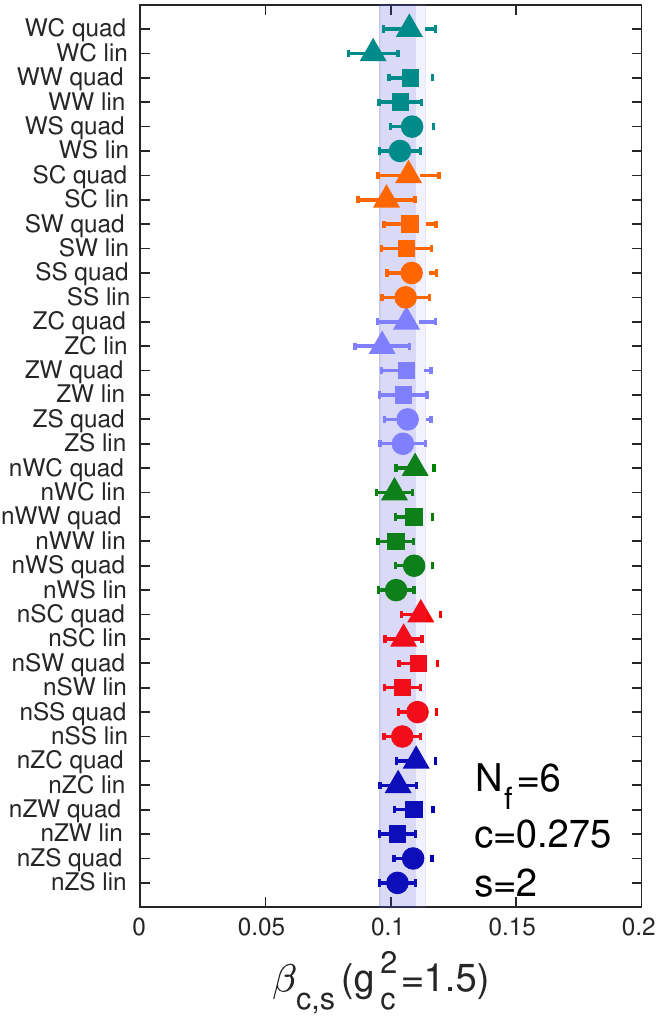}
  \includegraphics[height=0.298\textheight]{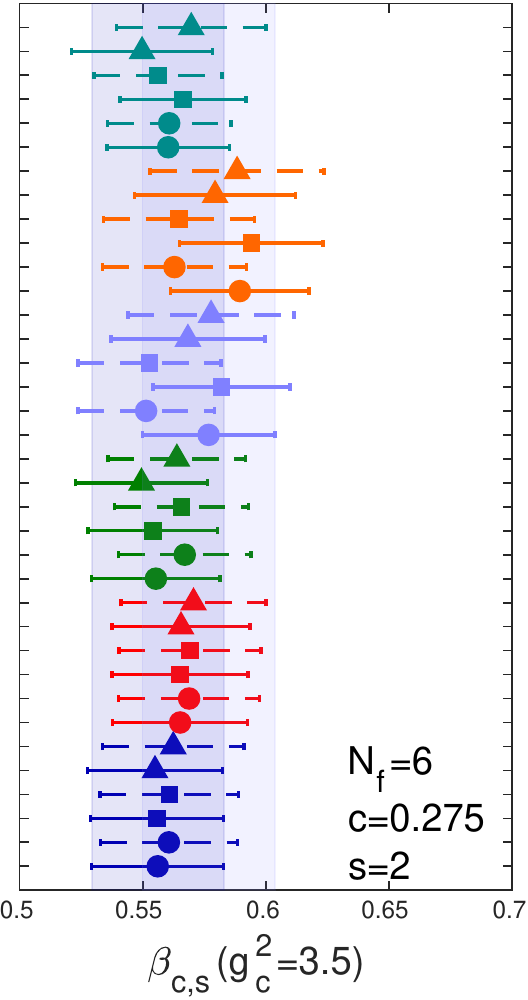}
  \includegraphics[height=0.298\textheight]{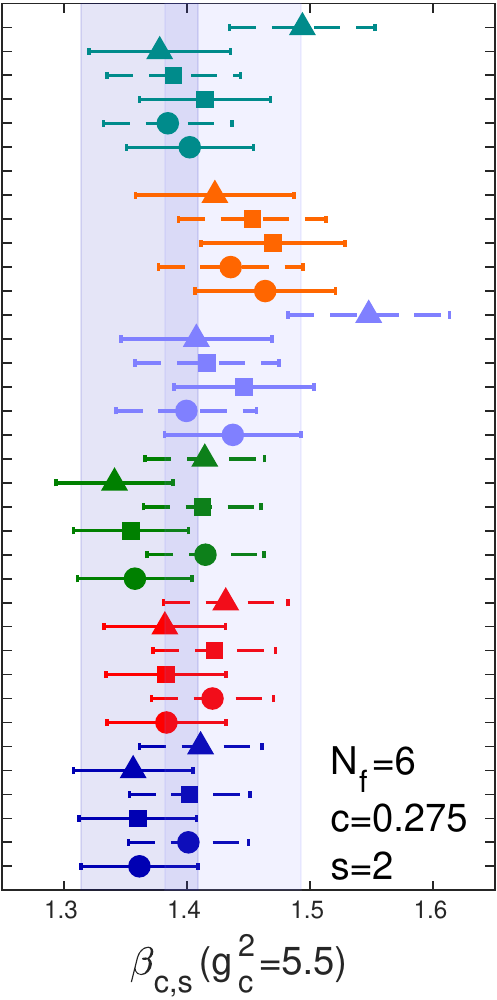}
  \includegraphics[height=0.298\textheight]{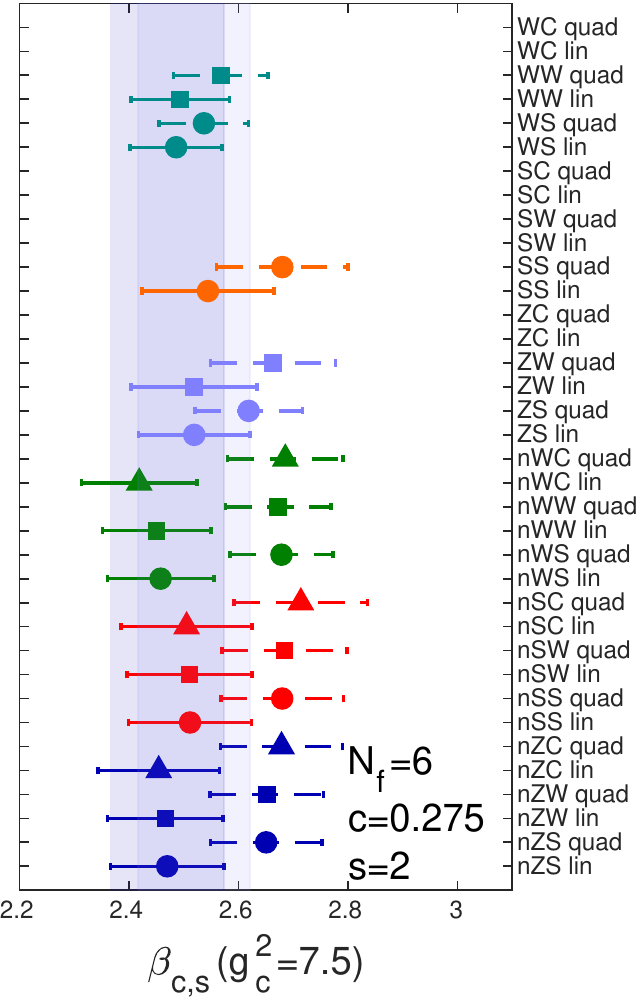}\\
  \includegraphics[height=0.298\textheight]{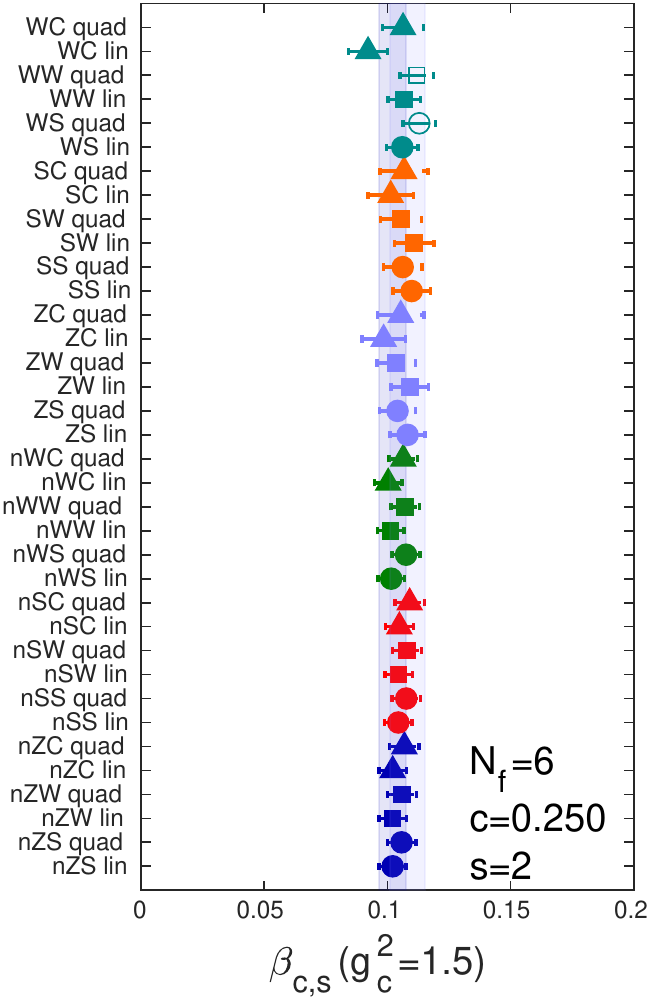}
  \includegraphics[height=0.298\textheight]{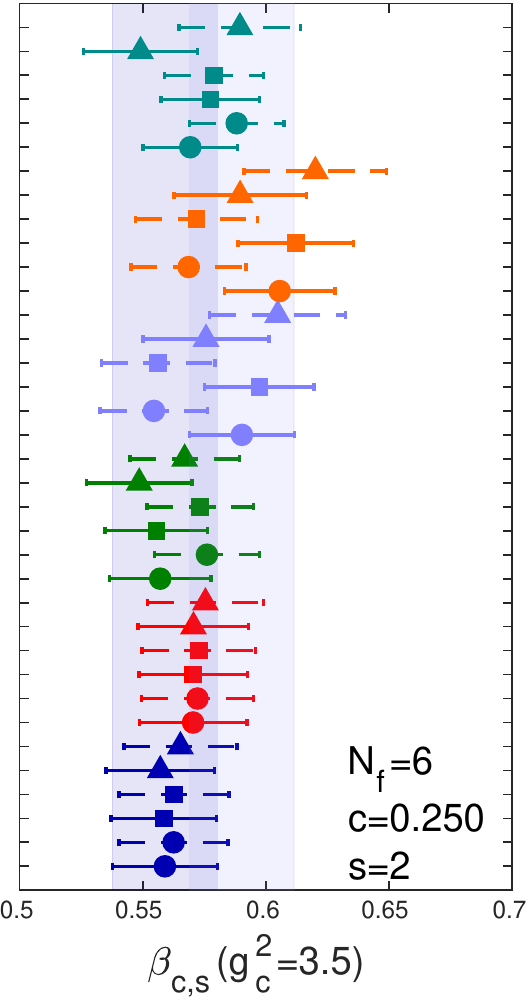}
  \includegraphics[height=0.298\textheight]{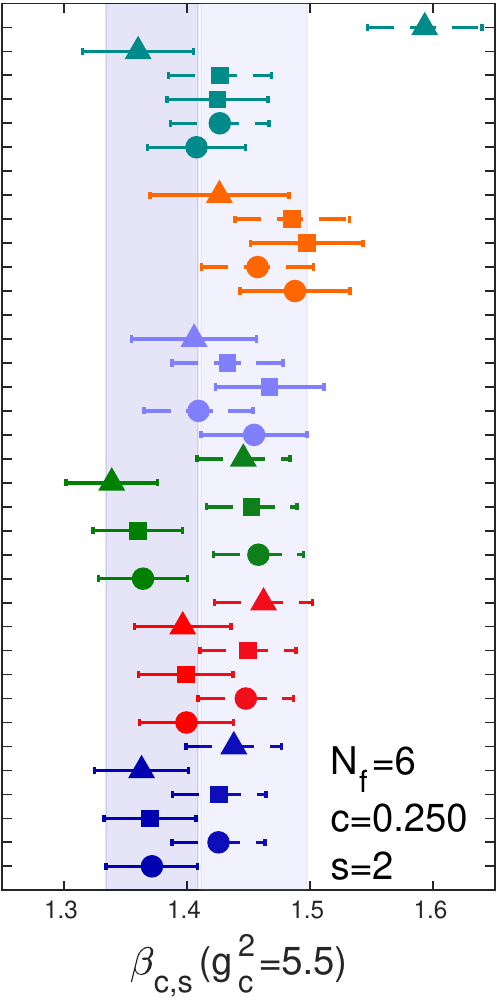}
  \includegraphics[height=0.298\textheight]{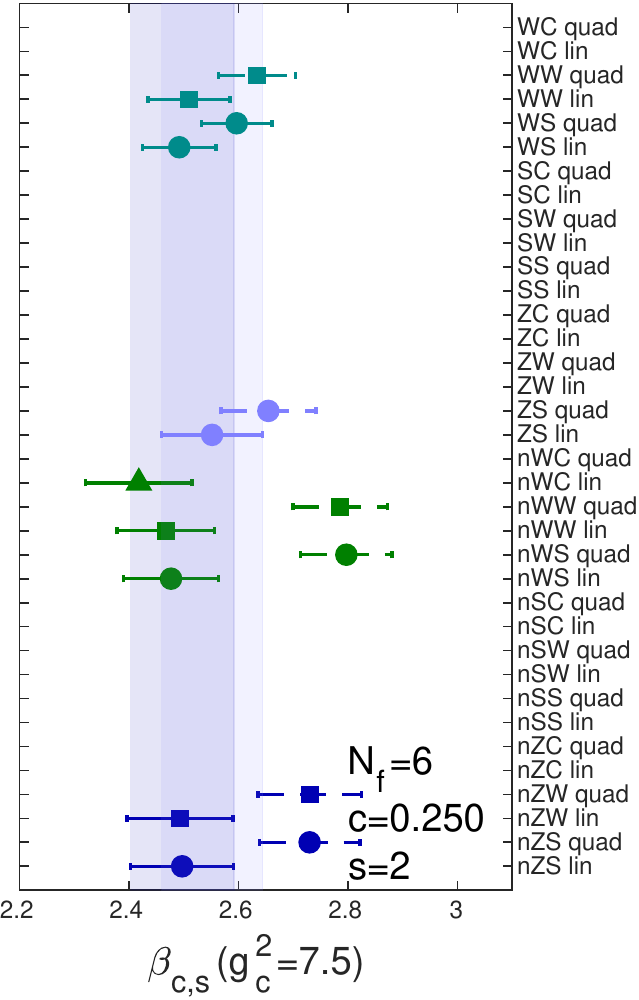}  
  \caption{Systematic effects on the $N_f=6$ results for $\beta_{c,s}(g_c^2)$ due to tree-level improvement and different flows and operators. In all cases we obtain the continuum limit considering a linear extrapolation to the three largest volume pairs and a quadratic extrapolation to all volume pairs. The columns show our continuum limit results at selective $g_c^2 = 1.5$, 3.5, 5.5, and 7.5; the rows correspond to renormalization schemes $c=0.300$, 0.275, 0.250. Open symbols indicate extrapolations with a $p$-value below 5\%. The vertical shaded bands highlight our preferred (n)ZS analysis.}
  \label{Fig.Nf6_beta_sys}
\end{figure*}

We repeat our analysis using the additional flow/operator combinations we have investigated in order to check for possible systematic effects. In total we have performed 18 different analyses considering Zeuthen, Wilson, and Symanzik gradient flow and each time determining the energy density $E(t)$ using the Wilson, Symanzik, and clover operator. Further we carry out each analysis with and without the use of the tree-level normalization. Using again four representative values of $g_c^2=1.5$, 3.5, 5.5, and 7.5 we show in Fig.~\ref{Fig.Nf6_beta_sys} comparison plots for the continuum limit values obtained for these 18 different analyses each time performing a linear extrapolation to our three largest volume pairs and a quadratic extrapolation to all five volume pairs. The plots in the top row show comparisons for $c=0.300$, in the middle row for $c=0.275$ and in the bottom row for $c=0.250$, whereas the columns align plots with $g_c^2=1.5$, 3.5, 5.5, and 7.5. In each plot our preferred analyses, linear continuum extrapolation for nZS and ZS, are highlighted by the shaded blue bands.  Alternative analysis based on Zeuthen flow are shown with blue symbols, whereas we use green symbols for Wilson flow and red symbols for Symanzik flow. As we have also observed in our determinations of the step-scaling $\beta$ function for SU(3) with $N_f=10$ \cite{Hasenfratz:2020ess} and $N_f=12$ \cite{Hasenfratz:2017qyr,Hasenfratz:2019dpr}, the reach in $g_c^2$ depends on the flow-operator combination. In particular when using the clover operator only a shorter range in $g_c^2$ is covered. This explains ``missing'' data points for some analysis in the $g_c^2=7.5$ panels.

Looking at renormalization schemes $c=0.300$ and $c=0.275$ we observe that {\it all}\/ analyses have overlapping error bars with the ones we prefer (nZS and ZS) highlighted by the blue bands. In the case of $c=0.250$ we count in total three outliers and note that only less than half of our analysis reach $g_c^2=7.5$. This suggests $g_c^2=7.5$ is a bit too large for $c=0.250$ and that systematic effects on our data set slightly increase for decreasing $c$ value.

In the following we use the envelope covering both of our preferred nZS and ZS analysis to quote our final result accounting in addition of statistical uncertainties also for systematic effects. These are shown in Fig.~\ref{Fig.FinalNf6}, further discussion is, however, postponed to Sec.~\ref{Sec.Discussion}.

%=================================================
\section{SU(3) with four flavors} \label{Sec.Nf4}
%=================================================
\begin{figure*}[t]
  \begin{minipage}{0.49\textwidth}
   \flushright 
   \includegraphics[width=0.96\textwidth]{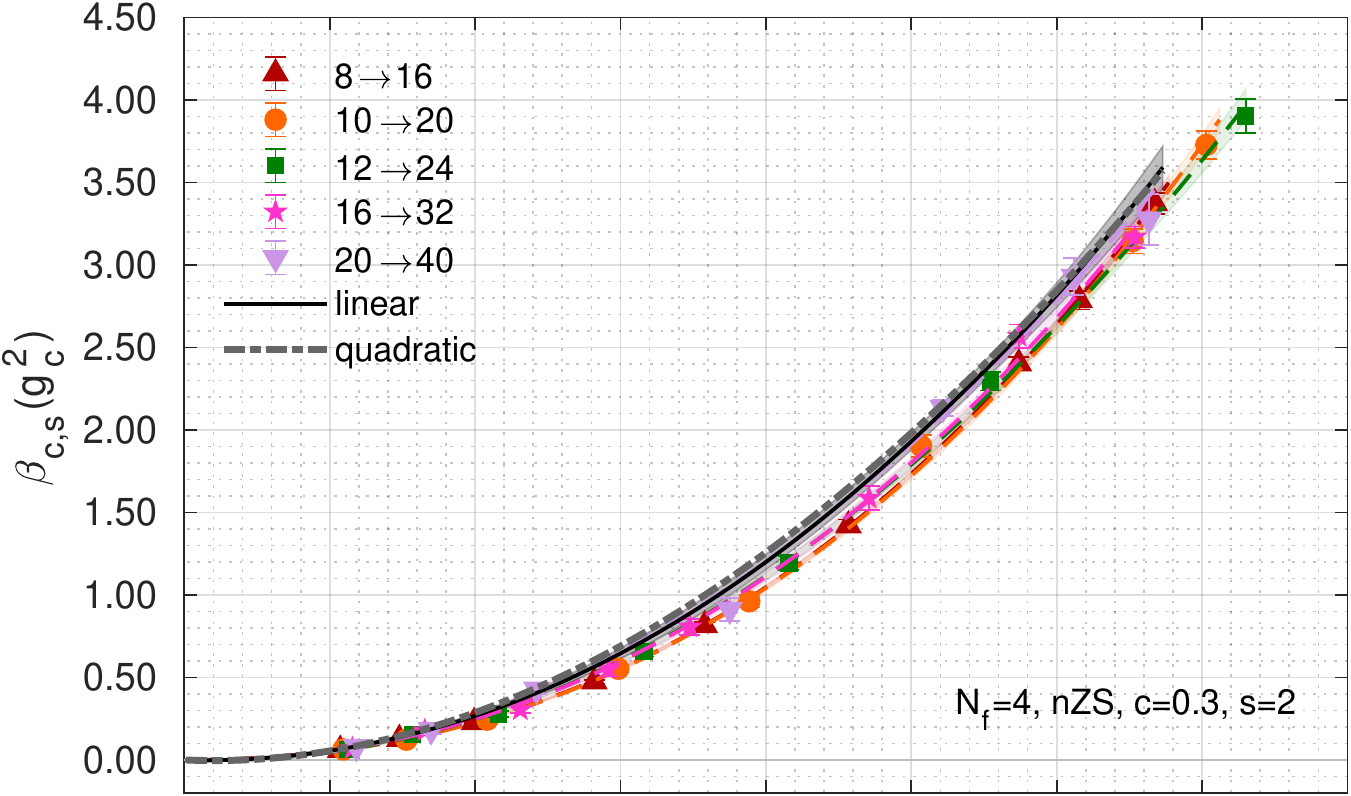}\\
   \includegraphics[width=0.937\textwidth]{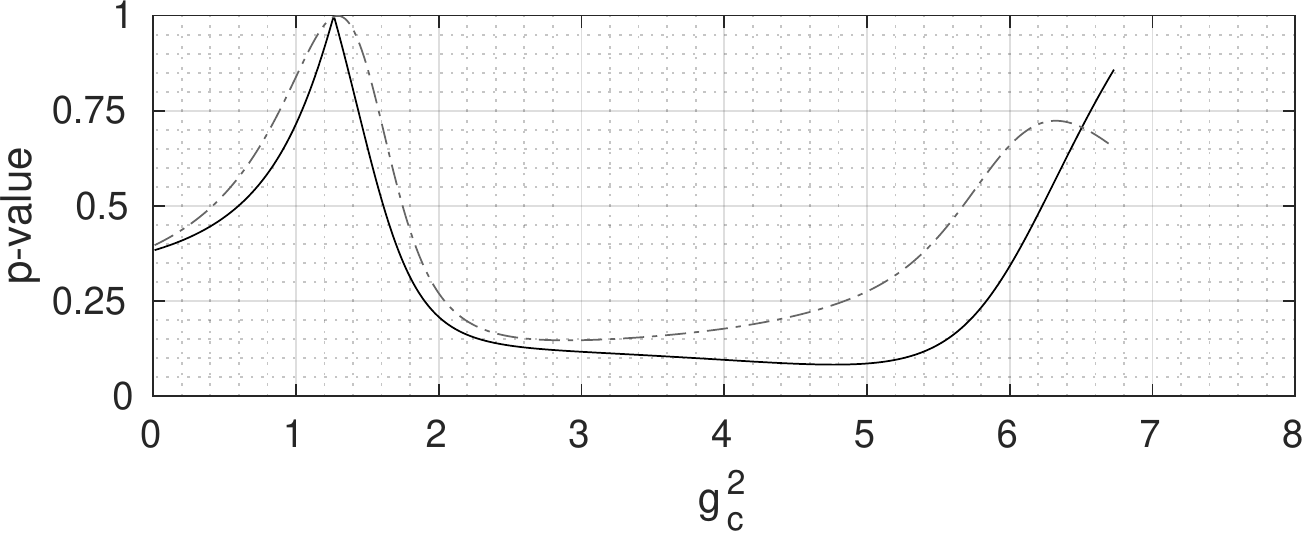} %\\[3mm]
   \includegraphics[width=0.96\textwidth]{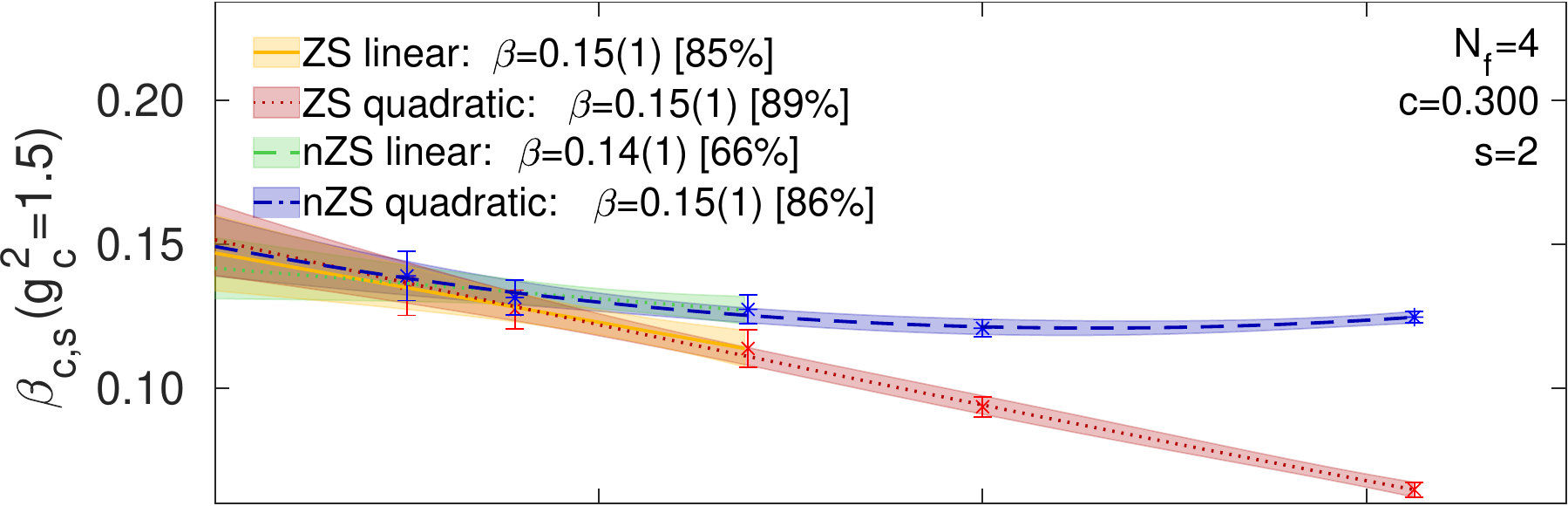}\\
   \includegraphics[width=0.96\textwidth]{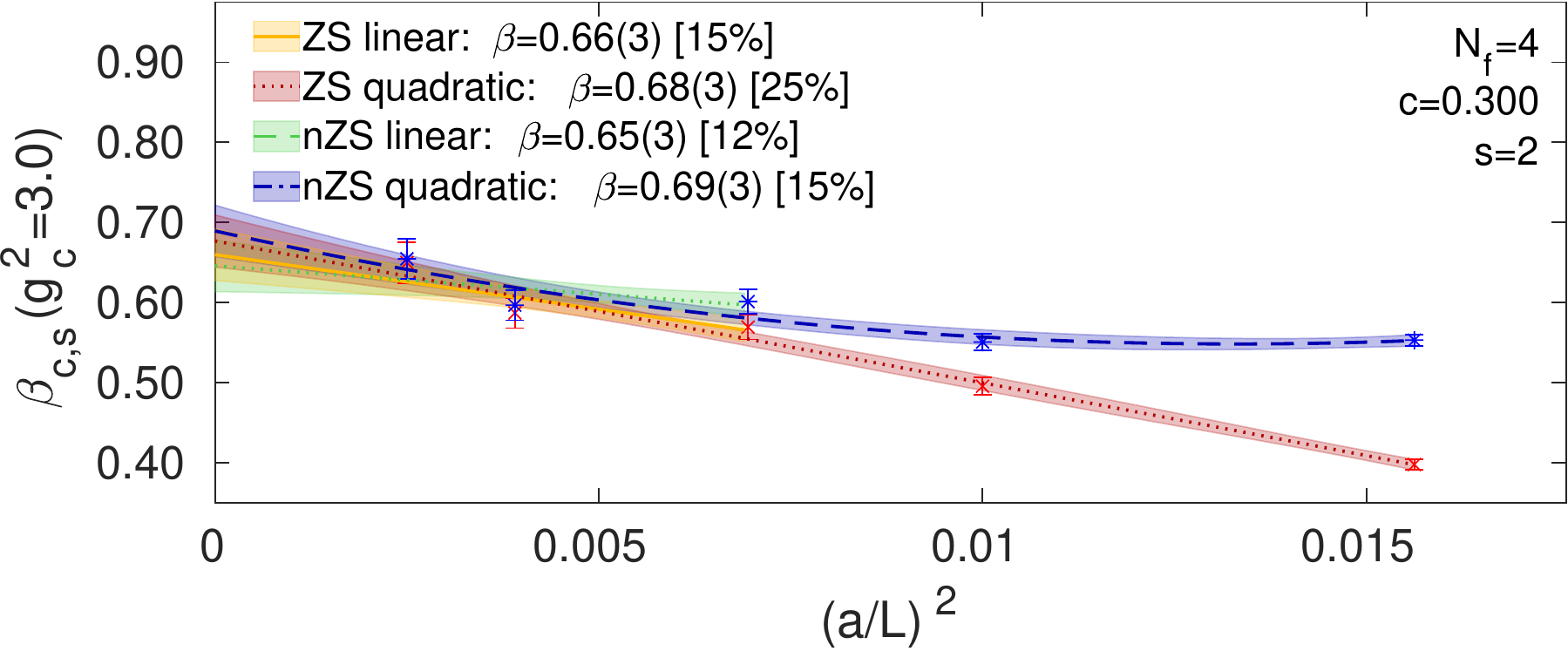}
  \end{minipage}
  \begin{minipage}{0.49\textwidth}
    \flushright
    \includegraphics[width=0.96\textwidth]{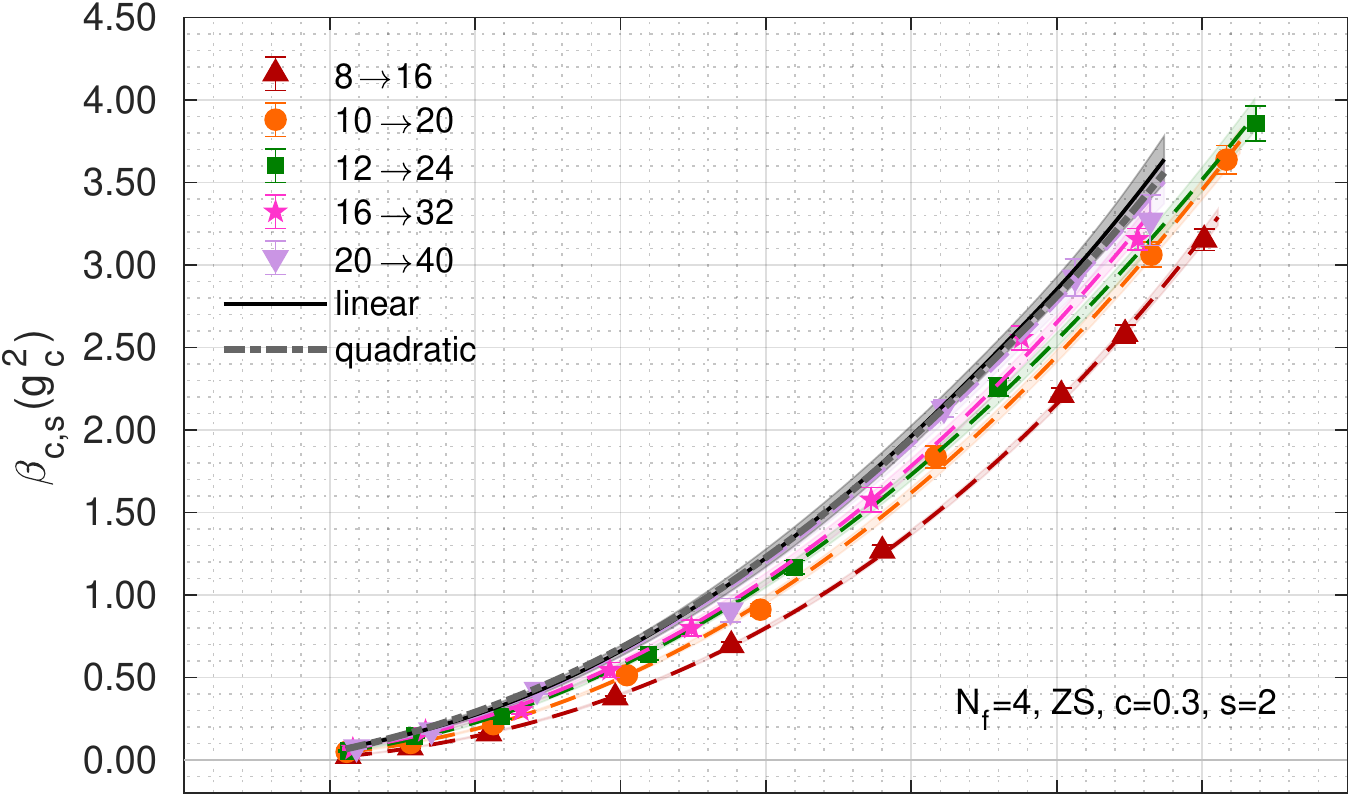}\\    
    \includegraphics[width=0.937\textwidth]{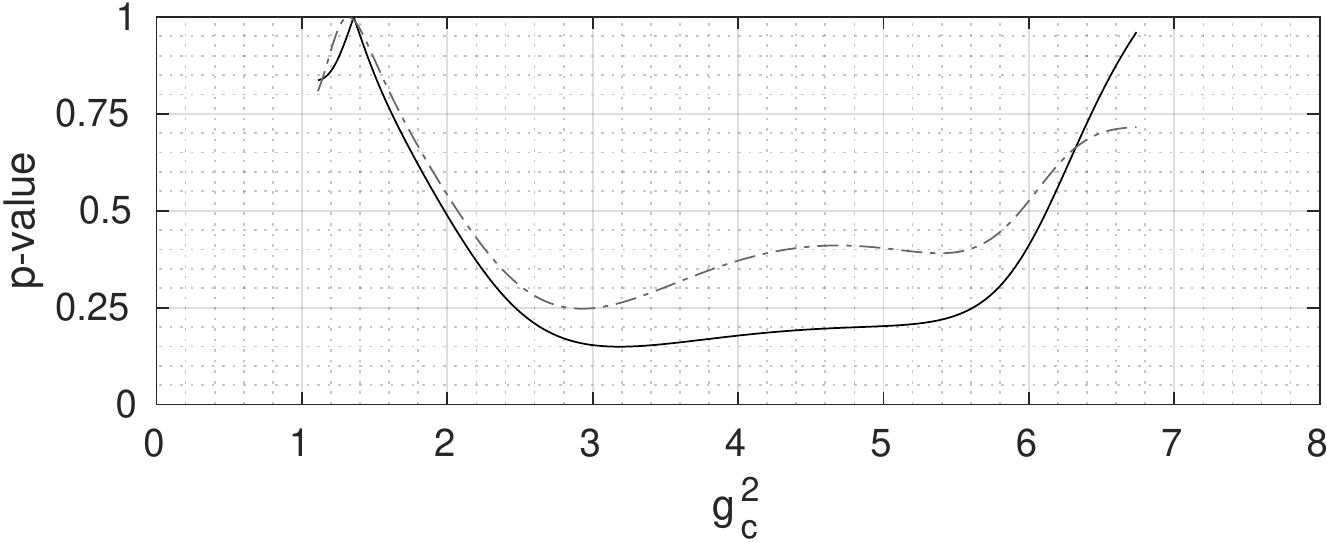} %\\[3mm]
    \includegraphics[width=0.96\textwidth]{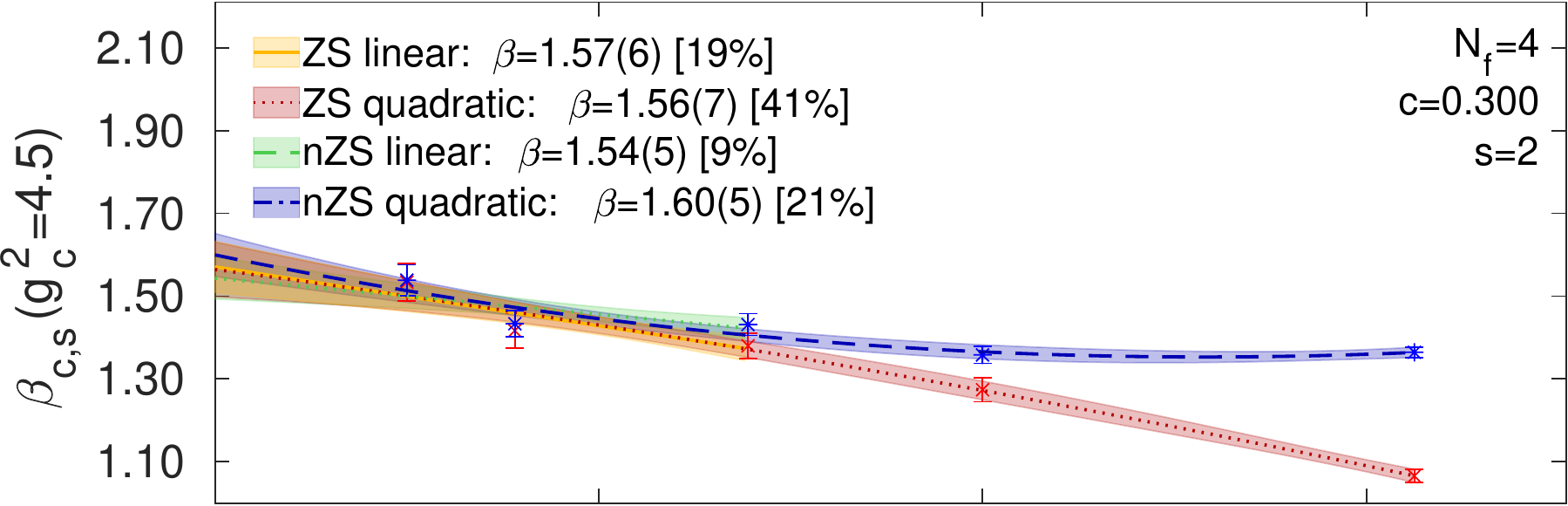}\\
    \includegraphics[width=0.96\textwidth]{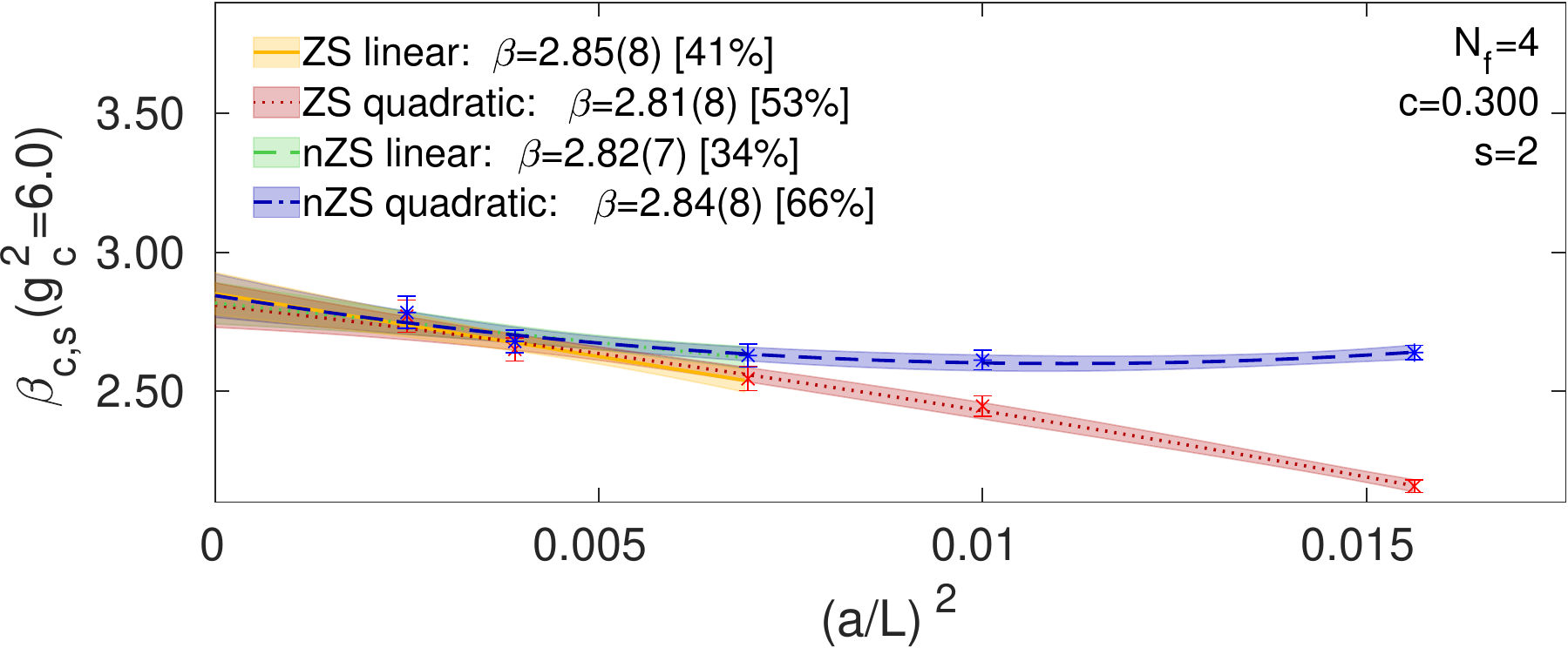}
  \end{minipage}
  \caption{Discrete step-scaling $\beta$-function for $N_f=4$ in the $c=0.300$ gradient flow scheme for our preferred nZS (left) and ZS (right) data sets. The symbols in the top row show our results for the finite volume discrete $\beta$ function with scale change $s=2$. The dashed lines with shaded error bands in the same color of the data points show the interpolating fits. We consider two continuum limits: a linear fit (black line with gray error band) in $a^2/L^2$ to the three largest volume pairs and a quadratic fit to all volume pairs (black dash-dotted line). The $p$-values of the continuum extrapolation fits are shown in the plots in the second row. Further details of the continuum extrapolation at selected $g_c^2$ values are presented in the small panels at the bottom where the legend lists the extrapolated values in the continuum limit with $p$-values in brackets. Only statistical errors are shown.}
  \label{Fig.Nf4_beta_c300}
\end{figure*}

Our analysis for $N_f=4$ proceeds following the same steps as for $N_f=6$. We list the renormalized couplings for our preferred (n)ZS analyses in Tab.~\ref{Tab.Nf4_nZS_ZS}. These values are the input to determine the discrete step-scaling function $\beta_{c,s}$ shown by the colored symbols in Fig.~\ref{Fig.Nf4_beta_c300} for the renormalization scheme $c=0.300$ and in Appendix \ref{Sec.c0275_c0250} for schemes $c=0.275$ and 0.250. To interpolate theses data points in $g_c^2$ we again use a third order polynomial and constrain the intercept to vanish when using tln. The outcome of these interpolation fits are summarized in Tab.~\ref{Tab.interpolationsNf4}. Subsequently, we perform the continuum limit extrapolation using the interpolating curves for all five volume pairs which are continuous in $g_c^2$. To check for systematic effects, we again consider a linear fit in $a^2/L^2$ to the three largest volume pairs as well as a quadratic fit in $a^2/L^2$ using all five volume pairs. Using the same convention as for $N_f=6$, the linear (quadratic) fits are shown by a solid black line with gray error band (a dash-dotted black line) in Figs.~\ref{Fig.Nf4_beta_c300}. The overall quality ($p$-value) of the fits is very good and the resulting continuum limits are very close to each other and fall mostly within the $1\sigma$ statistical error band.

\begin{figure*}[p]
  \includegraphics[height=0.298\textheight]{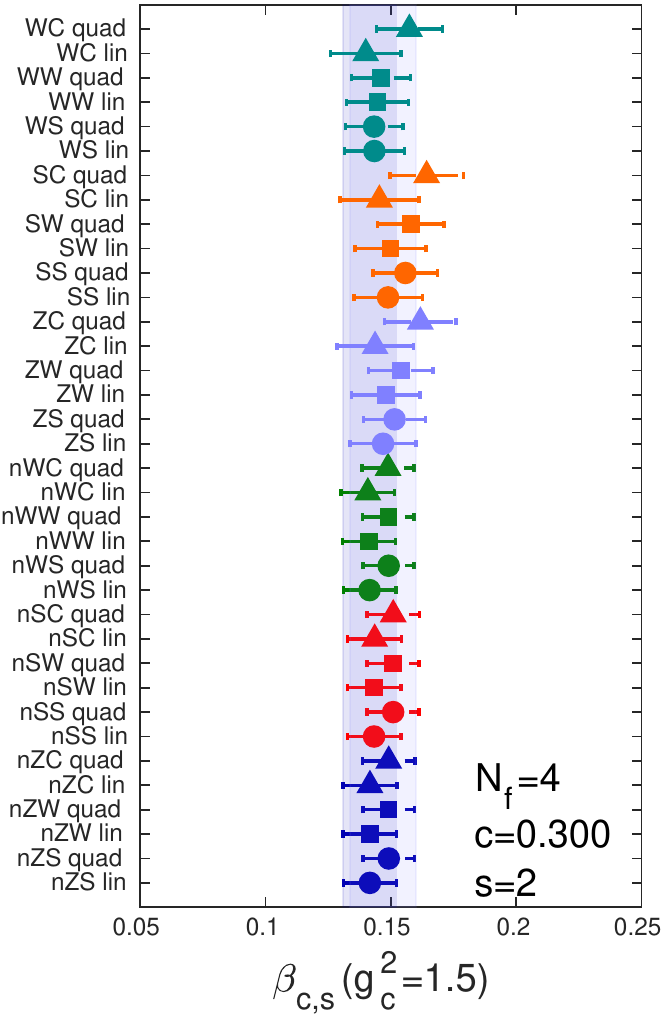}
  \includegraphics[height=0.298\textheight]{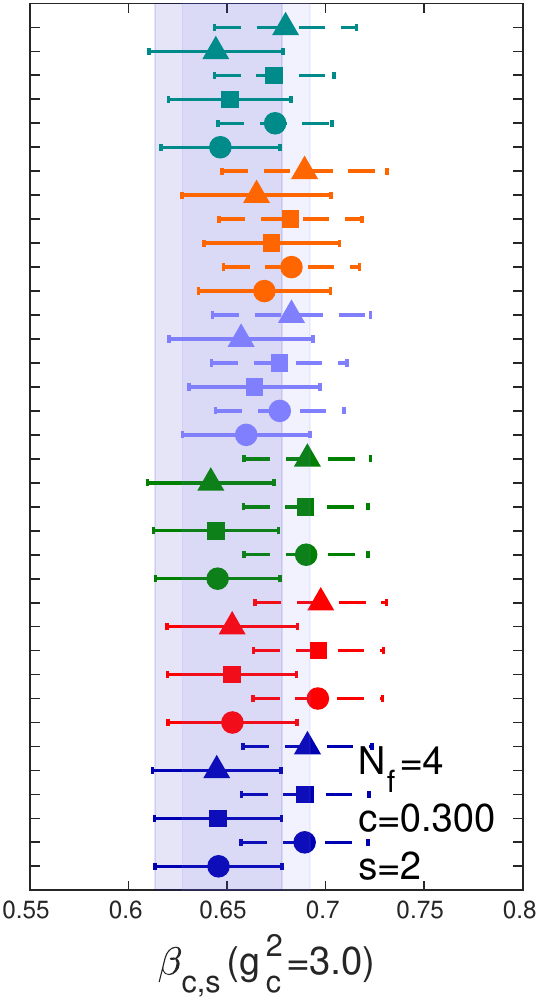}
  \includegraphics[height=0.298\textheight]{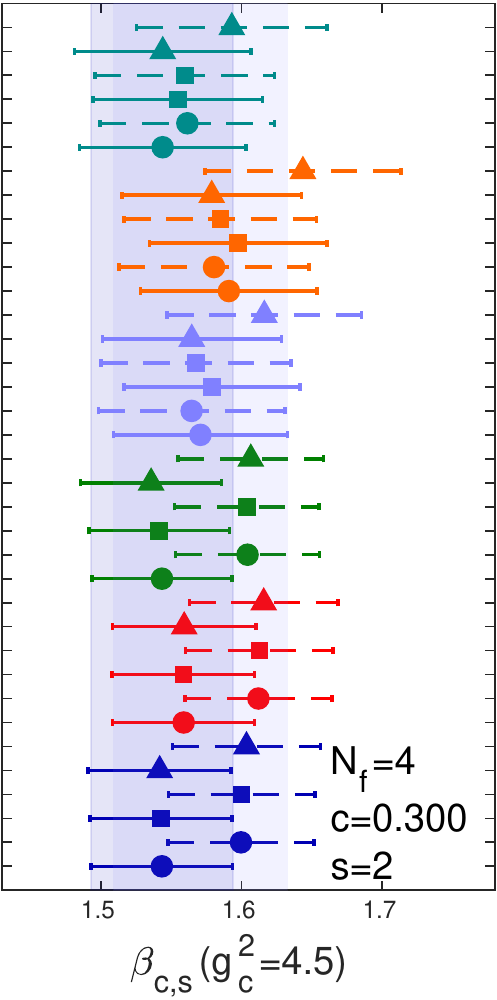}
  \includegraphics[height=0.298\textheight]{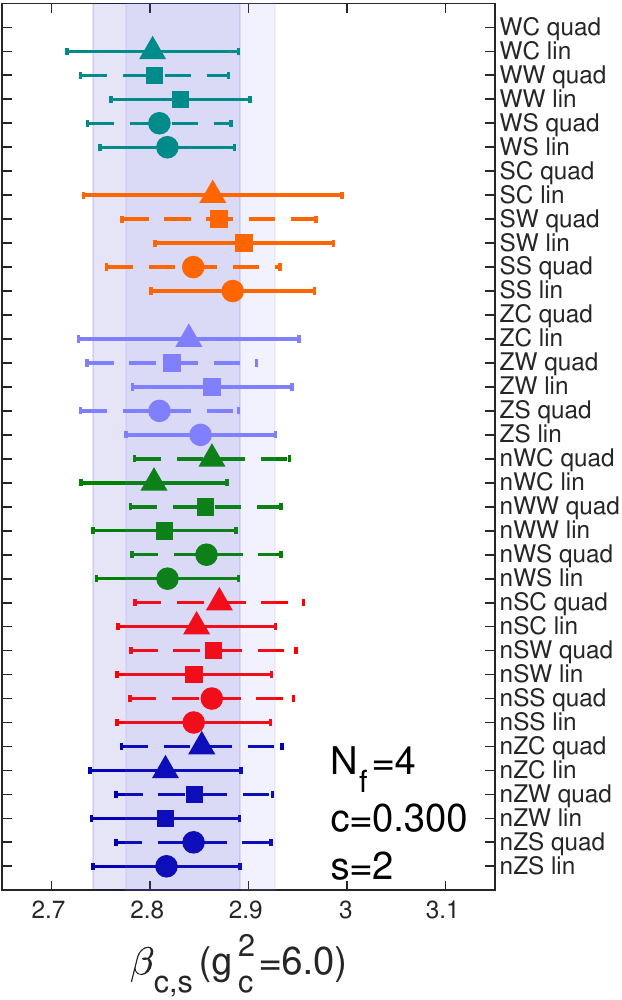}\\  
  \includegraphics[height=0.298\textheight]{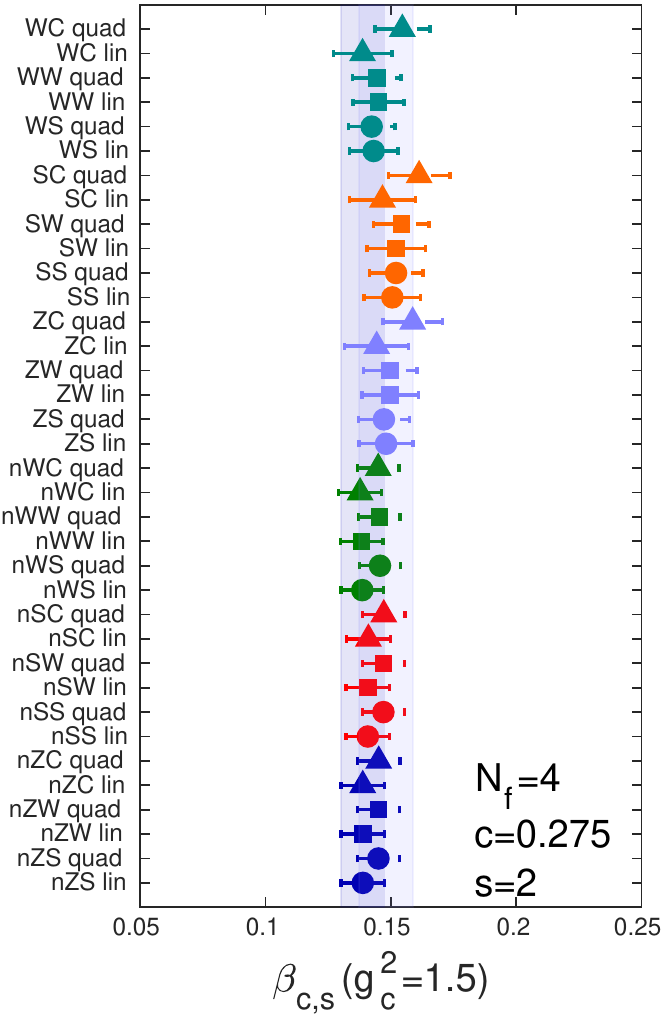}
  \includegraphics[height=0.298\textheight]{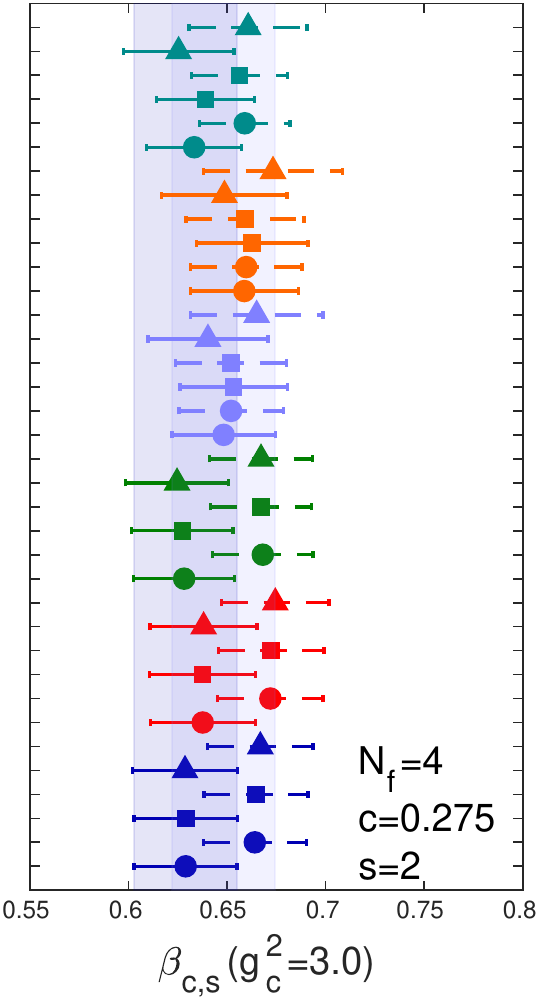}
  \includegraphics[height=0.298\textheight]{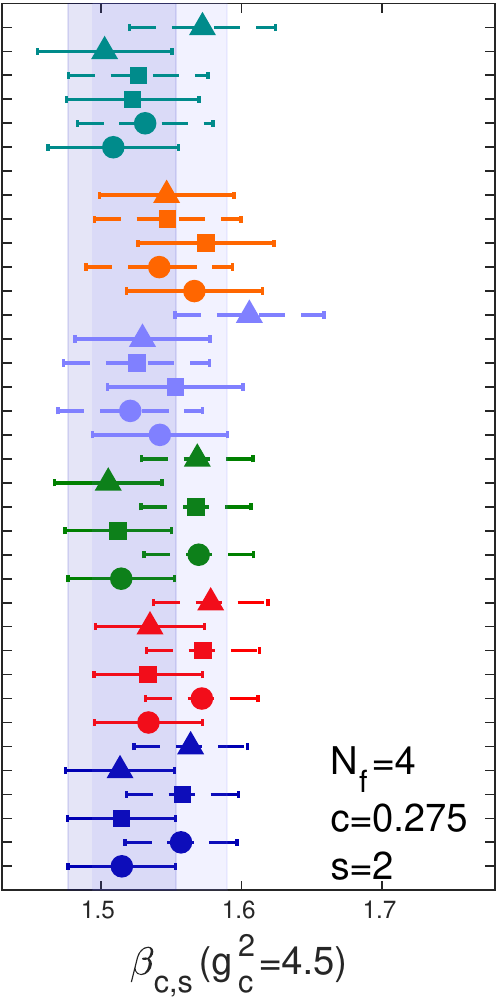}
  \includegraphics[height=0.298\textheight]{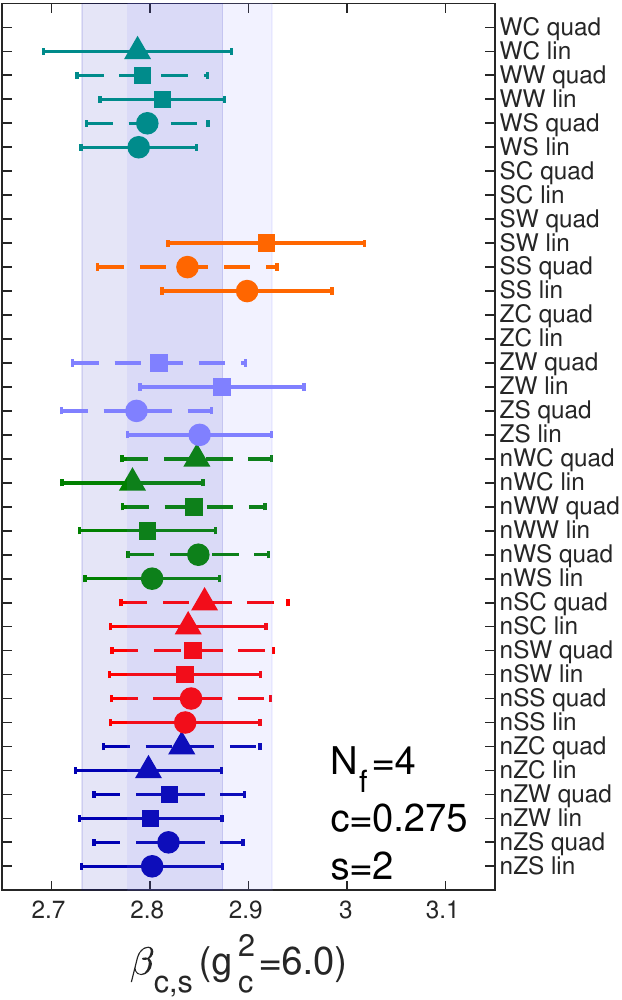}\\
  \includegraphics[height=0.298\textheight]{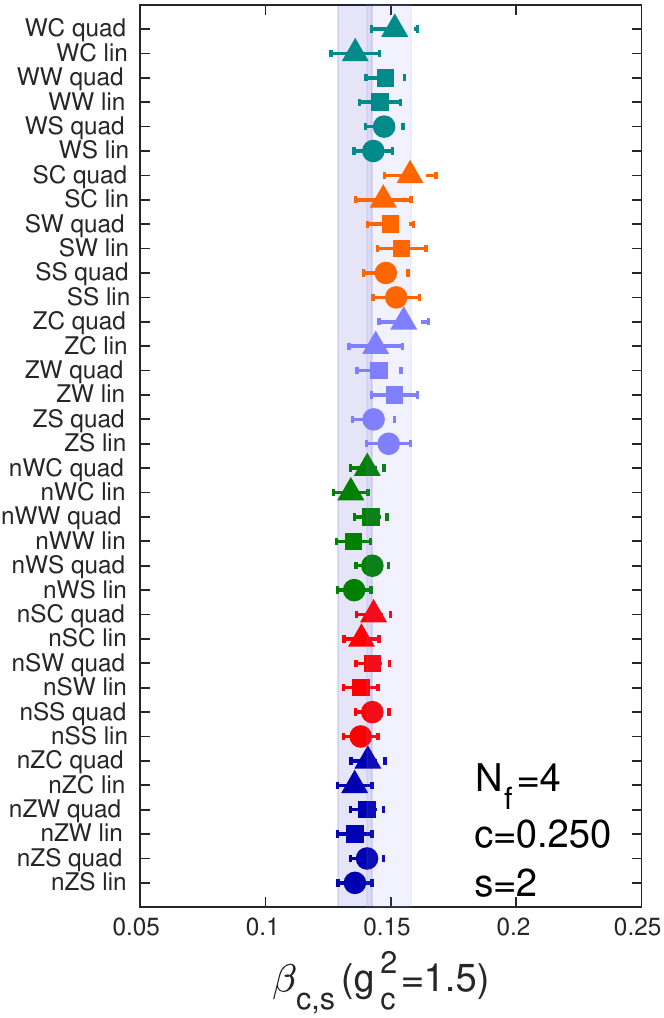}
  \includegraphics[height=0.298\textheight]{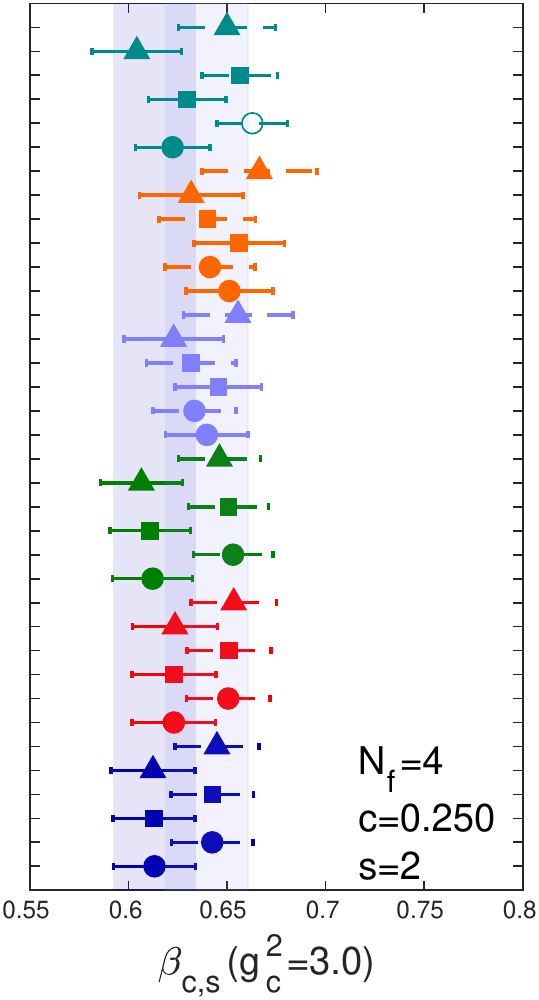}
  \includegraphics[height=0.298\textheight]{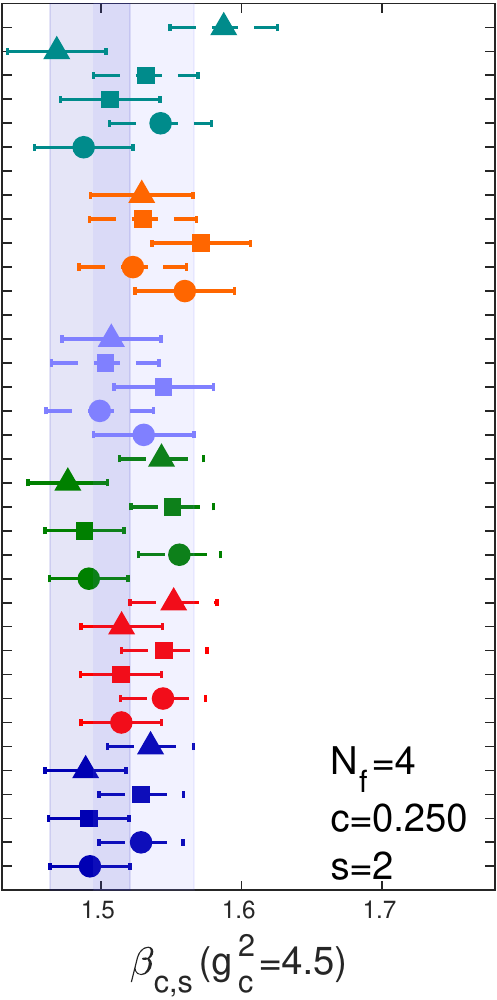}
  \includegraphics[height=0.298\textheight]{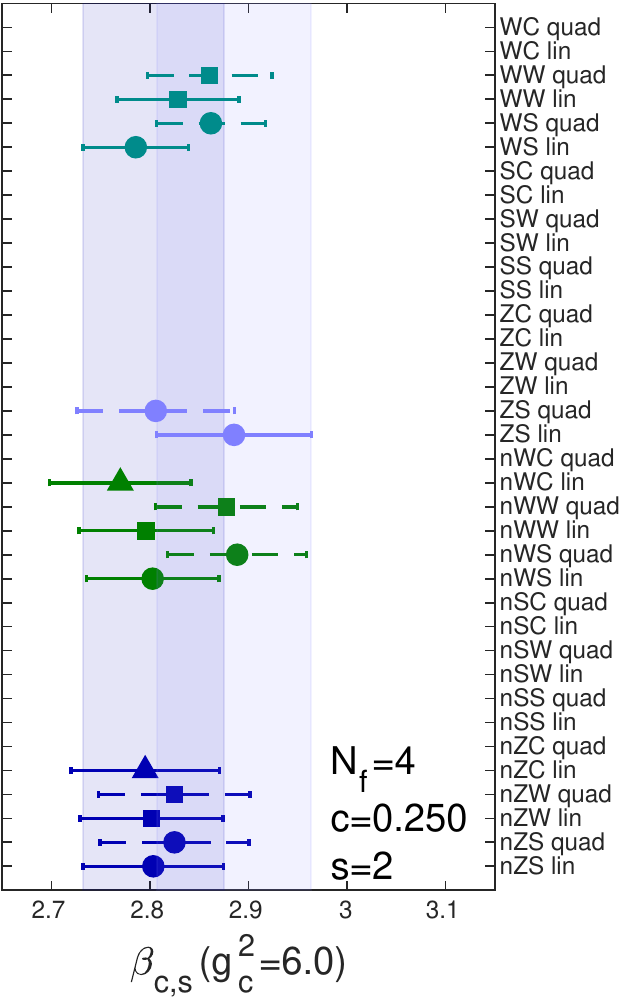}
  \caption{Systematic effects on the $N_f=4$ results for $\beta_{c,s}(g_c^2)$ due to tree-level improvement and different flows and operators. In all cases we obtain the continuum limit considering a linear extrapolation to the three largest volume pairs and a quadratic extrapolation to all volume pairs. The columns show our continuum limit results at selective $g_c^2 = 1.5$, 3.0, 4.5, and 6.0; the rows correspond to renormalization schemes $c=0.300$, 0.275, 0.250. Open symbols indicate extrapolations with a $p$-value below 5\%. The vertical shaded bands highlight our preferred (n)ZS analysis.}  
  \label{Fig.Nf4_beta_sys}
\end{figure*}

Repeating the analysis for our additional data obtained with different operators to estimate the energy density or different kernels to perform the gradient flow, we compare again a total of 18 different analyses as is shown in Fig.~\ref{Fig.Nf4_beta_sys} for representative values of $g_c^2=1.5$, 3.0, 4.5, and 6.0. Similar to the case of six fundamental flavors, we do not observe large variations and use the envelope of ZS and nZS to account for systematic effects. Our final result for $N_f=4$ is finally presented in Fig.~\ref{Fig.FinalNf4}.

%=================================================
\section{Discussion} \label{Sec.Discussion}
%=================================================
%=================================================
\subsection{Comparison to perturbative predictions}
\begin{figure*}[p]
  \begin{minipage}{0.49\textwidth}
    \includegraphics[width=0.98\textwidth]{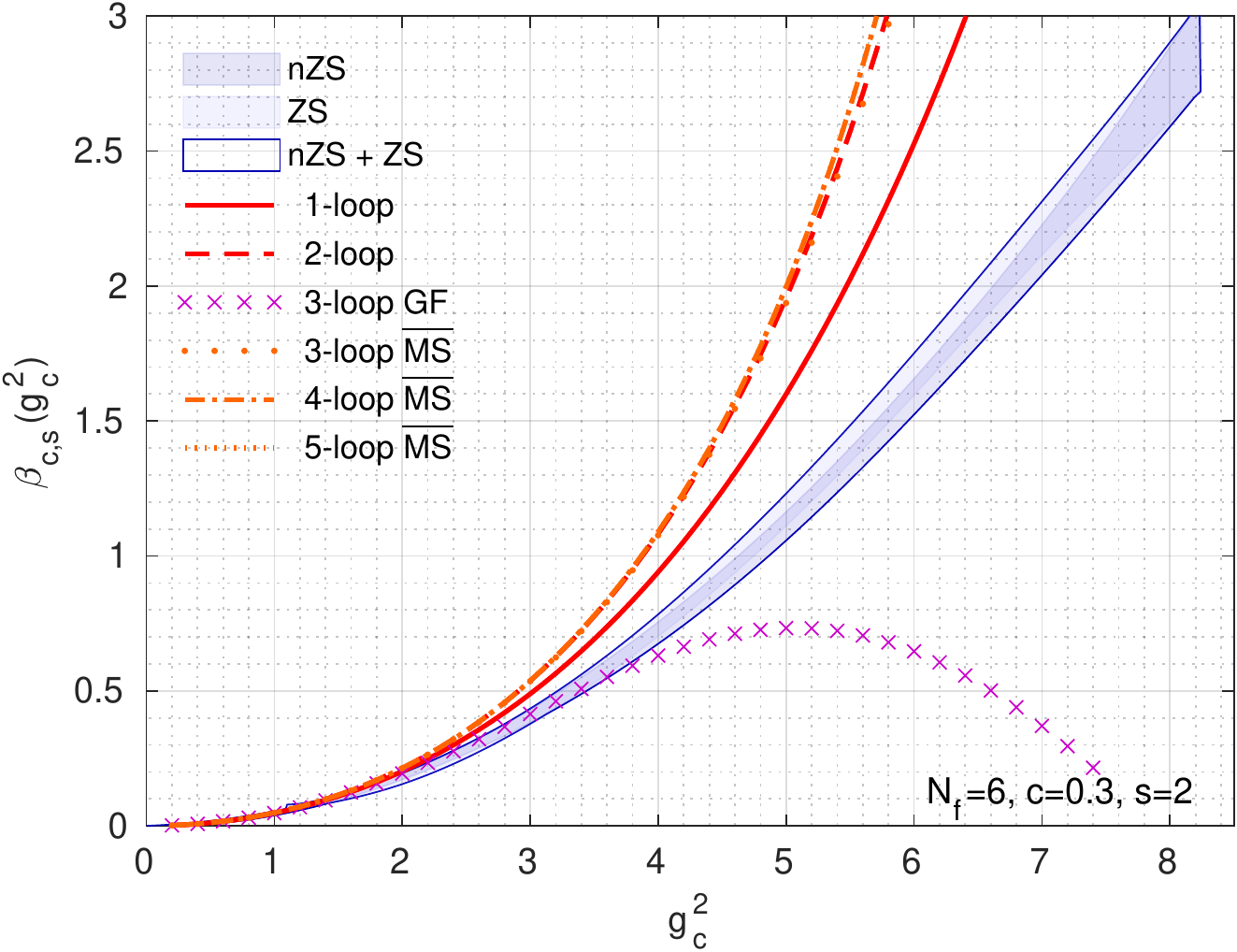}
    \includegraphics[width=0.98\textwidth]{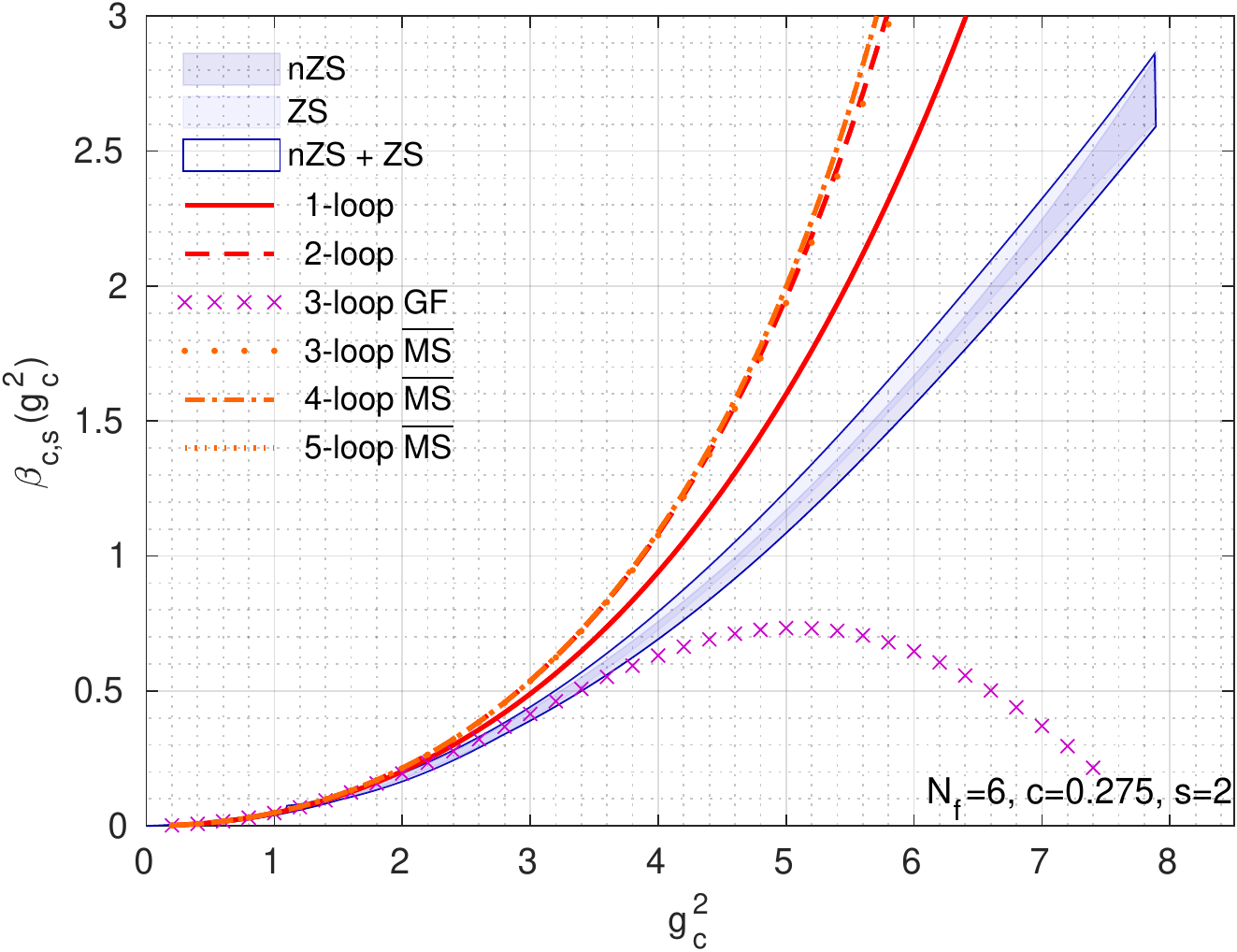} 
    \includegraphics[width=0.98\textwidth]{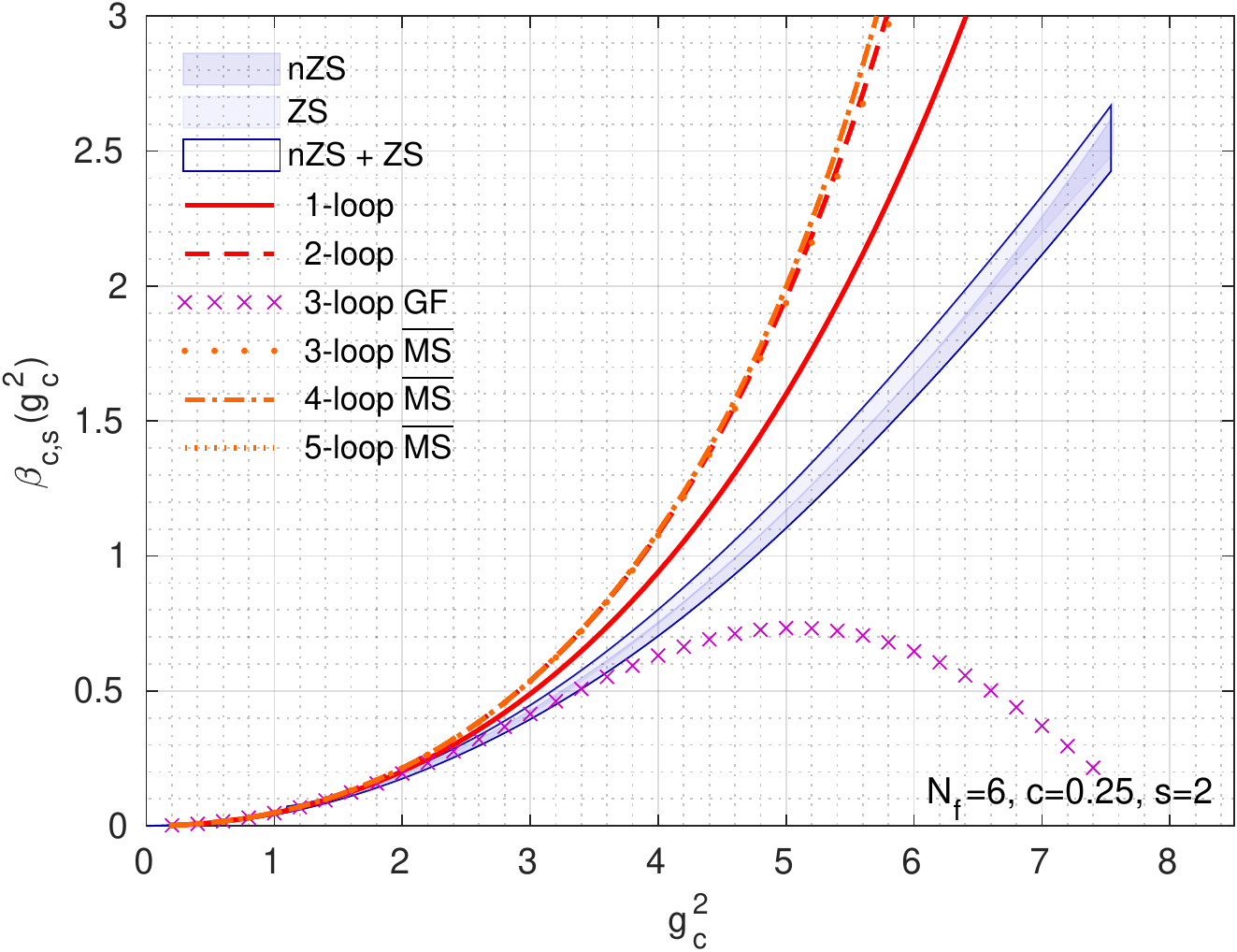}
  \end{minipage}
  \begin{minipage}{0.49\textwidth}
     \includegraphics[width=0.98\textwidth]{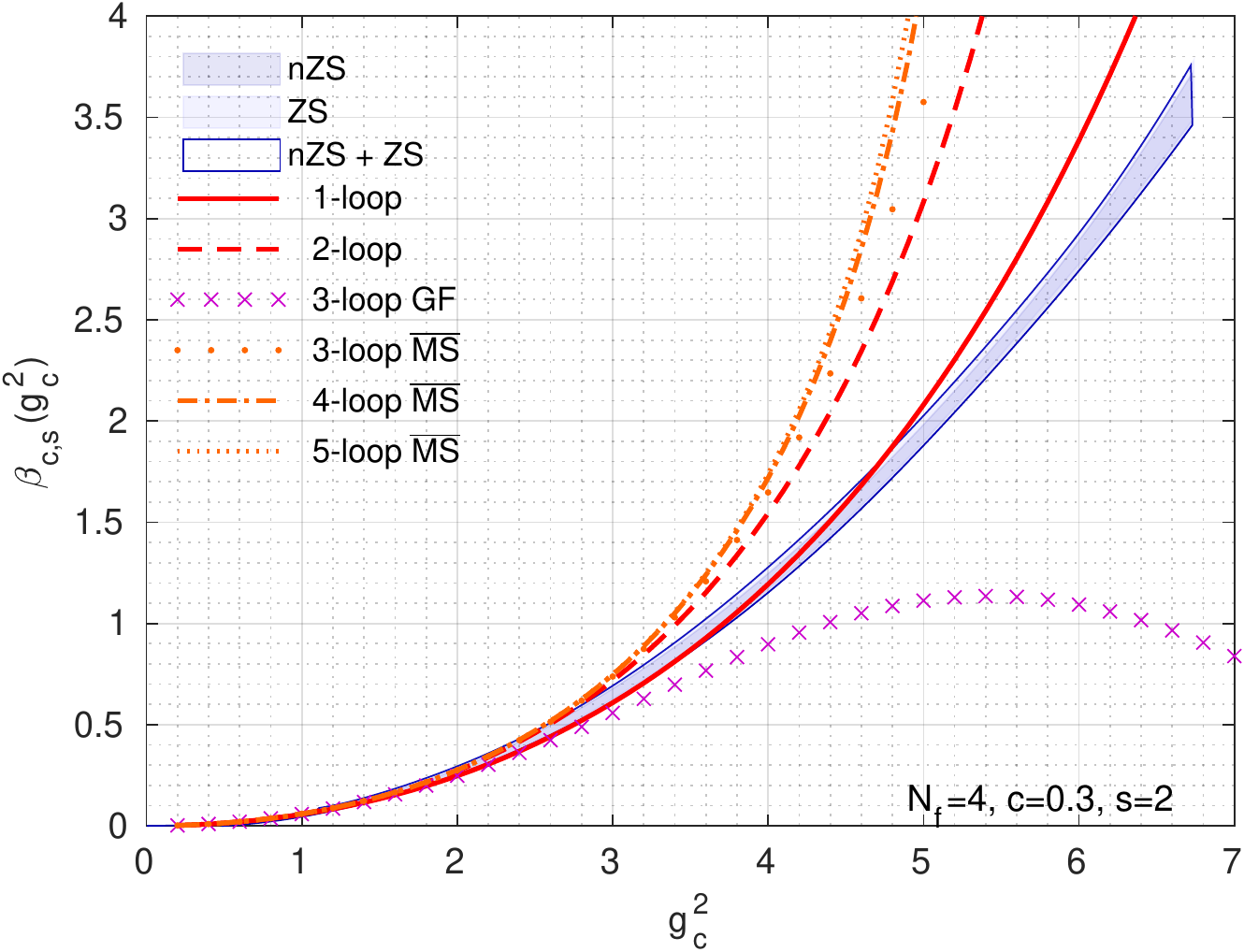}
     \includegraphics[width=0.98\textwidth]{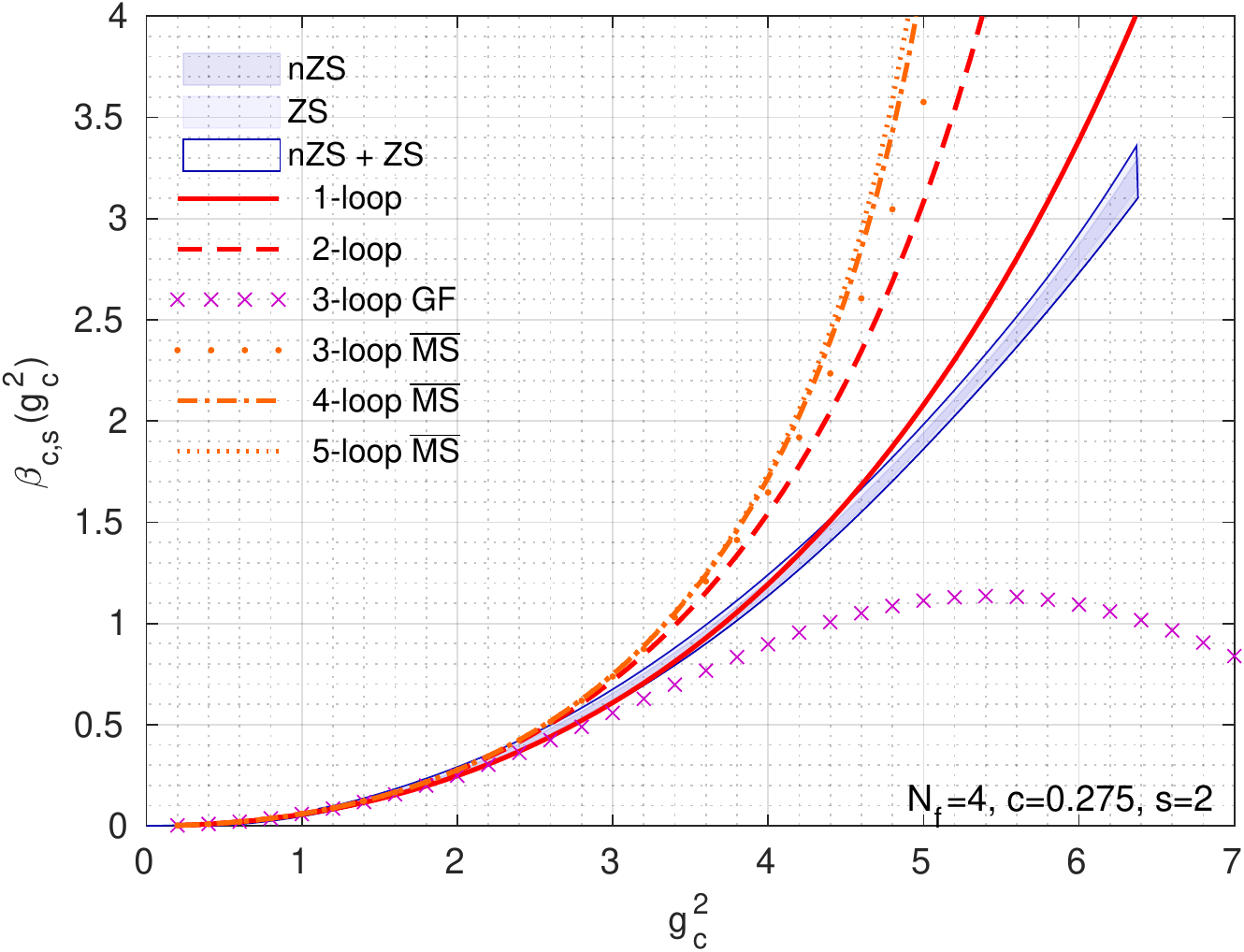}
     \includegraphics[width=0.98\textwidth]{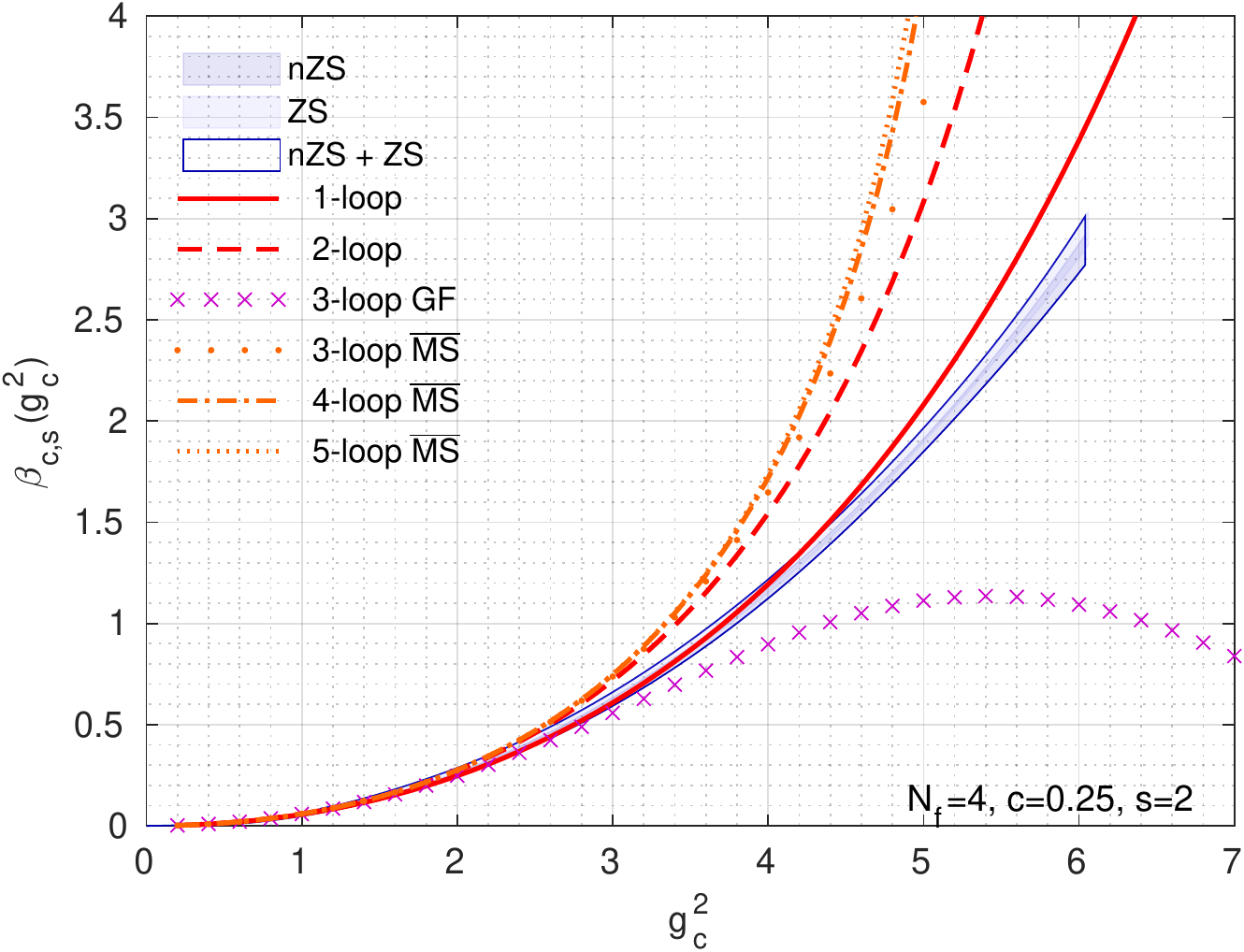}
     \end{minipage}
  \caption{Comparison of our final $N_f=6$ (left) and $N_f=4$ (right) continuum results obtained from our preferred (n)ZS data set for $c = 0.300$ (top), 0.275 (middle), and 0.250 (bottom) to universal 1- and 2-loop perturbative predictions (red), 3-loop perturbative predictions in the gradient flow scheme (purple) and  3-, 4-, and 5-loop $\overline{\textrm{MS}}$ scheme predictions (orange).}
  \label{Fig.FinalNf6}  
  \label{Fig.FinalNf4}
\end{figure*}

%=================================================
Figure \ref{Fig.FinalNf6} shows our final results for step-scaling $\beta$-function for SU(3) with $N_f=6$ (left) or $N_f=4$ (right) fundamental flavors. In comparison to our nonperturbative results for the renormalization schemes $c=0.300$ (top), 0.275 (center), and 0.250 (bottom) we show perturbative predictions. In red the scheme independent results at 1-loop (solid line) and 2-loop (dashed line) are shown, purple crosses denote the 3-loop result in the gradient flow scheme \cite{Harlander:2016vzb}, while the 3-loop (dots), 4-loop (dashed-dotted line), and 5-loop (dotted line) in orange show predictions in the $\overline{\textrm{MS}}$ scheme \cite{Baikov:2016tgj,Ryttov:2016ner}.

In the case of $N_f=6$ the 3-loop GF prediction follows our nonperturbative prediction to about $g_c^2\sim 3.5$  For stronger coupling, however, the 3-loop GF prediction shows a qualitatively different behavior: it turns around pointing to an IRFP. That behavior appears to be common to 3-loop GF for all flavor numbers and may suggest poor convergence of the GF perturbative series. 
On the other hand, the  $\overline{\textrm{MS}}$ predictions show good convergence; the 2 - 5 loop values are very close throughout the investigated $g^2$ regime.
Notably, our nonperturbative results exhibits a noticeably slower running of the $\beta$-function even than the 1-loop perturbative prediction.

For the system with $N_f=4$ flavors our nonperturbative result happens to follow the universal 1-loop prediction up to $g_c^2\sim 4.5$, while again 2-loop and 3- to 5-loop in the $\overline{\textrm{MS}}$ scheme predict a faster running compared to our results. Similar to the case of $N_f=6$ flavors, the 3-loop GF result shows a different qualitatively behavior deviation from our nonperturbative result at about $g_c^2\sim 3$.

%=================================================
\subsection{Comparison to other nonperturbative determinations}
%=================================================
\begin{figure}[t]
  \includegraphics[width=0.98\columnwidth]{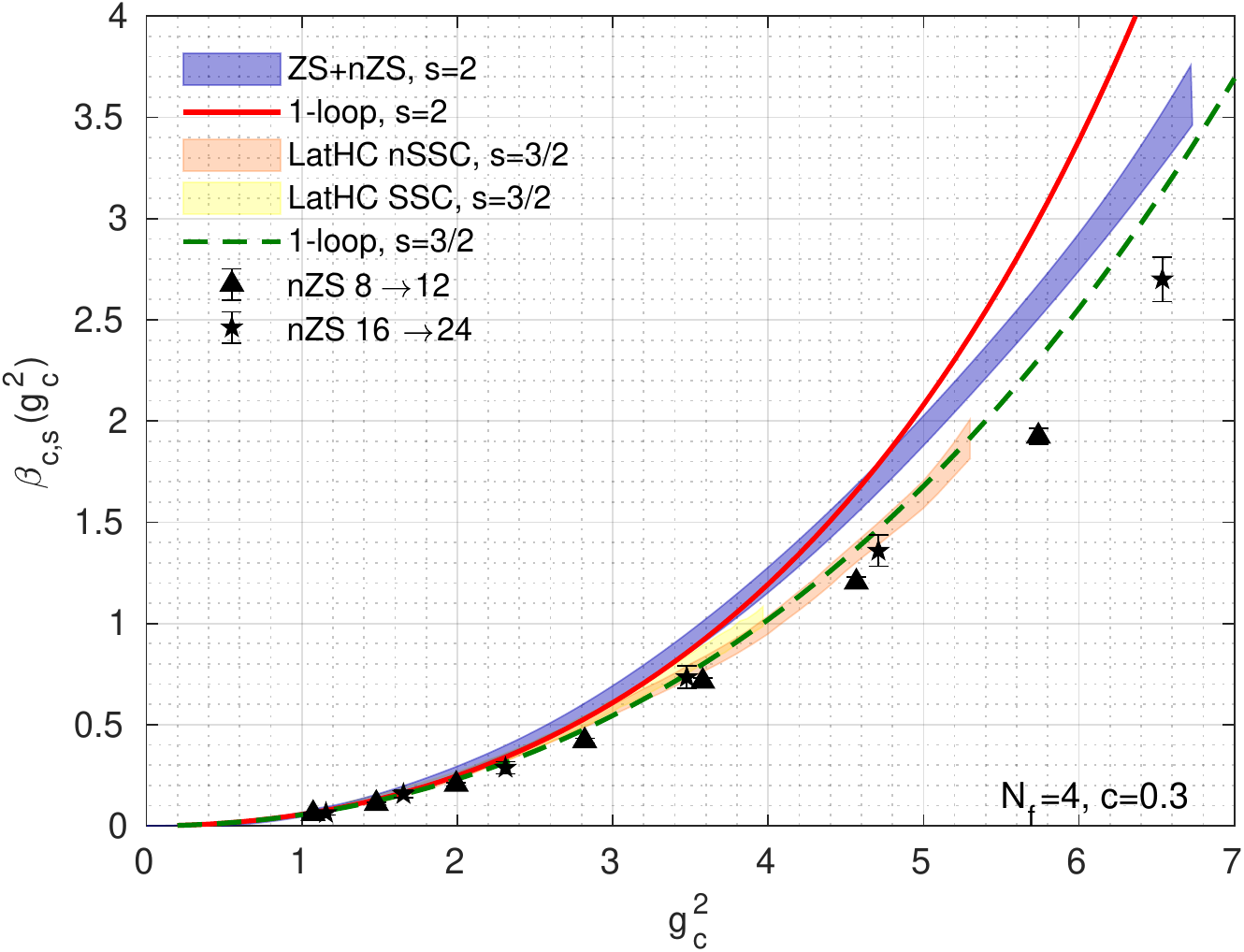}
  \caption{Comparison of our $N_f=4$ continuum results obtained from our preferred (n)ZS data set for $c = 0.300$ to the non-perturbative results obtained by the Lattice Higgs Collaboration (LatHC) \cite{Fodor:2012td,Fodor:2014cpa}. While our results are based on the scale change $s=2$, LatHC uses $s=3/2$. }
  \label{Fig.Nf4cmp}
\end{figure}

To the best of our knowledge, no nonperturbative results on the RG $\beta$ function have been reported to date for the SU(3) gauge system with $N_f=6$ flavors, although $N_f=6$ has been considered to determine e.g.~the $S$-parameter \cite{LSD:2010rei}. For $N_f=4$ the Lattice Higgs Collaboration (LatHC) has presented two results based on analyzing gauge field configurations generated with Symanzik gauge action and staggered fermions. They use Symanzik flow and the clover operator  and analyze their data without tree-level normalization \cite{Fodor:2012td} and with tree-level normalization \cite{Fodor:2014cpa}. In both analyses a scale change $s=3/2$ is used and we refer to their analysis as SSC and nSSC, respectively. 

For the fast running $N_f=4$ system, the choice of the scale change $s$ has a significant impact on the step scaling function at large coupling.  
In Fig.~\ref{Fig.Nf4cmp} we compare our nonperturbative results and the corresponding $s=2$ 1-loop perturbative  line to the LatHC results and the corresponding $s=1.5$ perturbative line. In both cases the nonperturbative results follow the 1-loop values up to $g^2\approx 5$, suggesting consistency between the numerical predictions.

Given the difference in $s$ a direct comparison between our determination and the determination by LatHC is hence not meaningful. Unfortunately, our existing lattice volumes do not allow to perform a full, alternative analysis using $s=3/2$. Out of the eight different $(L/a)^4$ volumes simulated only two volume pairs with $s=3/2$ can be formed, $8 \to 12$ and $16 \to 24$, which is insufficient to take a meaningful continuum limit. Since our nZS analysis for $s=2$ exhibits very small cutoff effects and different volumes sit on top of each other, we can, however, demonstrate consistency by adding the values of the corresponding finite volume step-scaling function to the plot (black triangles and stars). The symbols for these finite volume step-scaling function indeed sit on top of the LatHC prediction and imply perfect consistency of both results.

%=================================================
\subsection{Comparison of different \texorpdfstring{$N_f$}{Nf} = 4, 6, 10, 12}
%=================================================
Finally we return to Fig.~\ref{Fig.All} where we show our newly obtained results for $N_f=4$ and 6 in comparison to our published results for $N_f=10$ and 12. As we have already discussed in the Introduction, all step scaling functions run slower than perturbatively predicted. Existing data for $N_f=2$ suggest the same is true for smaller $N_f$ as well. In addition, the GF scheme 3-loop perturbative $\beta$ function shows poor convergence in the numerically interesting strong coupling regime. These observations underline the necessity to use fully nonperturbative running when comparing lattice data to continuum values, like the matching renormalization group $Z$ factors of matrix elements.

%=================================================
\section{Summary}
\label{Sec.summary}
%=================================================
In this paper we continued our quest to map the renormalization group $\beta$ function of SU(3) gauge theories with no fermions to the conformal regime, $N_f=0 - 12$. As Fig.~\ref{Fig.All}  illustrates the GF step scaling function approach can predict the nonperturbative $\beta$ function from
 the weak coupling regime where perturbation theory is applicable to strong coupling where chiral symmetry breaking blocks its applicability ($N_f\le 6$), or to the occurrence of a bulk phase transition that prevents the investigation of stronger couplings ($N_f\ge 8$). Once our ongoing $N_f=8$ and $N_f=0$ studies are complete, all even flavor numbers have been considered. To go beyond the presently accessible range in the QCD-like systems we need to extend the simulations to the confining regime where finite fermion mass will be required. Within the conformal window the biggest challenge is the bulk phase transition that  is possibly avoidable, or at least controllable, by improving the action by e.g.~including unphysical Pauli-Villars bosons \cite{Hasenfratz:2021zsl}. These studies are among our future plans.

Since the step-scaling function approach is justified only in the chirally symmetric small volume regime, it cannot be applied once chiral symmetry breaking introduces an infrared scale.  In a forthcoming paper we analyze the data using the continuous $\beta$ function approach to demonstrate the consistency between the two methods. That opens the possibility to determine the non-perturbative $\beta$ function from the perturbative ultraviolet to the strongly coupled infrared regimes.

%=================================================
\begin{acknowledgments}

  We are very grateful to Peter Boyle, Guido Cossu, Anontin Portelli, and Azusa Yamaguchi who develop the \texttt{GRID} software library providing the basis of this work and who assisted us in installing and running \texttt{GRID} on different architectures and computing centers. A.H.~acknowledges support by DOE grant No.~DE-SC0010005 and C.R.~by DOE Grant No.~DE-SC0015845.

  Computations for this work were carried out in part on facilities of the USQCD Collaboration, which are funded by the Office of Science of the U.S.~Department of Energy, the RMACC Summit supercomputer \cite{UCsummit}, which is supported by the National Science Foundation (awards No.~ACI-1532235 and No.~ACI-1532236), the University of Colorado Boulder, and Colorado State University, and the compute cluster \texttt{OMNI} of the University of Siegen. This work used the Extreme Science and Engineering Discovery Environment (XSEDE), which is supported by National Science Foundation grant number ACI-1548562 \cite{xsede} through allocation TG-PHY180005 on the XSEDE resource \texttt{stampede2}.  This research also used resources of the National Energy Research Scientific Computing Center (NERSC), a U.S. Department of Energy Office of Science User Facility operated under Contract  No. DE-AC02-05CH11231. This document was prepared using the resources of the USQCD Collaboration at the Fermi National Accelerator Laboratory (Fermilab), a U.S. Department of Energy, Office of Science, HEP User Facility. Fermilab is managed by Fermi Research Alliance, LLC (FRA), acting under Contract No.~DE-AC02-07CH11359. We thank  Brookhaven National Laboratory (BNL), Fermilab,  Jefferson Lab, NERSC, the University of Colorado Boulder, the University of Siegen, TACC, the NSF, and the U.S.~DOE for providing the facilities essential for the completion of this work.
\end{acknowledgments}
%=================================================

\clearpage
\appendix

\setlength{\LTcapwidth}{\textwidth}
\section{Renormalized couplings \texorpdfstring{$g_c^2$}{gc2} and details of the polynomial interpolation}
\label{Sec.RenCouplings}

\begin{longtable*}{cccccccccccc}
  \caption{Details of our preferred analysis for $N_f=6$ based on Zeuthen flow and Symanzik operator. For each ensemble specified by the spatial extent $L/a$ and bare gauge coupling $\beta$ we list the number of measurements $N$ as well as the renormalized couplings $g_c^2$ for the analysis with (nZS) and without tree-level improvement (ZS) for the three renormalization schemes $c=0.300$, 0.275 and 0.250. In addition the integrated autocorrelation times estimated using the $\Gamma$-method \cite{Wolff:2003sm} are listed in units of 10 MDTU.}
  \label{Tab.Nf6_nZS_ZS}\\
  
  \hline\hline
      &         &     & \multicolumn{3}{c}{$c=0.300$}&\multicolumn{3}{c}{$c=0.275$}&\multicolumn{3}{c}{$c=0.250$}\\
  $L/a$ & $\beta$ & $N$ & $g_c^2$(nZS) & $g_c^2$(ZS) & $\tau_\text{int}$& $g_c^2$(nZS) &  $g_c^2$(ZS)  & $\tau_\text{int}$ &$g_c^2$(nZS)  &$g_c^2$(ZS)  & $\tau_\text{int}$\\
  \hline
  \endfirsthead

  \hline
      &         &     & \multicolumn{3}{c}{$c=0.300$}&\multicolumn{3}{c}{$c=0.275$}&\multicolumn{3}{c}{$c=0.250$}\\
  $L/a$ & $\beta$ & $N$ & $g_c^2$(nZS) &  $g_c^2$(ZS) & $\tau_\text{int}$& $g_c^2$(nZS) &$g_c^2$(ZS) & $\tau_\text{int}$ & $g_c^2$(nZS) & $g_c^2$(ZS) & $\tau_\text{int}$\\
  \hline
  \endhead

  \hline
  \endfoot

  \hline \hline
  \endlastfoot
\input{gcSq_Nf6_nZS_ZS}
\end{longtable*}

\begin{longtable*}{cccccccccccc}
  \caption{Details of our preferred analysis for $N_f=4$ based on Zeuthen flow and Symanzik operator. For each ensemble specified by the spatial extent $L/a$ and bare gauge coupling $\beta$ we list the number of measurements $N$ as well as the renormalized couplings $g_c^2$ for the analysis with (nZS) and without tree-level improvement (ZS) for the three renormalization schemes $c=0.300$, 0.275 and 0.250. In addition the integrated autocorrelation times estimated using the $\Gamma$-method \cite{Wolff:2003sm} are listed in units of 10 MDTU.}
  \label{Tab.Nf4_nZS_ZS}\\
  
  \hline\hline
      &         &     & \multicolumn{3}{c}{$c=0.300$}&\multicolumn{3}{c}{$c=0.275$}&\multicolumn{3}{c}{$c=0.250$}\\
  $L/a$ & $\beta$ & $N$ & $g_c^2$(nZS) & $g_c^2$(ZS) & $\tau_\text{int}$& $g_c^2$(nZS) &  $g_c^2$(ZS)  & $\tau_\text{int}$ &$g_c^2$(nZS)  &$g_c^2$(ZS)  & $\tau_\text{int}$\\
  \hline
  \endfirsthead

  \hline
      &         &     & \multicolumn{3}{c}{$c=0.300$}&\multicolumn{3}{c}{$c=0.275$}&\multicolumn{3}{c}{$c=0.250$}\\
  $L/a$ & $\beta$ & $N$ & $g_c^2$(nZS) &  $g_c^2$(ZS) & $\tau_\text{int}$& $g_c^2$(nZS) &$g_c^2$(ZS) & $\tau_\text{int}$ & $g_c^2$(nZS) & $g_c^2$(ZS) & $\tau_\text{int}$\\
  \hline
  \endhead

  \hline
  \endfoot

  \hline \hline
  \endlastfoot
  \input{gcSq_Nf4_nZS_ZS}
\end{longtable*}

%\clearpage
%\section{Details of the polynomial interpolation}
\begin{table*}[p]
  \caption{Results of the interpolation fits for the five $N_f=6$ lattice volume pairs for our preferred ZS (top half) and nZS (bottom half) analysis using renormalization schemes $c = 0.300$, $0.275$, and $0.250$. Since discretization effects are sufficiently small for nZS, we constrain the constant term $b_0 = 0$ in Eq.~(\ref{Eq.fit_form}) whereas for ZS the intercept $b_0$ is fitted. In addition we list the degree of freedom (d.o.f.), $\chi^2/\text{d.o.f.}$~as well as the $p$-value.}
 \label{Tab.interpolationsNf6}
  \begin{tabular}{c@{~~~~}ccc@{~~}c@{~~}ccccc}
    \hline \hline
    & analysis &    $c$ &d.o.f. &$\chi^2$/d.o.f.&$p$-value&    $b_3$        & $b_2$     & $b_1$      & $b_0$ \\ \hline
    $8\to 16$  & ZS & 0.300 & 7 & 0.780 &  0.604 &-0.00268(58) & 0.0681(70) &-0.110(23) & 0.053(20) \\
    $10\to 20$ & ZS & 0.300 & 6 & 0.077 &  0.998 &-0.00319(97) & 0.071(11)  &-0.076(36) & 0.029(30) \\
    $12\to 24$ & ZS & 0.300 & 5 & 0.555 &  0.734 &-0.0002(12)  & 0.039(14)  & 0.035(42) &-0.050(34) \\
    $16\to 32$ & ZS & 0.300 & 4 & 0.585 &  0.673 &-0.0016(12)  & 0.057(16)  &-0.024(53) & 0.013(46) \\
    $20\to 40$ & ZS & 0.300 & 3 & 1.273 &  0.282 &-0.0018(23)  & 0.066(27)  &-0.070(86) & 0.053(72)\\
    \hline
    $8\to 16$  & ZS & 0.275 & 7 & 0.940 & 0.474 &-0.00195(48) &0.0595(58) &-0.108(19) & 0.041(16)\\ 
    $10\to 20$ & ZS & 0.275 & 6 & 0.093 & 0.997 &-0.00304(82) &0.0681(94) &-0.076(29) & 0.027(24)\\ 
    $12\to 24$ & ZS & 0.275 & 5 & 0.644 & 0.666 &-0.0005(10)  &0.042(11)  & 0.017(34) &-0.035(27)\\ 
    $16\to 32$ & ZS & 0.275 & 4 & 0.751 & 0.557 &-0.0018(11)  &0.059(13)  &-0.034(43) & 0.019(37)\\ 
    $20\to 40$ & ZS & 0.275 & 3 & 1.346 & 0.257 &-0.0016(21)  &0.063(23)  &-0.057(72) & 0.041(59)\\
    \hline
    $8\to 16$  & ZS & 0.250 & 7 & 1.219 & 0.288 &-0.00082(40) &0.0476(47) &-0.109(15) & 0.025(14) \\
    $10\to 20$ & ZS & 0.250 & 6 & 0.154 & 0.988 &-0.00278(69) &0.0645(77) &-0.081(23) & 0.026(19) \\
    $12\to 24$ & ZS & 0.250 & 5 & 0.780 & 0.564 &-0.00090(88) &0.0459(93) &-0.004(26) &-0.020(21) \\
    $16\to 32$ & ZS & 0.250 & 4 & 0.970 & 0.423 &-0.00196(95) &0.061(11)  &-0.042(34) & 0.023(29) \\
    $20\to 40$ & ZS & 0.250 & 3 & 1.404 & 0.239 &-0.0014(19)  &0.060(20)  &-0.045(60) & 0.030(48) \\
    \hline \hline
    $8\to 16$  & nZS & 0.300 & 8 & 1.594  &0.121 &-0.00158(35) & 0.0561(29) &-0.0177(46) & ---\\
    $10\to 16$ & nZS & 0.300 & 7 & 0.199 & 0.986 &-0.00251(49) & 0.0634(43) &-0.0299(66) & ---\\
    $12\to 16$ & nZS & 0.300 & 6 & 0.821 & 0.553 &-0.00176(60) & 0.0587(50) &-0.0187(75) & ---\\
    $16\to 16$ & nZS & 0.300 & 5 & 0.484 & 0.788 &-0.00134(61) & 0.0536(57) &-0.0075(93) & ---\\
    $20\to 16$ & nZS & 0.300 & 4 & 1.090 & 0.359 &-0.0003(12)  & 0.0478(95) &-0.007(13)  & ---\\
    \hline
    $8\to 16$  & nZS & 0.275 & 8 & 1.620 & 0.113 &-0.00115(30) & 0.0533(25) &-0.0162(38) & ---\\
    $10\to 16$ & nZS & 0.275 & 7 & 0.263 & 0.968 &-0.00241(43) & 0.0615(36) &-0.0265(53) & ---\\
    $12\to 16$ & nZS & 0.275 & 6 & 0.818 & 0.556 &-0.00175(53) & 0.0574(42) &-0.0172(61) & ---\\
    $16\to 16$ & nZS & 0.275 & 5 & 0.653 & 0.659 &-0.00134(55) & 0.0536(49) &-0.0092(77) & ---\\
    $20\to 16$ & nZS & 0.275 & 4 & 1.135 & 0.338 &-0.0003(11)  & 0.0479(81) &-0.006(11)  & ---\\
    \hline
    $8\to 16$  & nZS & 0.250 & 8 & 1.560 & 0.131 &-0.00025(27) & 0.0487(21) &-0.0130(32) & ---\\ 
    $10\to 20$ & nZS & 0.250 & 7 & 0.409 & 0.898 &-0.00219(37) & 0.0591(30) &-0.0232(43) & ---\\ 
    $12\to 24$ & nZS & 0.250 & 6 & 0.807 & 0.564 &-0.00174(45) & 0.0563(34) &-0.0165(47) & ---\\ 
    $16\to 32$ & nZS & 0.250 & 5 & 0.907 & 0.475 &-0.00132(48) & 0.0534(40) &-0.0105(61) & ---\\
    $20\to 40$ & nZS & 0.250 & 4 & 1.154 & 0.329 &-0.00034(93) & 0.0480(68) &-0.0055(88) & ---\\
    \hline\hline    
  \end{tabular}
\end{table*}

\begin{table*}[t]
  \caption{Results of the interpolation fits for the five $N_f=4$ lattice volume pairs for our preferred ZS (top half) and nZS (bottom half) analysis using renormalization schemes $c = 0.300$, $0.275$, and $0.250$. Since discretization effects are sufficiently small for nZS, we constrain the constant term $b_0 = 0$ in Eq.~(\ref{Eq.fit_form}) whereas for ZS the intercept $b_0$ is fitted. In addition we list the degree of freedom (d.o.f.), $\chi^2/\text{d.o.f.}$~as well as the $p$-value.}
  \label{Tab.interpolationsNf4}
  \begin{tabular}{c@{~~~~}ccc@{~~}c@{~~}ccccc}  
    \hline \hline
    & analysis &    $c$ & d.o.f.&$\chi^2$/d.o.f.&$p$-value&    $b_3$        & $b_2$     & $b_1$      & $b_0$ \\ \hline
    $8\to 16$  & ZS & 0.300 & 5 & 0.496 & 0.779 & 0.0044(12) & 0.035(12) & -0.005(32) & -0.021(24)\\
    $10\to 20$ & ZS & 0.300 & 4 & 0.776 & 0.540 & 0.0009(17) & 0.076(17) & -0.087(45) & 0.051(34)\\
    $12\to 24$ & ZS & 0.300 & 3 & 1.473 & 0.220 &-0.0001(21) & 0.080(22) & -0.053(63) & 0.015(50)\\ 
    $16\to 32$ & ZS & 0.300 & 4 & 2.516 & 0.039 & 0.0016(32) & 0.068(32) & -0.023(87) & 0.005(68)\\
    $20\to 40$ & ZS & 0.300 & 3 & 2.015 & 0.109 &-0.0009(47) & 0.091(46) & -0.05(13)  & 0.01(10)\\
    \hline
    $8\to 16$  & ZS & 0.275 & 5 & 0.589 & 0.709 & 0.00403(94)& 0.0345(92)&-0.029(25) & -0.015(19)\\
    $10\to 20$ & ZS & 0.275 & 4 & 0.822 & 0.511 & 0.0012(15) & 0.070(15) &-0.080(38) & 0.041(28)\\
    $12\to 24$ & ZS & 0.275 & 3 & 1.731 & 0.158 & 0.0003(19) & 0.076(19) &-0.054(53) & 0.016(41)\\
    $16\to 32$ & ZS & 0.275 & 4 & 2.155 & 0.071 & 0.0023(30) & 0.061(28) &-0.011(76) & -0.005(58)\\
    $20\to 40$ & ZS & 0.275 & 3 & 1.702 & 0.164 & 0.0010(44) & 0.073(42) &-0.02(11)  &-0.006(85)\\
    \hline
    $8\to 16$  & ZS & 0.250 & 5 & 0.588 & 0.709 & 0.00384(74)& 0.0315(72)&-0.056(20) & -0.011(16)\\
    $10\to 10$ & ZS & 0.250 & 4 & 0.906 & 0.459 & 0.0015(13) & 0.064(12) &-0.078(31) & 0.033(23)\\
    $12\to 24$ & ZS & 0.250 & 3 & 2.141 & 0.093 & 0.0005(16) & 0.072(16) &-0.057(43) & 0.018(33)\\
    $16\to 32$ & ZS & 0.250 & 4 & 1.694 & 0.148 & 0.0028(28) & 0.057(25) &-0.006(66) & -0.008(49)\\ 
    $20\to 40$ & ZS & 0.250 & 3 & 1.403 & 0.240 & 0.0024(41) & 0.062(37) &-0.003(96) & -0.013(71)\\
    \hline\hline
    $8\to 16$  &nZS & 0.300 & 6 & 0.562 & 0.761 & 0.00393(57)& 0.0498(39)&-0.0005(50) & --- \\
    $10\to 20$ &nZS & 0.300 & 5 & 1.079 & 0.369 & 0.00339(75)& 0.0534(56)&-0.0073(77) & --- \\
    $12\to 24$ &nZS & 0.300 & 4 & 1.129 & 0.341 & 0.0005(10) & 0.0748(75)&-0.028(10)  & --- \\
    $16\to 32$ &nZS & 0.300 & 5 & 2.015 & 0.073 & 0.0019(14) & 0.066(11) &-0.015(15)  & --- \\
    $20\to 40$ &nZS & 0.300 & 4 & 1.515 & 0.195 &-0.0004(20) & 0.085(14) &-0.035(19)  & --- \\    
    \hline
    $8\to 16$  &nZS & 0.275 & 6 & 0.613 & 0.720 & 0.00416(48)& 0.0476(32)&-0.0005(41) & --- \\
    $10\to 20$ &nZS & 0.275 & 5 & 1.097 & 0.360 & 0.00345(66)& 0.0524(47)&-0.0068(62) & --- \\
    $12\to 24$ &nZS & 0.275 & 4 & 1.341 & 0.252 & 0.00095(90)& 0.0705(63)&-0.0240(84) & --- \\
    $16\to 32$ &nZS & 0.275 & 5 & 1.726 & 0.125 & 0.0021(13) & 0.0640(93)&-0.014(12)  & --- \\
    $20\to 40$ &nZS & 0.275 & 4 & 1.279 & 0.276 & 0.0007(18) & 0.076(13) &-0.026(17)  & --- \\
    \hline
    $8\to 16$  &nZS & 0.250 & 6 & 0.603 & 0.728 & 0.00463(42)& 0.0446(27) &-0.0002(33) & --- \\
    $10\to 20$ &nZS & 0.250 & 5 & 1.143 & 0.335 & 0.00359(58)& 0.0508(39) &-0.0061(49) & --- \\
    $12\to 24$ &nZS & 0.250 & 4 & 1.685 & 0.150 & 0.00136(79)& 0.0667(53) &-0.0207(68) & --- \\
    $16\to 32$ &nZS & 0.250 & 5 & 1.362 & 0.235 & 0.0024(12) & 0.0615(79) &-0.013(10)  & --- \\
    $20\to 40$ &nZS & 0.250 & 4 & 1.061 & 0.374 & 0.0017(16) & 0.069(11)  &-0.019(14)  & --- \\
    \hline \hline    
  \end{tabular}
\end{table*}

\clearpage
\section{Tree-level normalization factors}
\label{Sec.tree-level}

\begin{table}[h!]
  \caption{Tree-level normalization coefficients $C(c,L/a)$ for renormalization schemes $c=0.250$, 0.275, and 0.300 and $(L/a)^4=\{40^4,\,48^4\}$ lattices. Values are quotes for simulations with Symanzik gauge action and Zeuthen, Symanzik, or Wilson flow combined with Symanzik, Wilson plaquette, or clover operator.}  \label{Tab.tlnL40}
  \begin{tabular}{ccccc}
  \hline \hline
  flow/op.& $L/a$ & $C(0.250,L/a)$& $C(0.275,L/a)$& $C(0.300,L/a)$\\ \hline
  ZS &             40  & 0.987231 & 0.981324 & 0.973551 \\
  ZS &             48  & 0.973551 & 0.981254 & 0.973510 \\
  ZW &             40  & 0.982875 & 0.977615 & 0.970391 \\
  ZW &             48  & 0.970391 & 0.978671 & 0.971310 \\
  ZC &             40  & 0.969595 & 0.966514 & 0.960929 \\
  ZC &             48  & 0.960929 & 0.970935 & 0.964718 \\
  \hline
  SS &             40  & 0.984012 & 0.978590 & 0.971254 \\
  SS &             48  & 0.971254 & 0.979351 & 0.971912 \\
  SW &             40  & 0.997304 & 0.989589 & 0.980385 \\
  SW &             48  & 0.980385 & 0.984379 & 0.976126 \\
  SC &             40  & 0.966420 & 0.963868 & 0.958695 \\
  SC &             48  & 0.958695 & 0.969076 & 0.963151 \\
  \hline
  WS &             40  & 0.997304 & 0.989589 & 0.980516 \\
  WS &             48  & 0.980516 & 0.986993 & 0.978348 \\
  WW &             40  & 0.992773 & 0.985817 & 0.977312 \\
  WW &             48  & 0.977312 & 0.984379 & 0.976126 \\
  WC &             40  & 0.979219 & 0.974528 & 0.967716 \\
  WC &             48  & 0.967716 & 0.976552 & 0.969470 \\
  \hline \hline
\end{tabular}
\end{table}

\clearpage
%=================================================
\section{\texorpdfstring{Analyses for $c=0.275$ and $c=0.250$}{Analyses for c=0.275 and c=0.250}}
\label{Sec.c0275_c0250}
%=================================================
\hspace*{-0.525\textwidth}\begin{minipage}{\textwidth}
%\begin{figure*}[!b]
  \begin{minipage}{0.49\textwidth}
   \flushright 
   \includegraphics[width=0.96\textwidth]{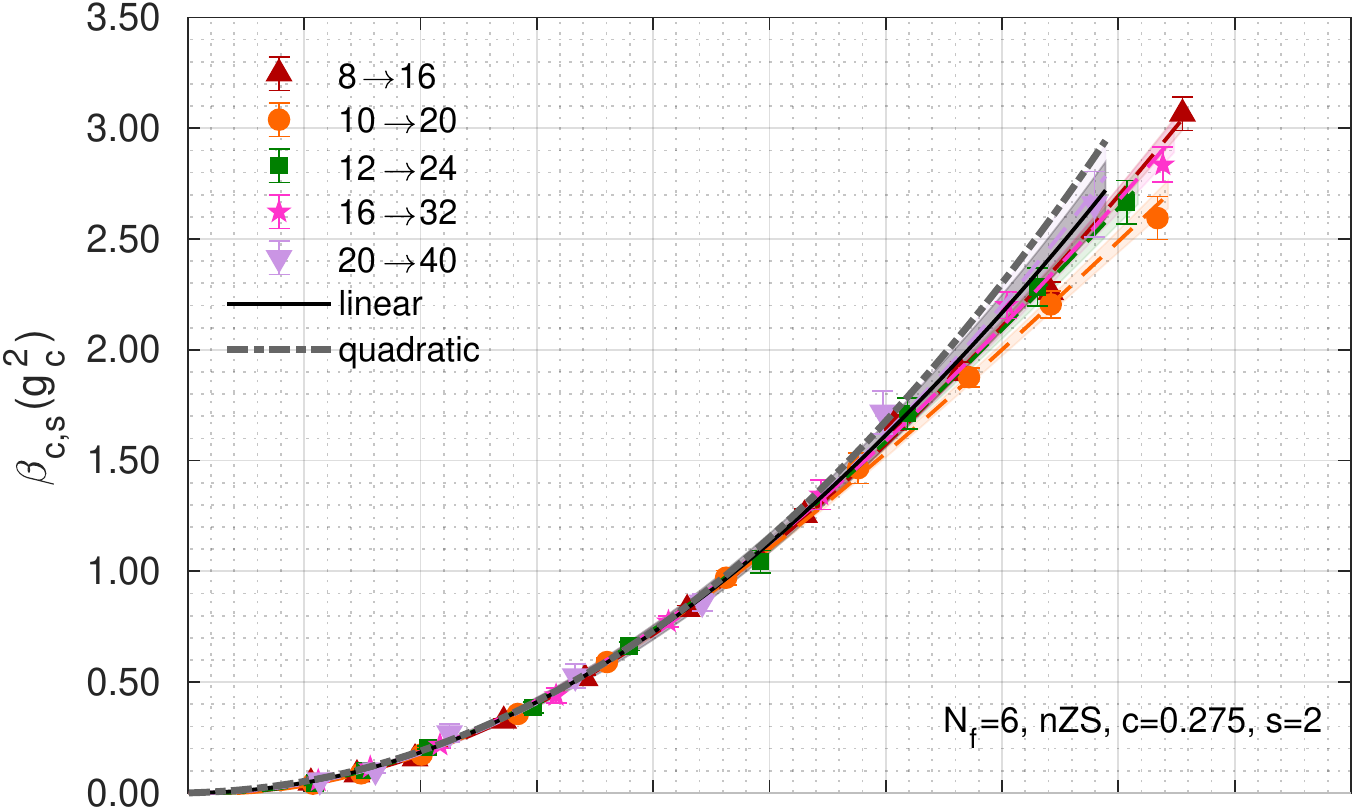}\\
   \includegraphics[width=0.937\textwidth]{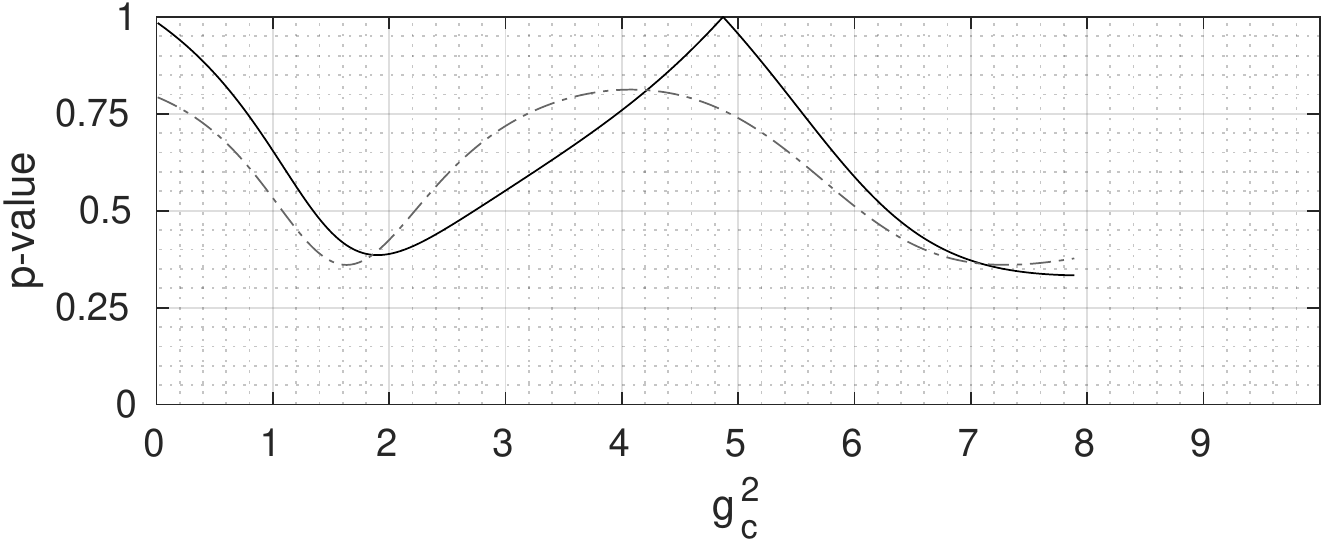} %\\[3mm]
   \includegraphics[width=0.96\textwidth]{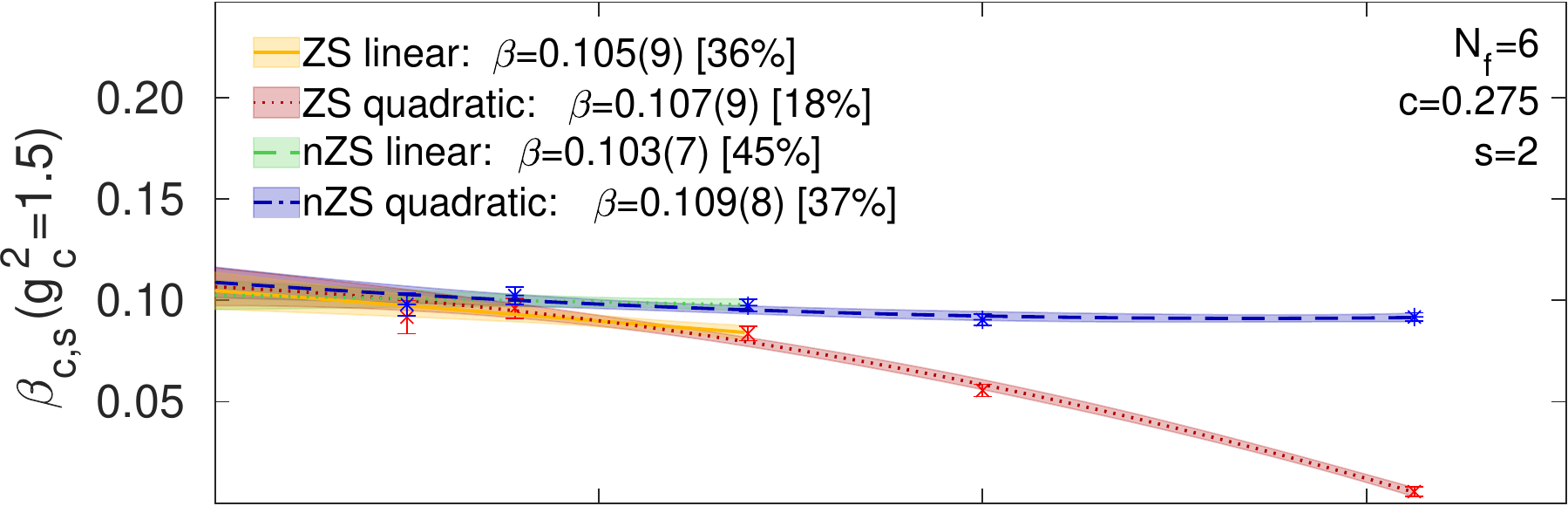}\\
   \includegraphics[width=0.96\textwidth]{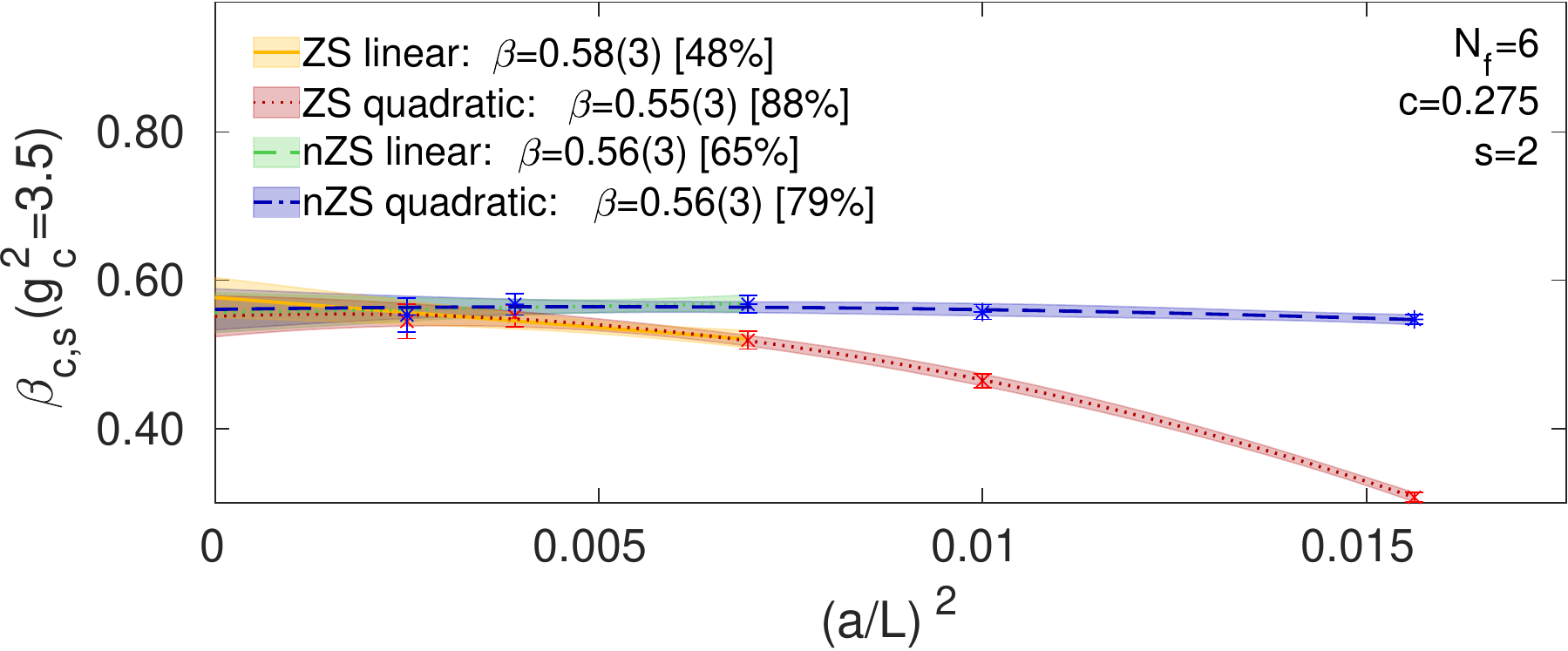}
  \end{minipage}
  \begin{minipage}{0.49\textwidth}
    \flushright
    \includegraphics[width=0.96\textwidth]{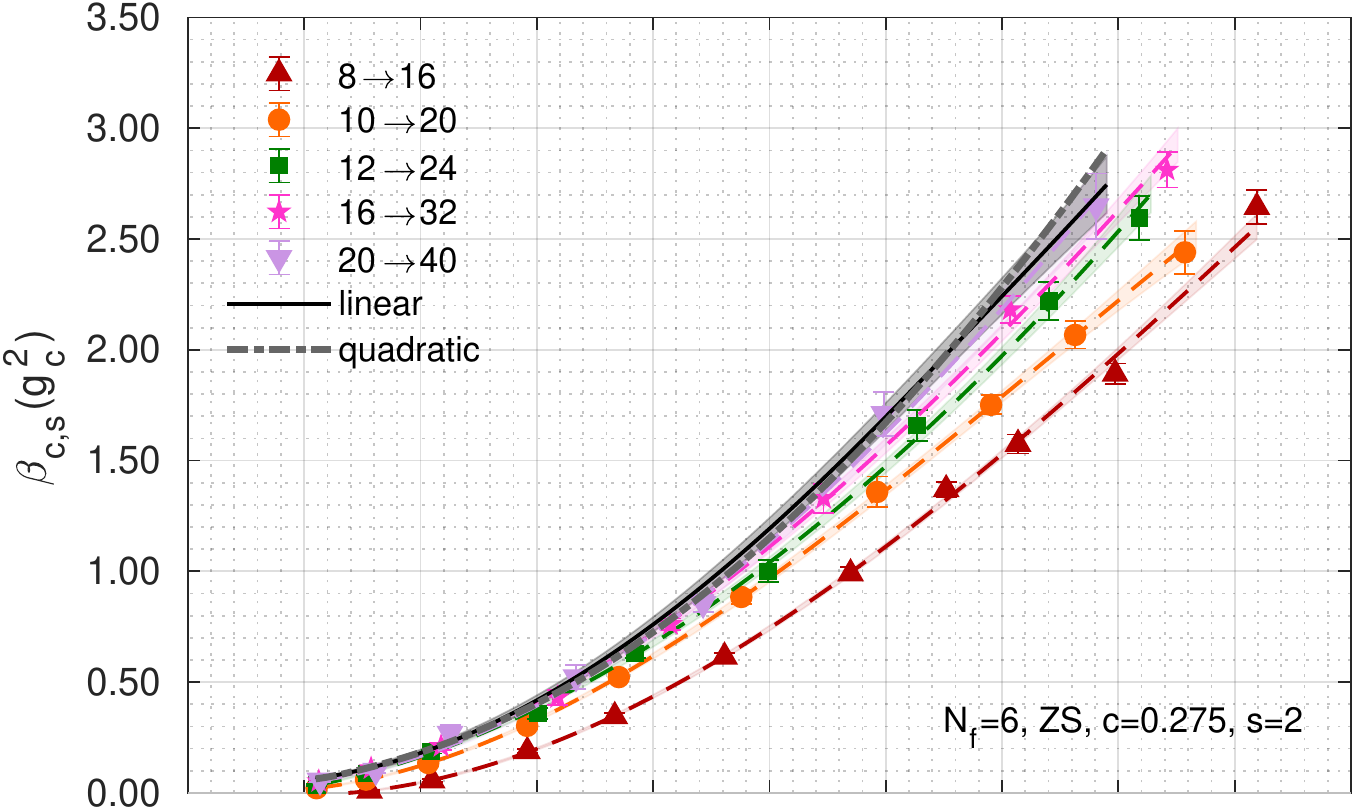}\\    
    \includegraphics[width=0.937\textwidth]{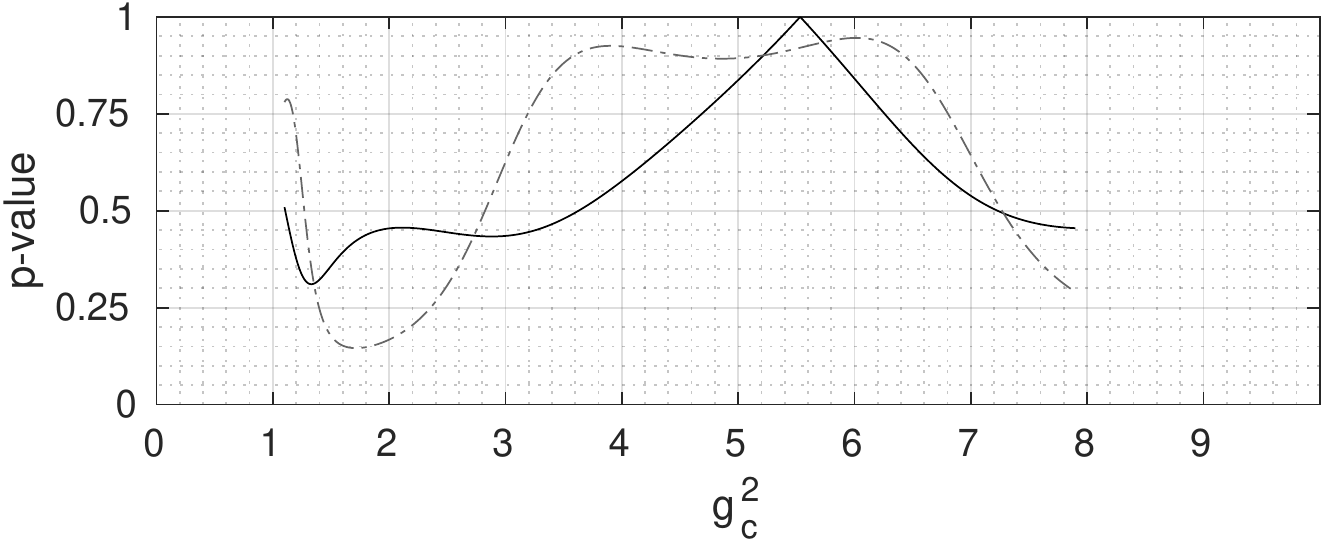} %\\[3mm]
    \includegraphics[width=0.96\textwidth]{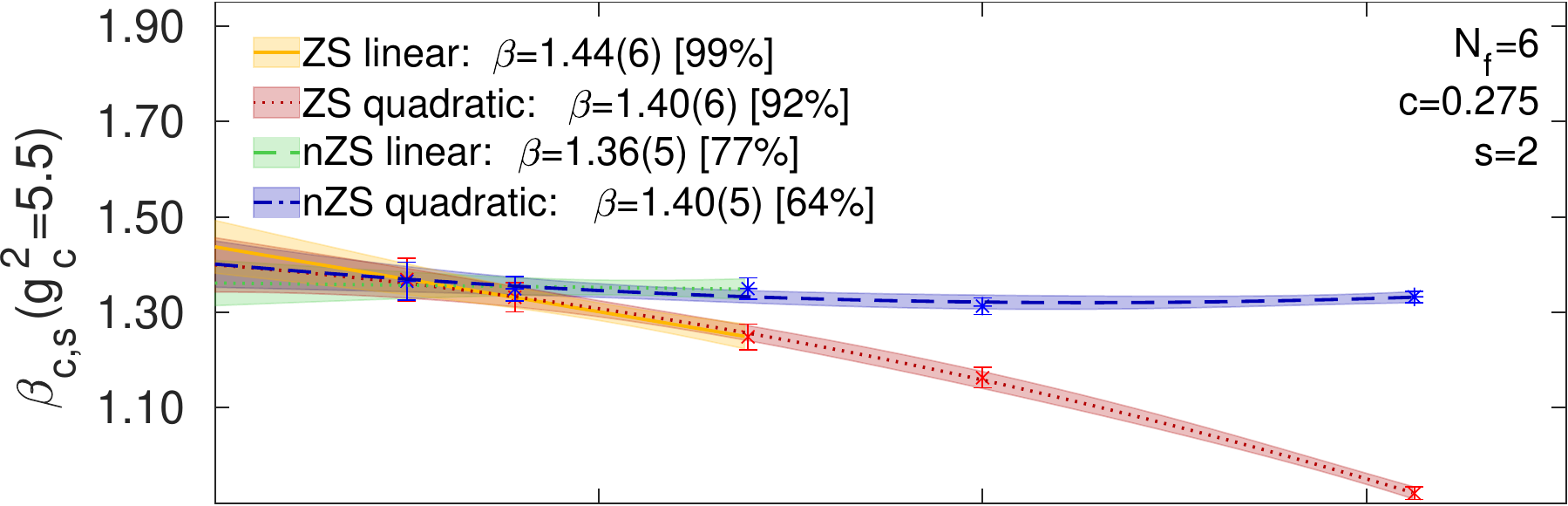}\\
    \includegraphics[width=0.96\textwidth]{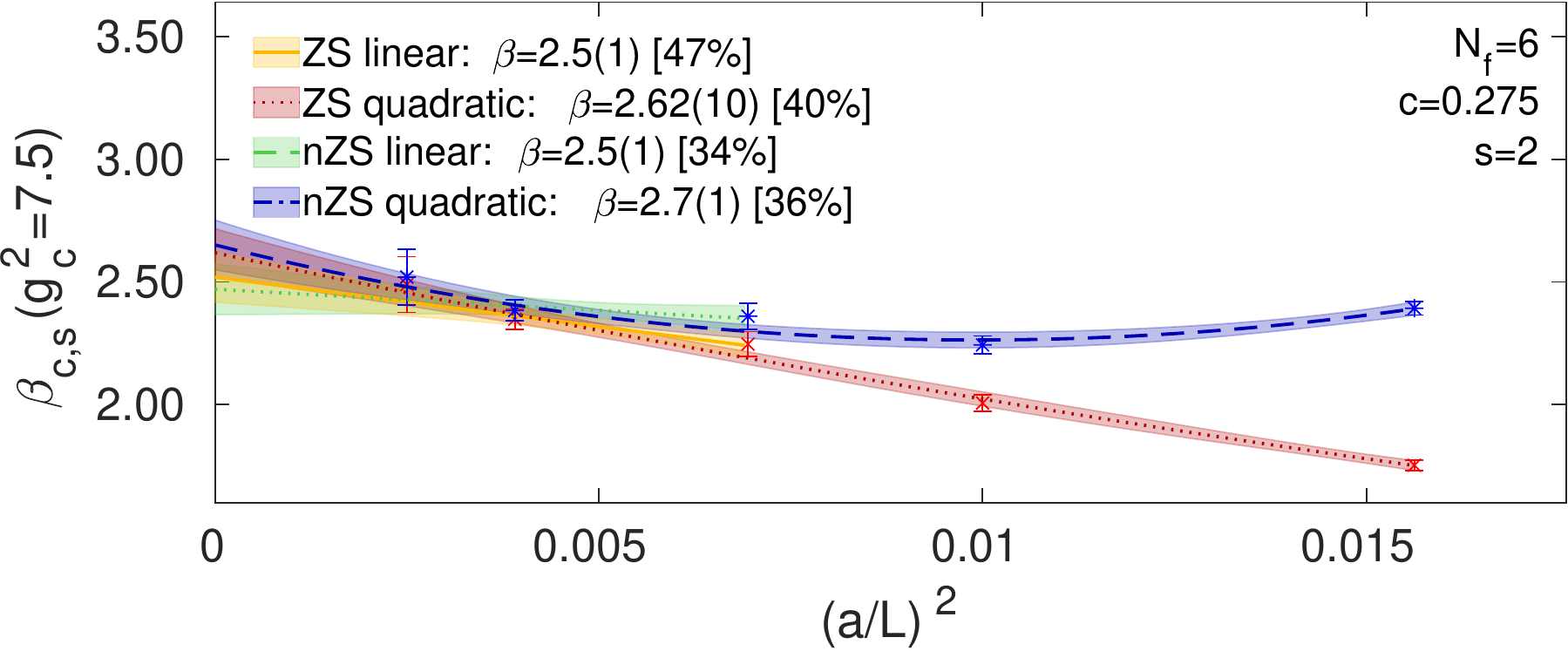}
  \end{minipage}
  \captionof{figure}{Discrete step-scaling $\beta$-function for $N_f=6$ in the $c=0.275$ gradient flow scheme for our preferred nZS (left) and ZS (right) data sets. The symbols in the top row show our results for the finite volume discrete $\beta$ function with scale change $s=2$. The dashed lines with shaded error bands in the same color of the data points show the interpolating fits. We consider two continuum limits: a linear fit (black line with gray error band) in $a^2/L^2$ to the three largest volume pairs and a quadratic fit to all volume pairs (black dash-dotted line). The $p$-values of the continuum extrapolation fits are shown in the plots in the second row. Further details of the continuum extrapolation at selected $g_c^2$ values are presented in the small panels at the bottom where the legend lists the extrapolated values in the continuum limit with $p$-values in brackets. Only statistical errors are shown.}    
  \label{Fig.Nf6_beta_c275}
 % \end{figure*}
\end{minipage}
\pagebreak

\begin{figure*}[t]
  \begin{minipage}{0.49\textwidth}
   \flushright 
   \includegraphics[width=0.96\textwidth]{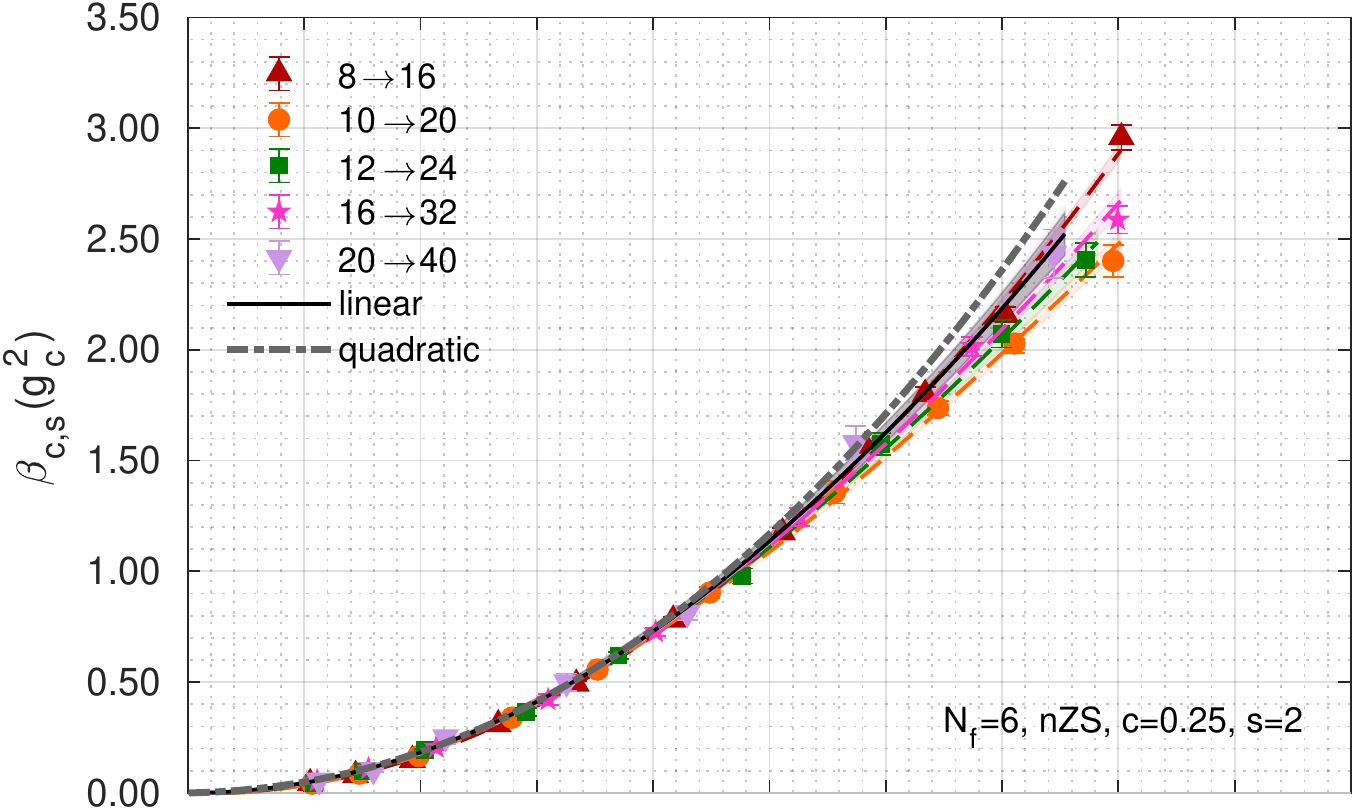}\\
   \includegraphics[width=0.937\textwidth]{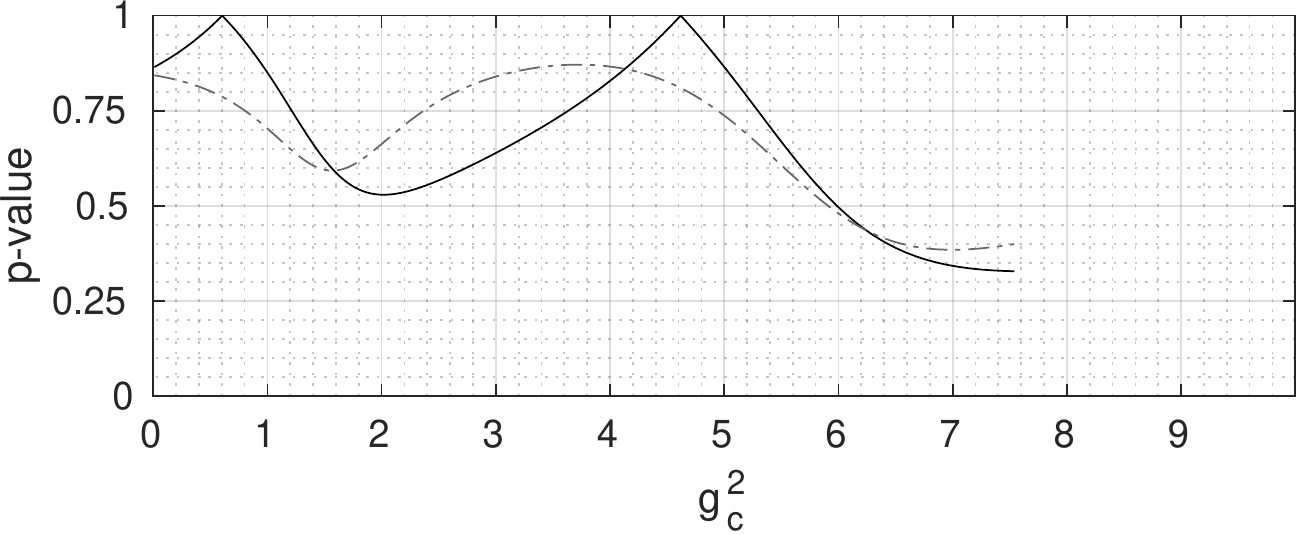} %\\[3mm]
   \includegraphics[width=0.96\textwidth]{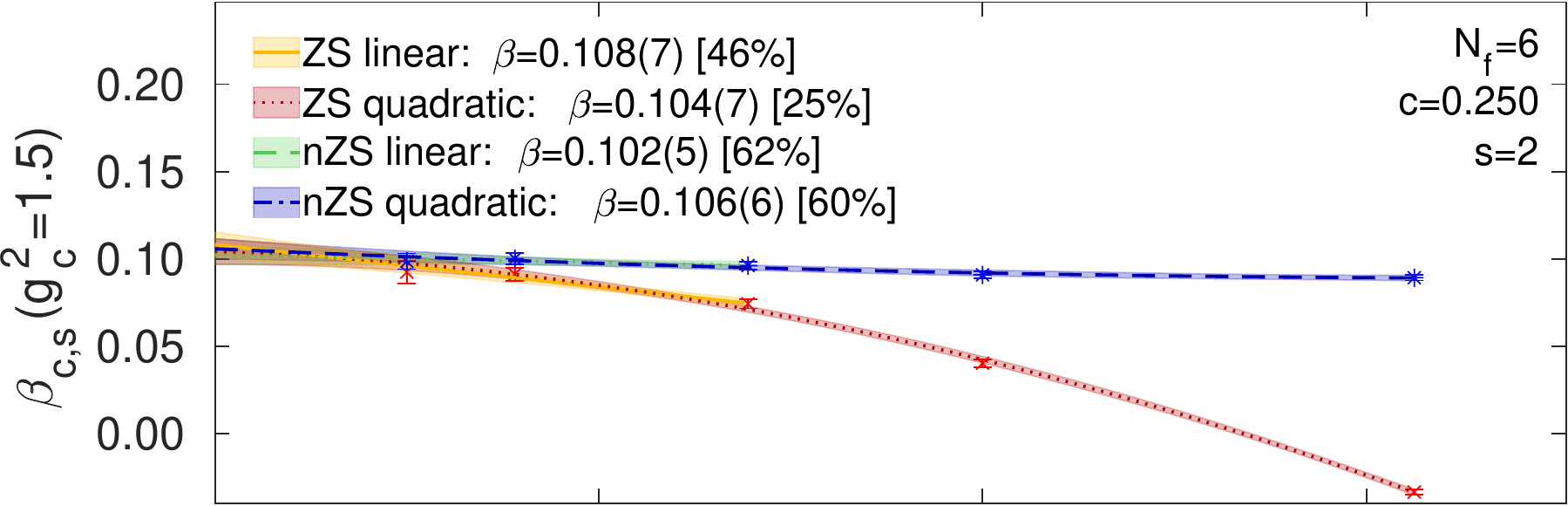}\\
   \includegraphics[width=0.96\textwidth]{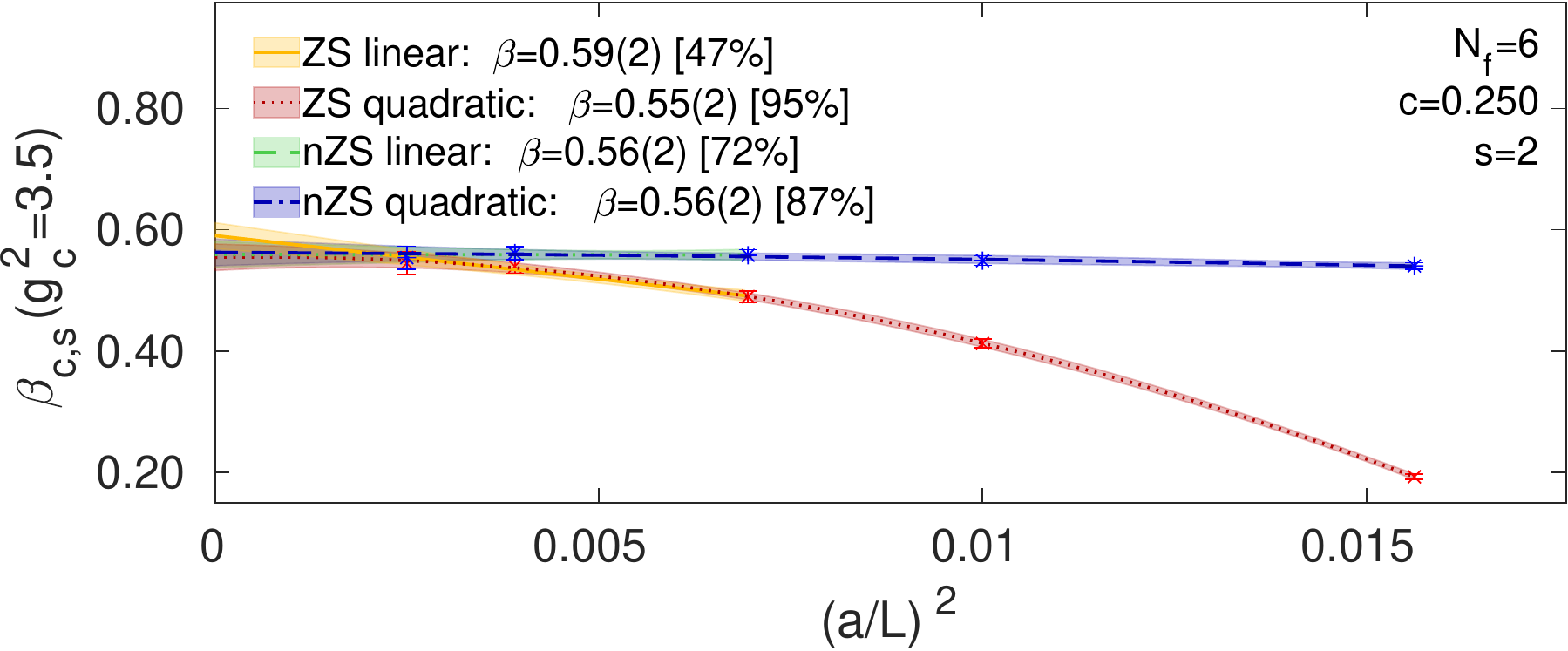}
  \end{minipage}
  \begin{minipage}{0.49\textwidth}
    \flushright
    \includegraphics[width=0.96\textwidth]{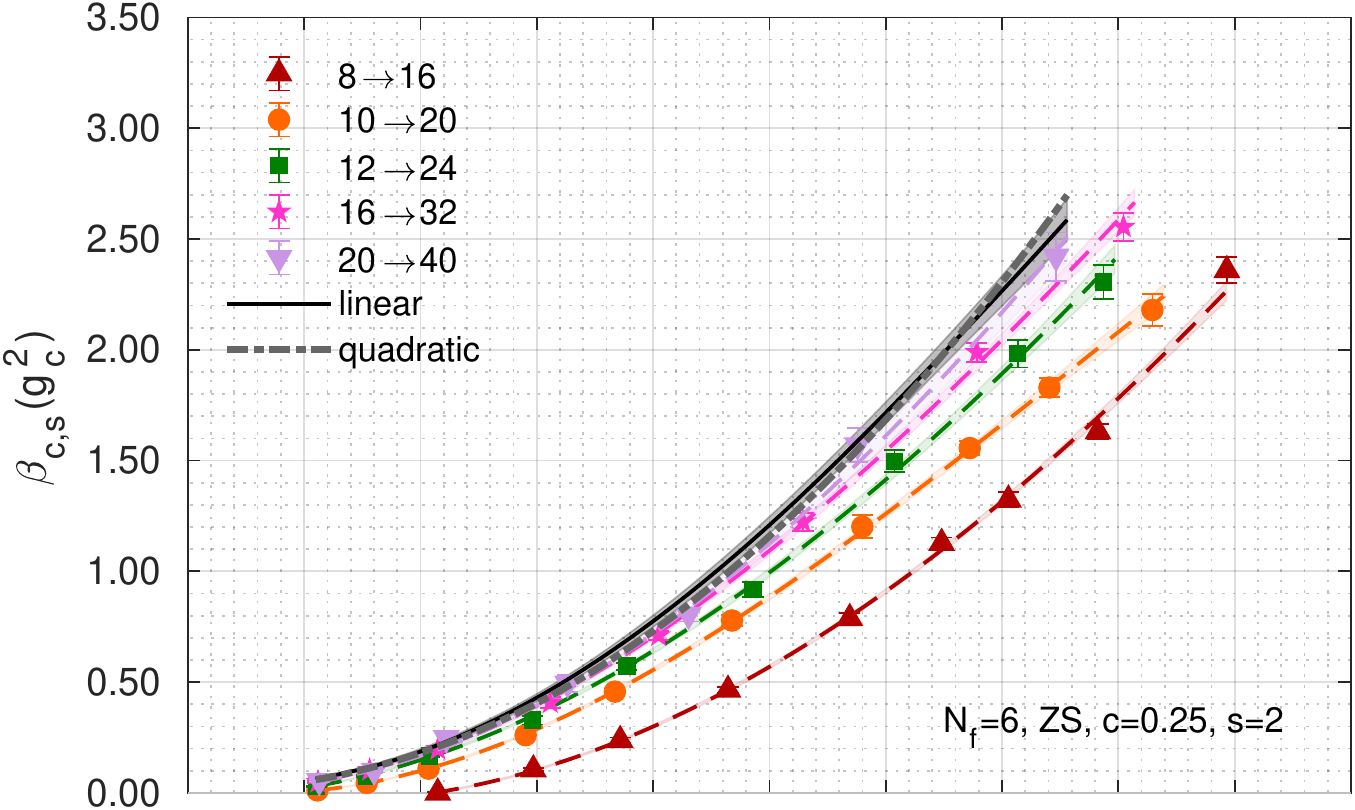}\\    
    \includegraphics[width=0.937\textwidth]{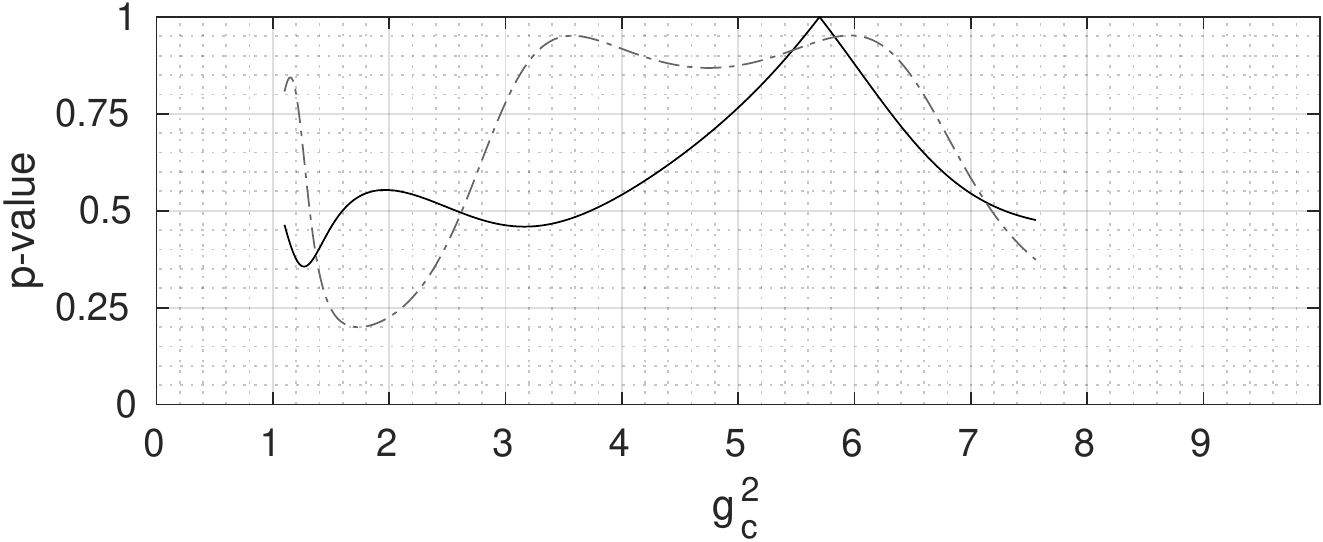} %\\[3mm]
    \includegraphics[width=0.96\textwidth]{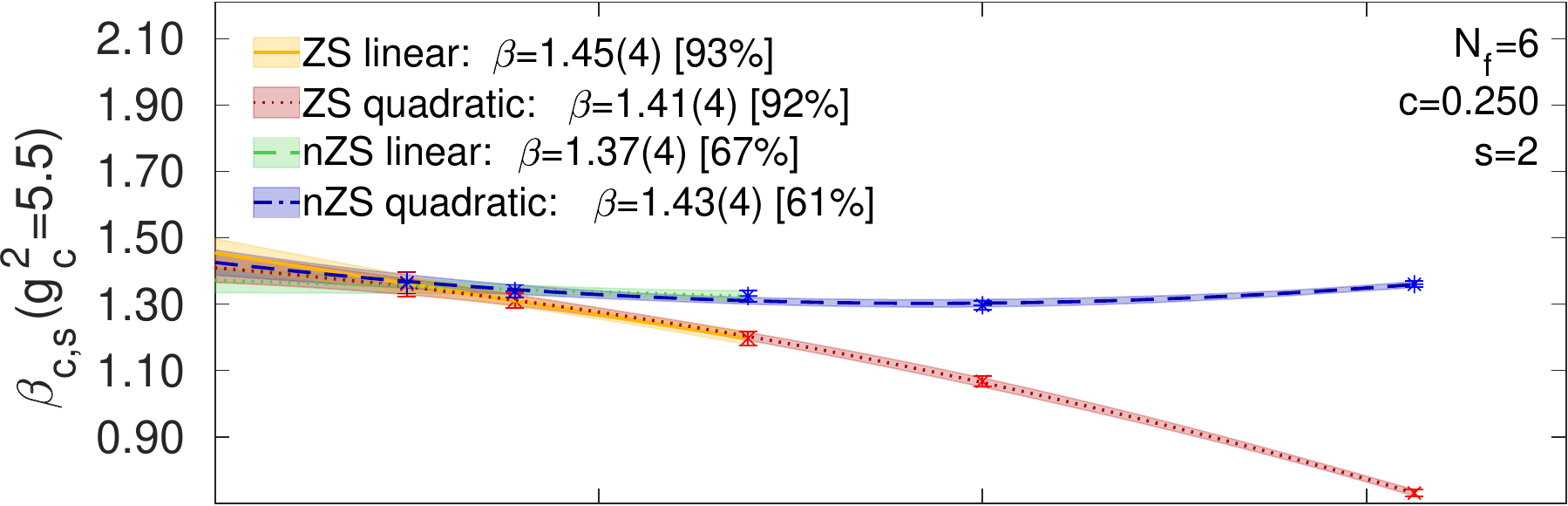}\\
    \includegraphics[width=0.96\textwidth]{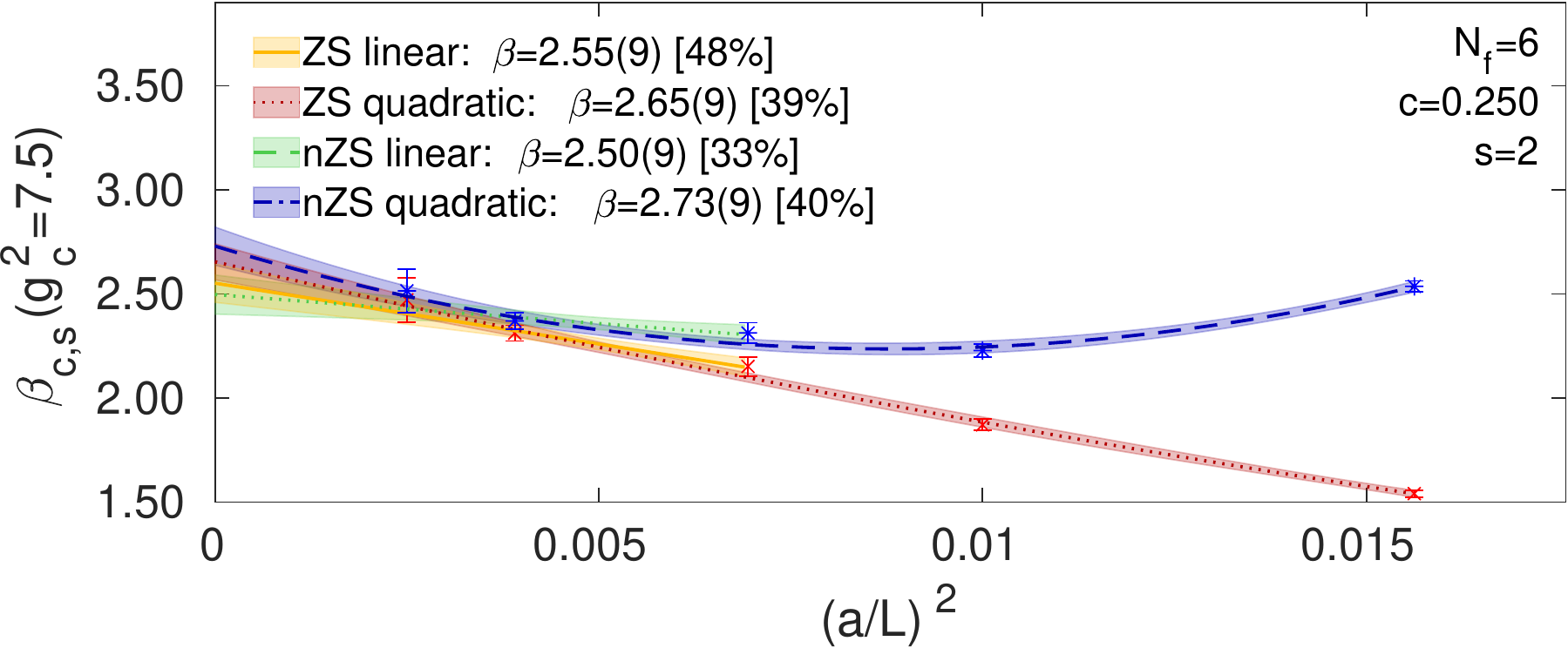}
  \end{minipage}
  \caption{Discrete step-scaling $\beta$-function for $N_f=6$ in the $c=0.250$ gradient flow scheme for our preferred nZS (left) and ZS (right) data sets. The symbols in the top row show our results for the finite volume discrete $\beta$ function with scale change $s=2$. The dashed lines with shaded error bands in the same color of the data points show the interpolating fits. We consider two continuum limits: a linear fit (black line with gray error band) in $a^2/L^2$ to the three largest volume pairs and a quadratic fit to all volume pairs (black dash-dotted line). The $p$-values of the continuum extrapolation fits are shown in the plots in the second row. Further details of the continuum extrapolation at selected $g_c^2$ values are presented in the small panels at the bottom where the legend lists the extrapolated values in the continuum limit with $p$-values in brackets. Only statistical errors are shown.}    
  \label{Fig.Nf6_beta_c250}
\end{figure*}

\begin{figure*}[t]
  \begin{minipage}{0.49\textwidth}
   \flushright 
   \includegraphics[width=0.96\textwidth]{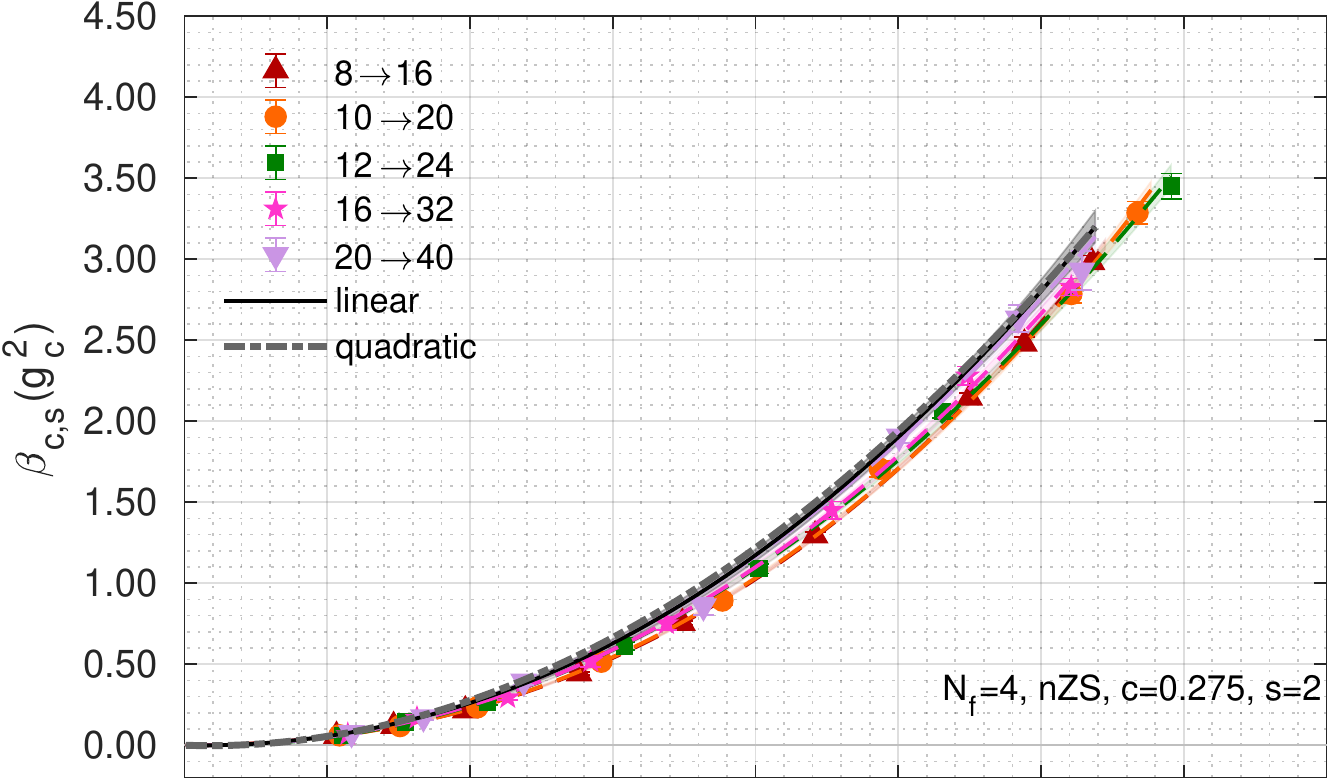}\\
   \includegraphics[width=0.937\textwidth]{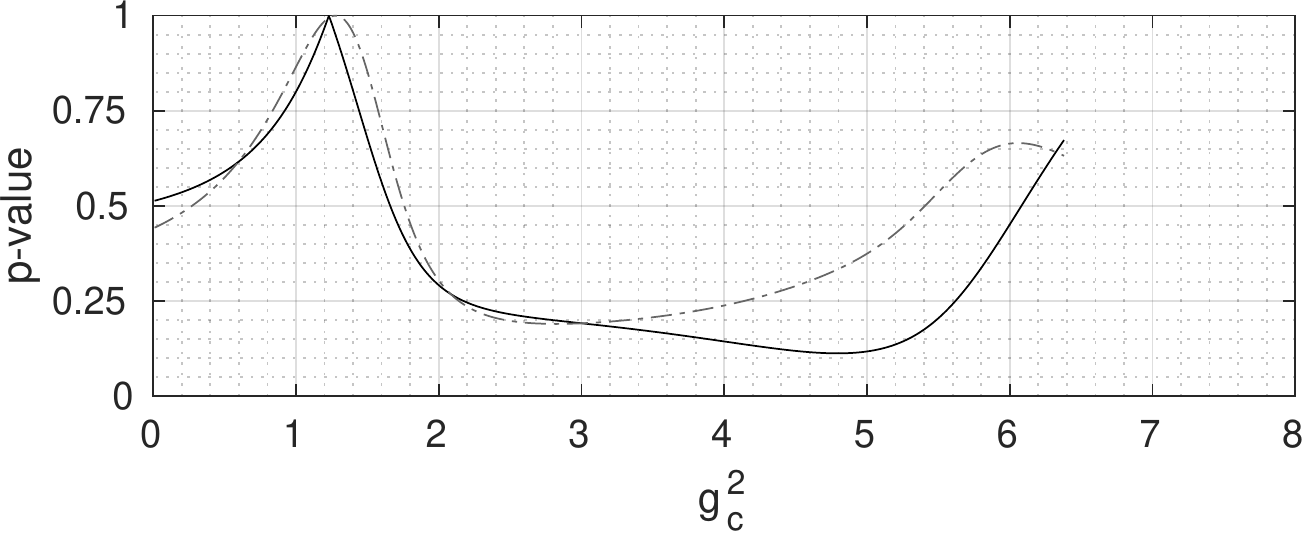} %\\[3mm]
   \includegraphics[width=0.96\textwidth]{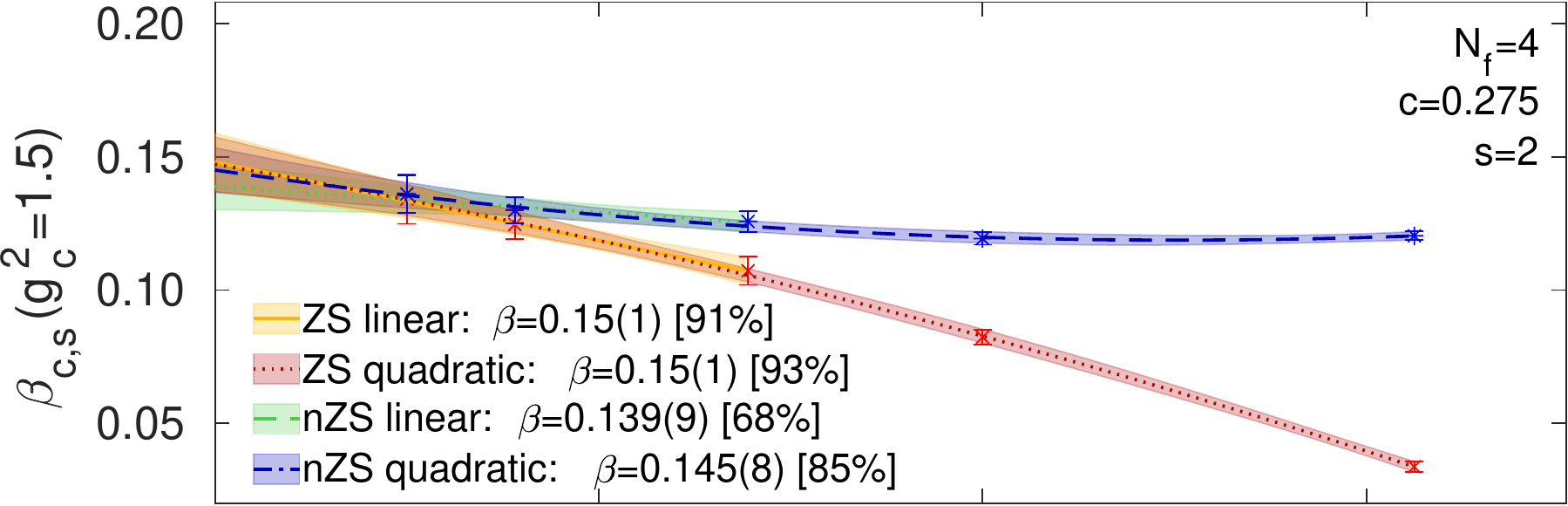}\\
   \includegraphics[width=0.96\textwidth]{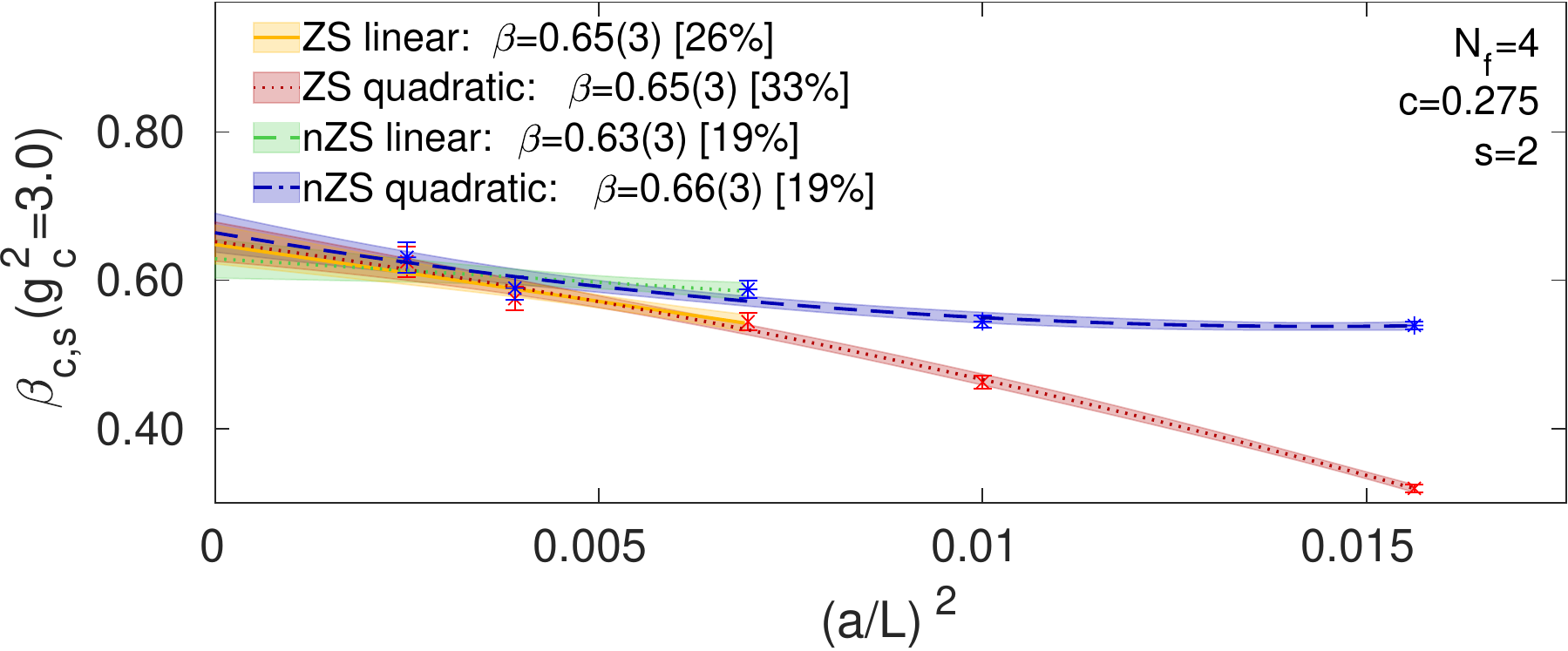}
  \end{minipage}
  \begin{minipage}{0.49\textwidth}
    \flushright
    \includegraphics[width=0.96\textwidth]{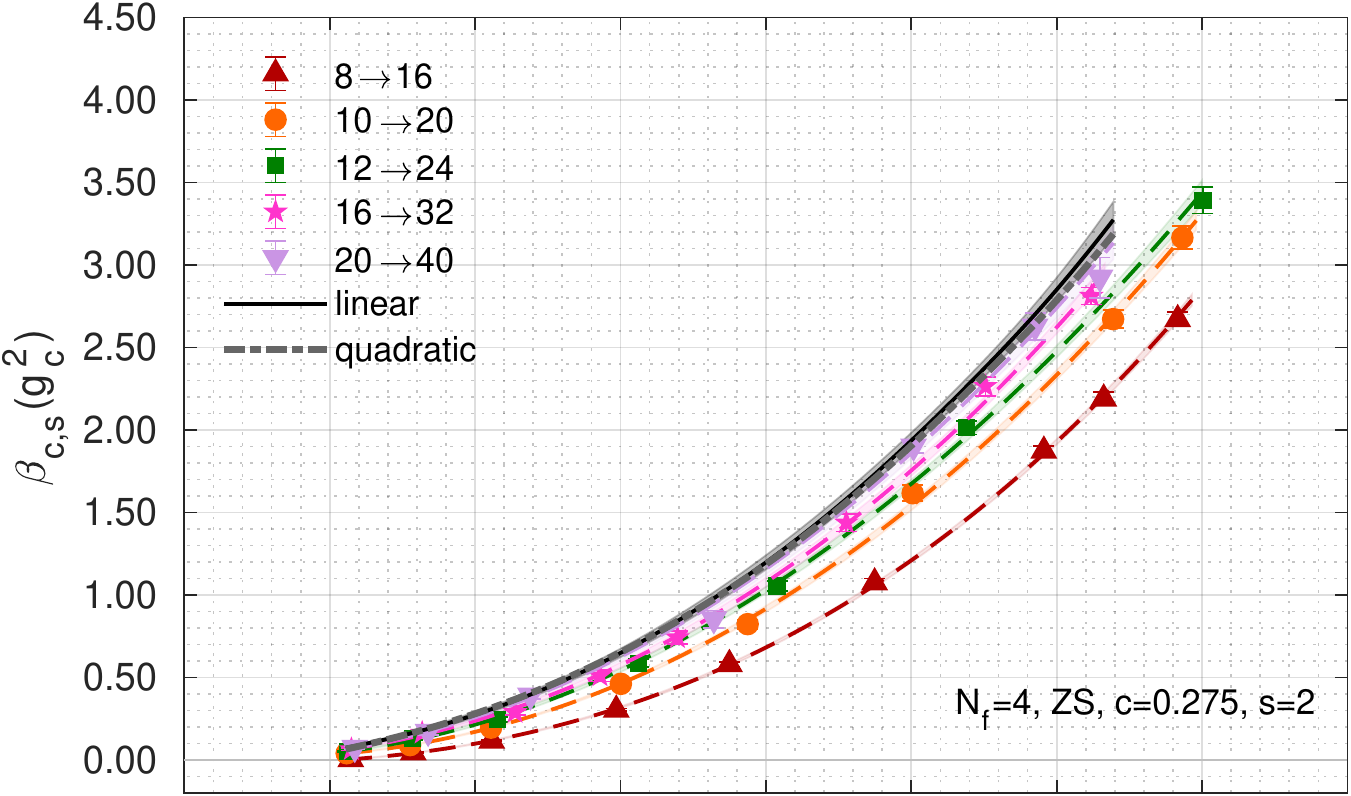}\\    
    \includegraphics[width=0.937\textwidth]{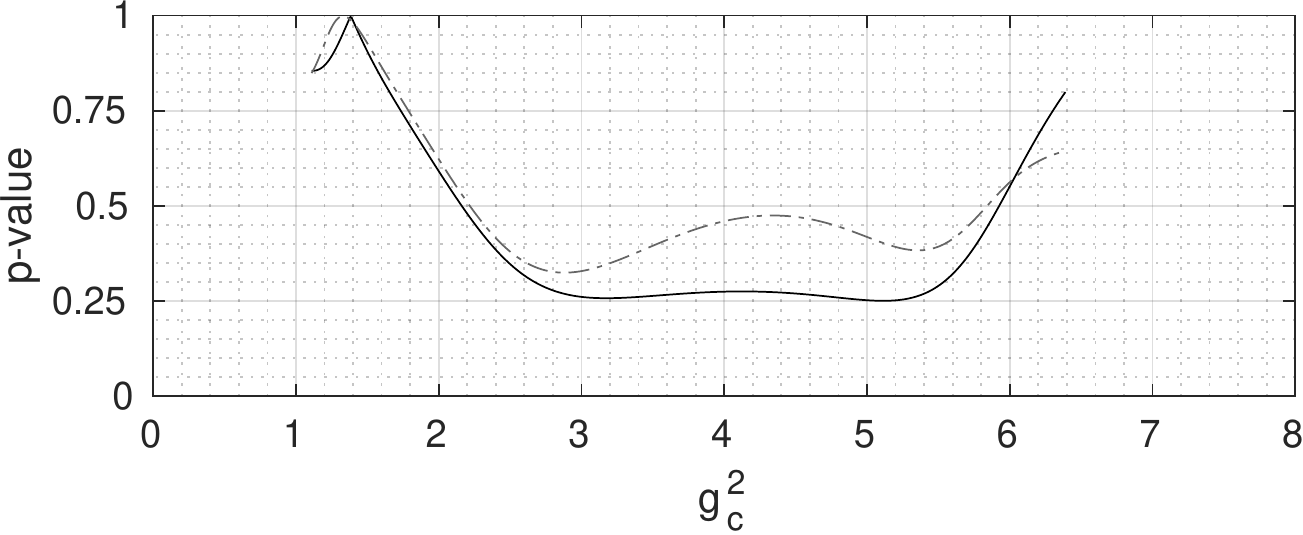} %\\[3mm]
    \includegraphics[width=0.96\textwidth]{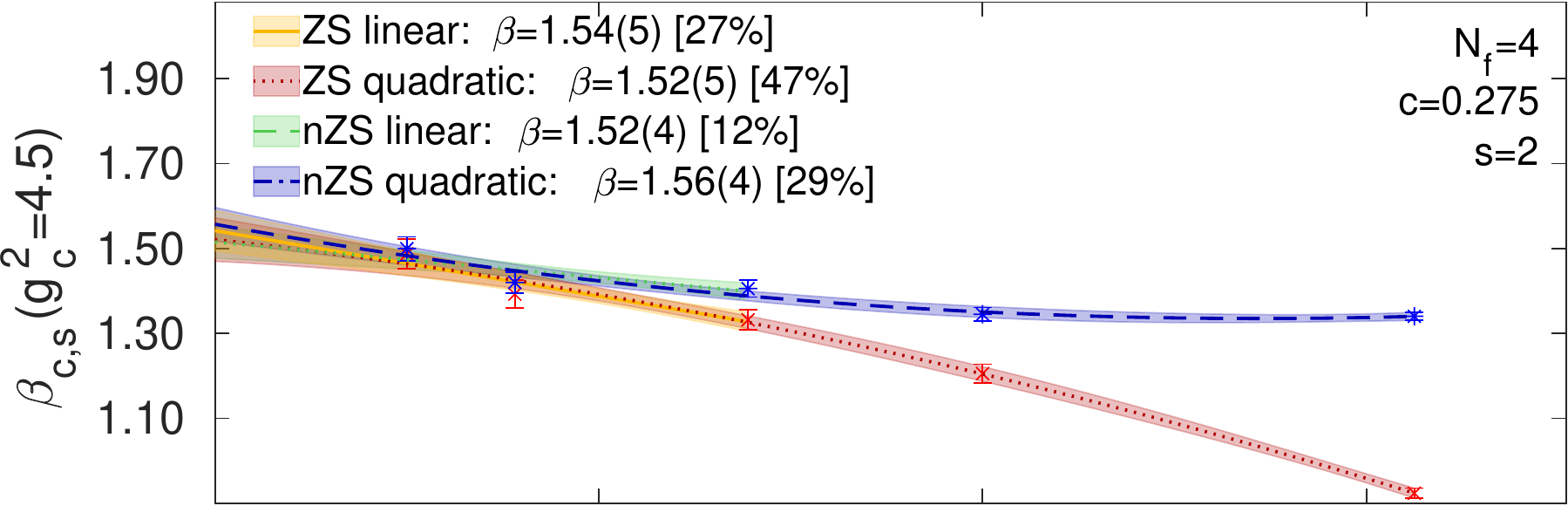}\\
    \includegraphics[width=0.96\textwidth]{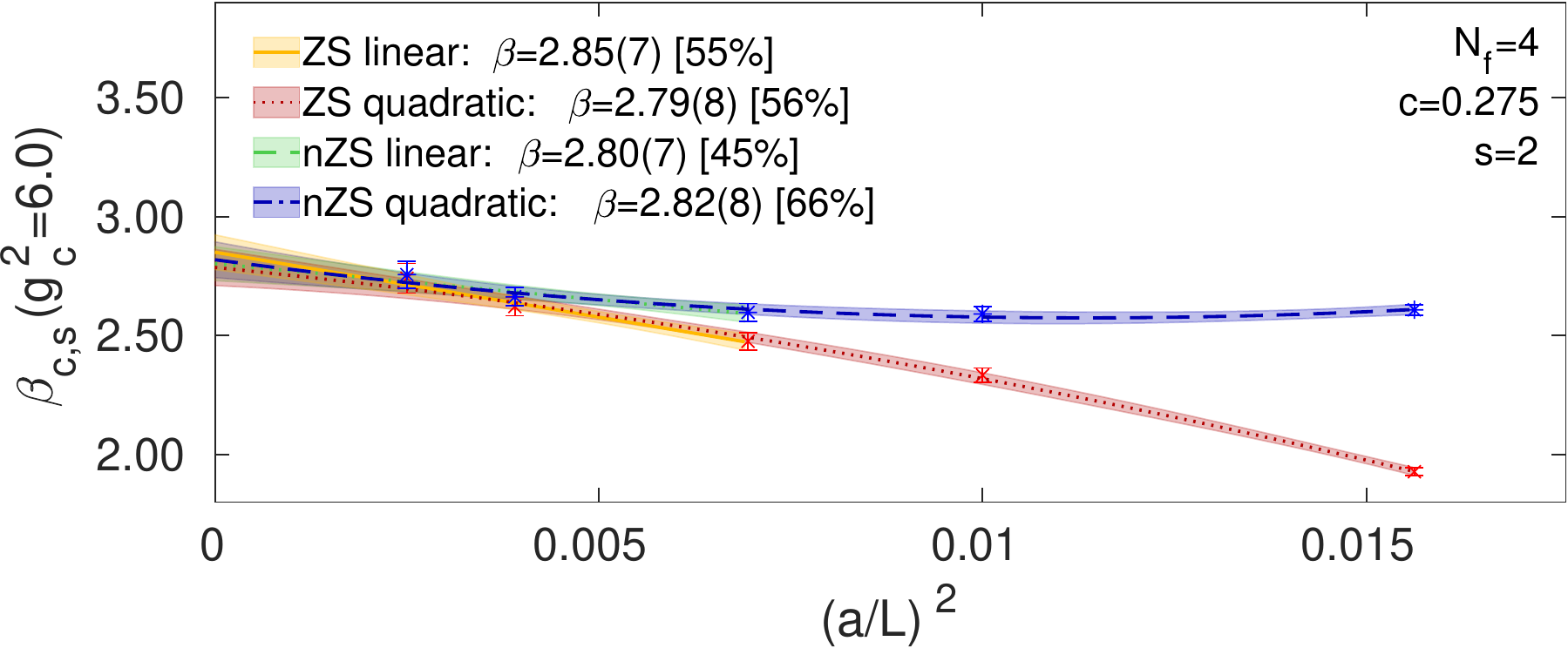}
  \end{minipage}
  \caption{Discrete step-scaling $\beta$-function for $N_f=4$ in the $c=0.275$ gradient flow scheme for our preferred nZS (left) and ZS (right) data sets. The symbols in the top row show our results for the finite volume discrete $\beta$ function with scale change $s=2$. The dashed lines with shaded error bands in the same color of the data points show the interpolating fits. We consider two continuum limits: a linear fit (black line with gray error band) in $a^2/L^2$ to the three largest volume pairs and a quadratic fit to all volume pairs (black dash-dotted line). The $p$-values of the continuum extrapolation fits are shown in the plots in the second row. Further details of the continuum extrapolation at selected $g_c^2$ values are presented in the small panels at the bottom where the legend lists the extrapolated values in the continuum limit with $p$-values in brackets. Only statistical errors are shown.}    
  \label{Fig.Nf4_beta_c275}
\end{figure*}

\begin{figure*}[t]
  \begin{minipage}{0.49\textwidth}
   \flushright 
   \includegraphics[width=0.96\textwidth]{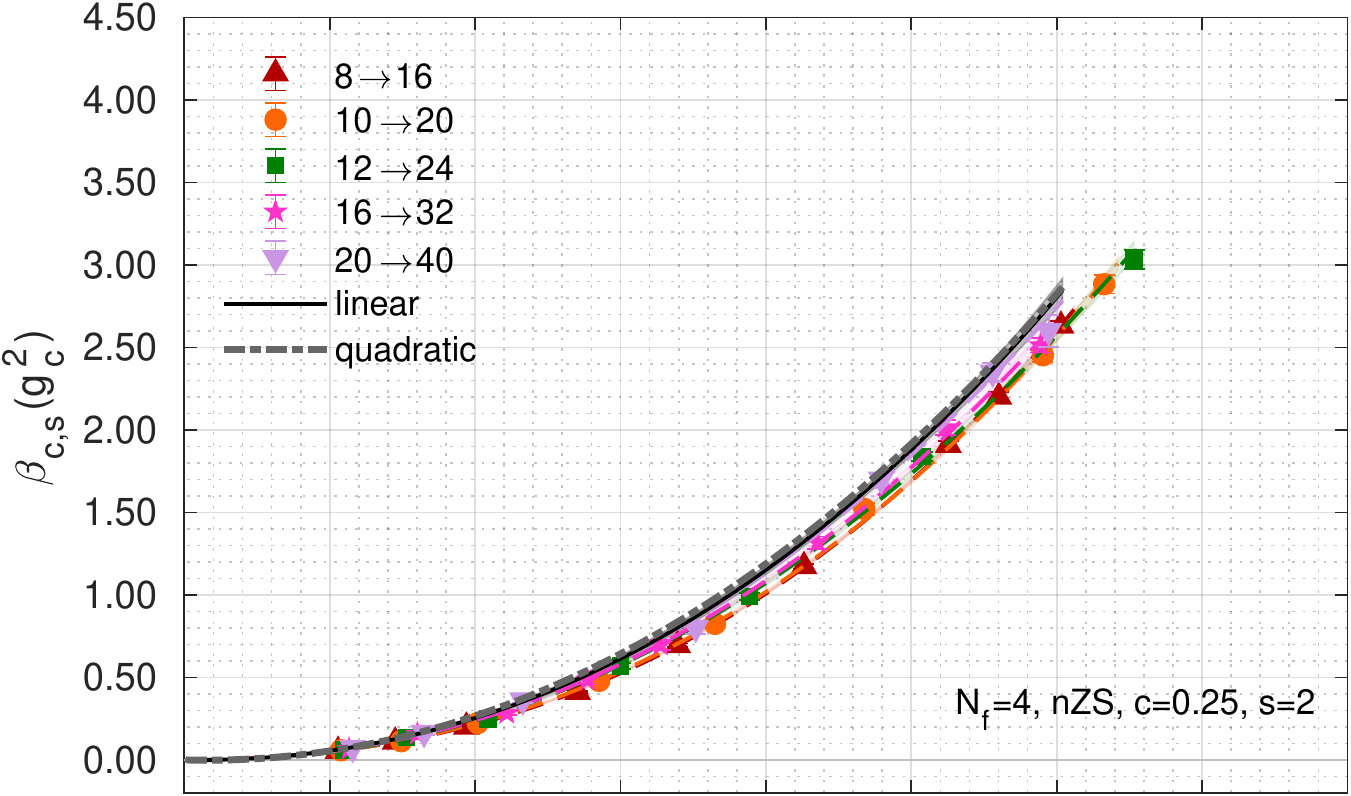}\\
   \includegraphics[width=0.937\textwidth]{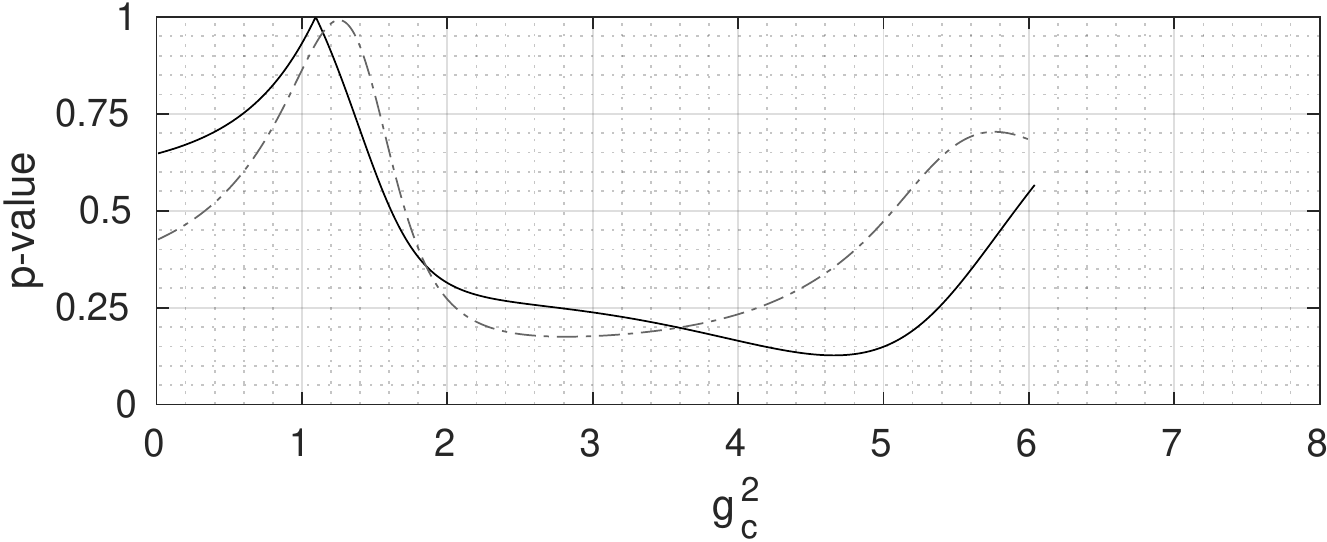} %\\[3mm]
   \includegraphics[width=0.96\textwidth]{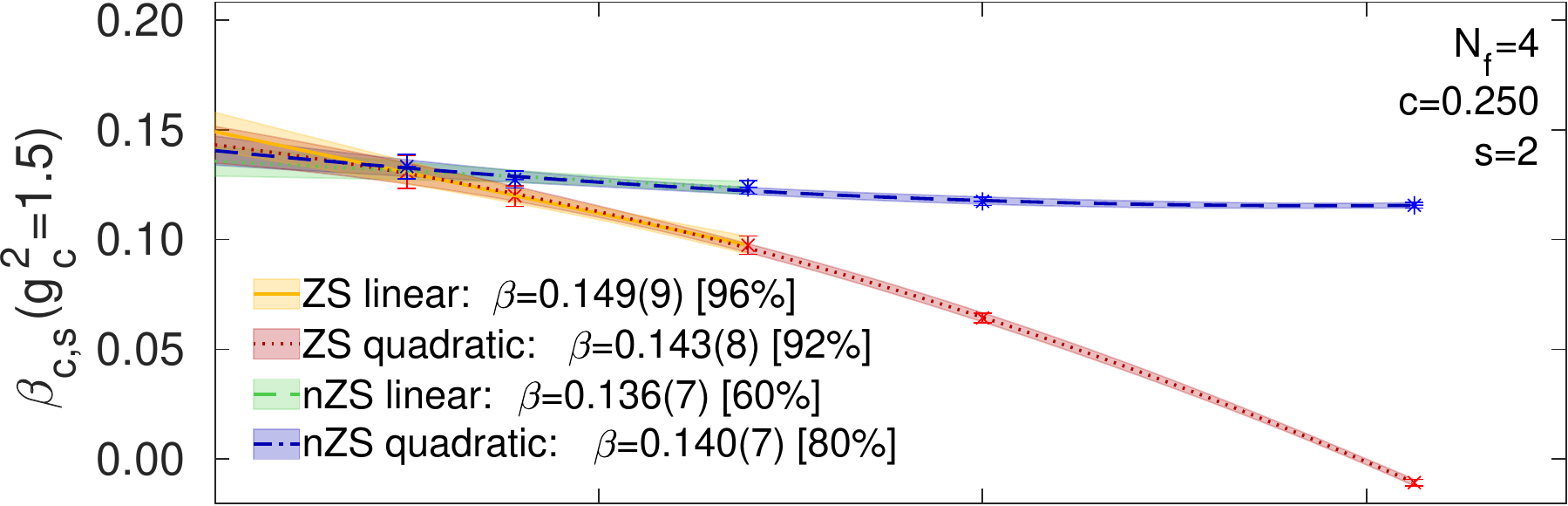}\\
   \includegraphics[width=0.96\textwidth]{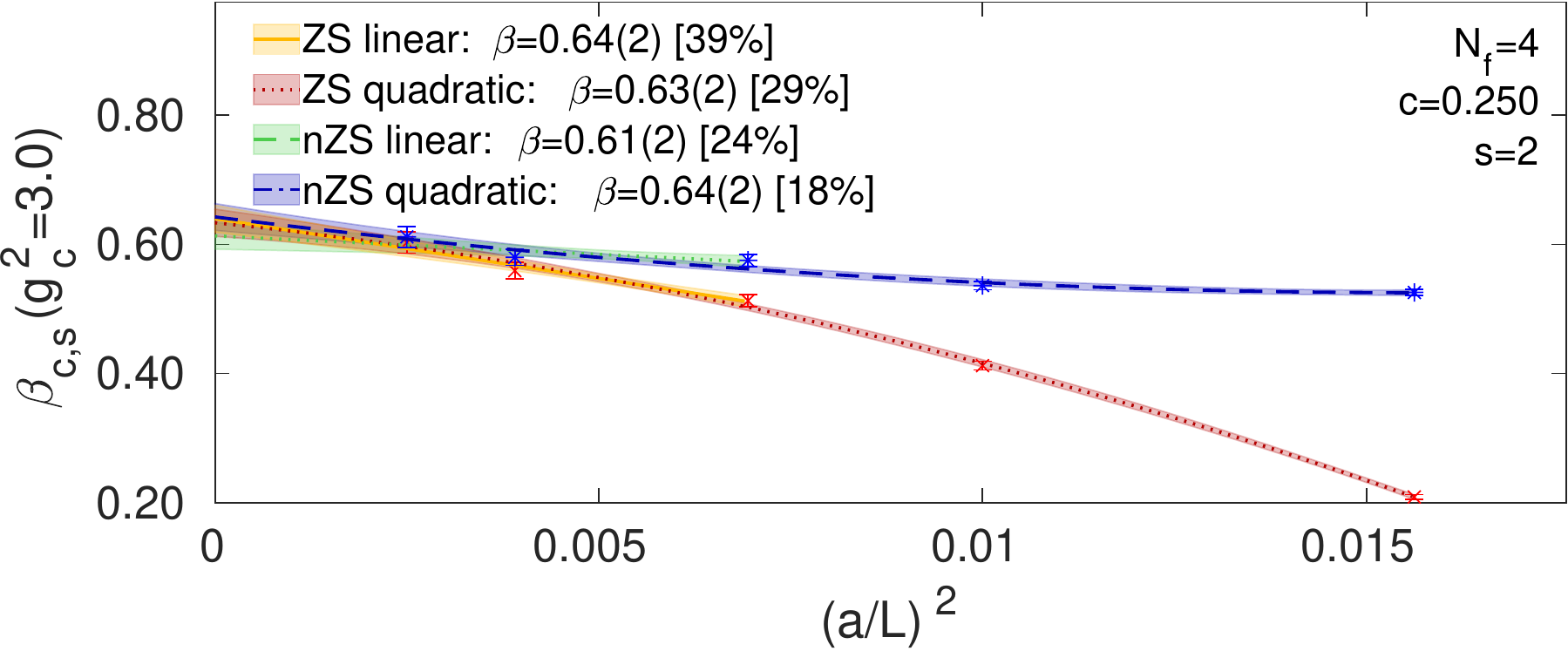}
  \end{minipage}
  \begin{minipage}{0.49\textwidth}
    \flushright
    \includegraphics[width=0.96\textwidth]{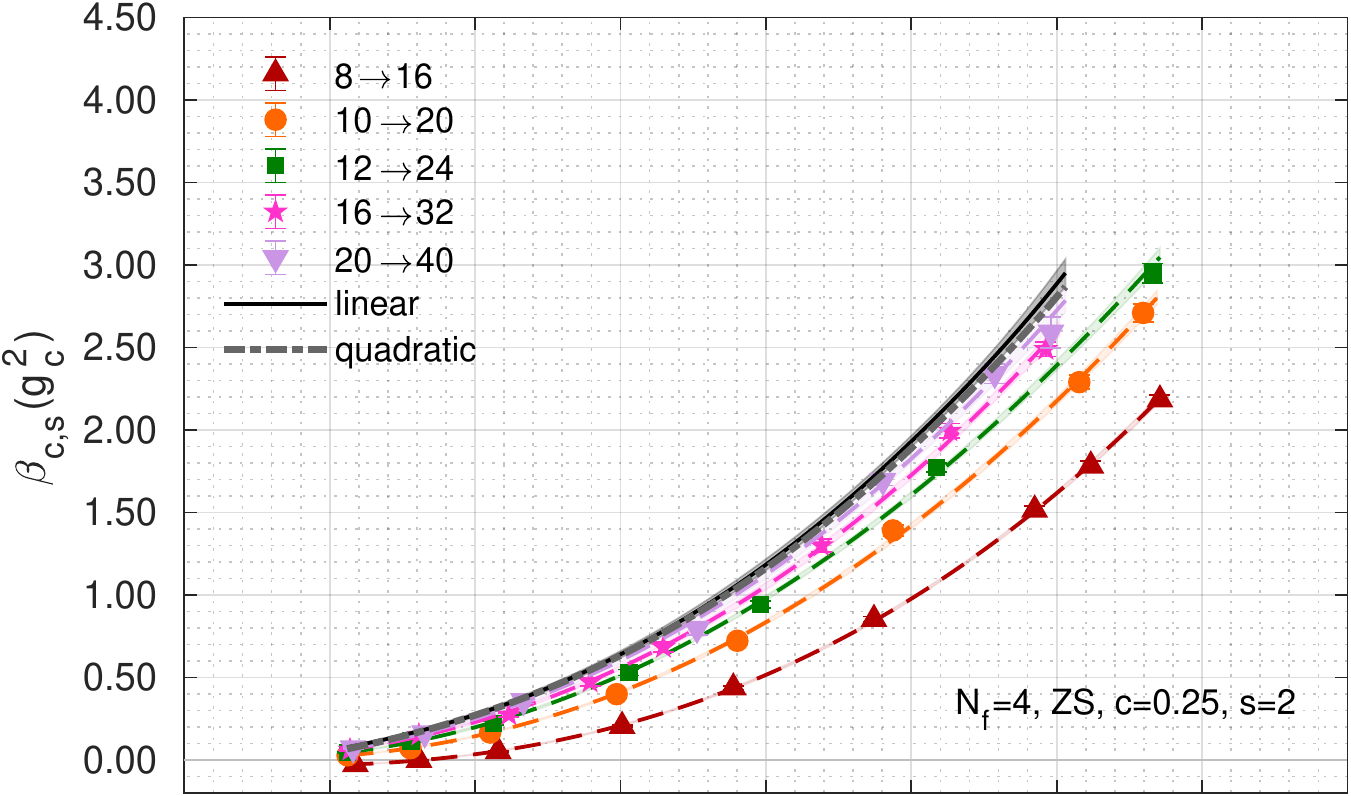}\\    
    \includegraphics[width=0.937\textwidth]{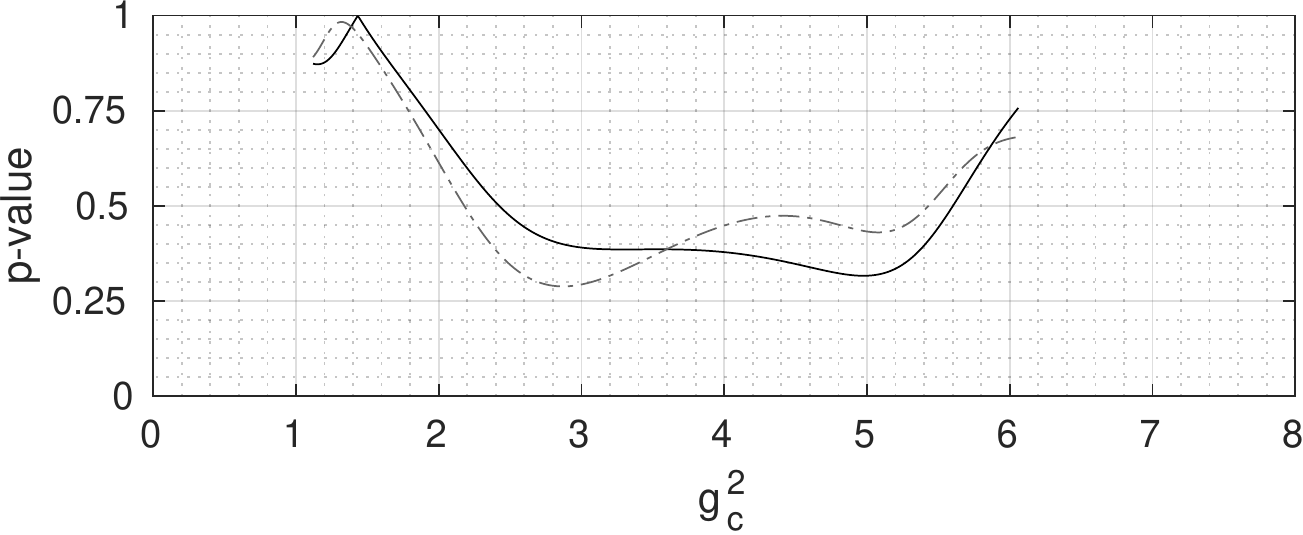} %\\[3mm]
    \includegraphics[width=0.96\textwidth]{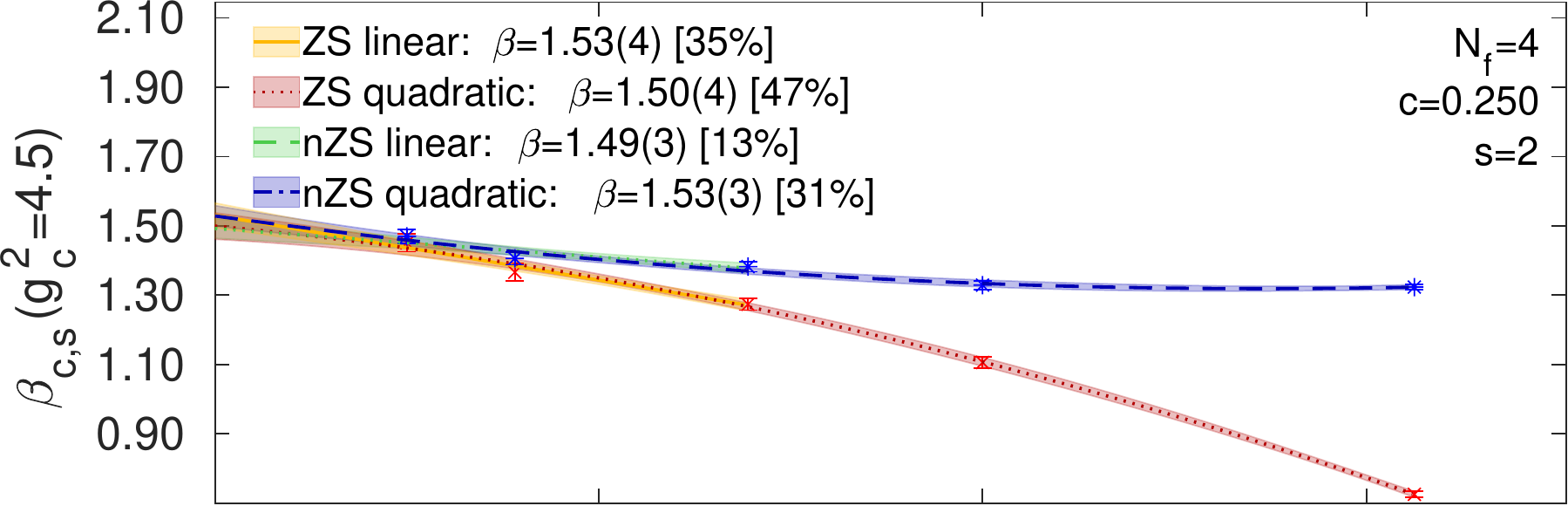}\\
    \includegraphics[width=0.96\textwidth]{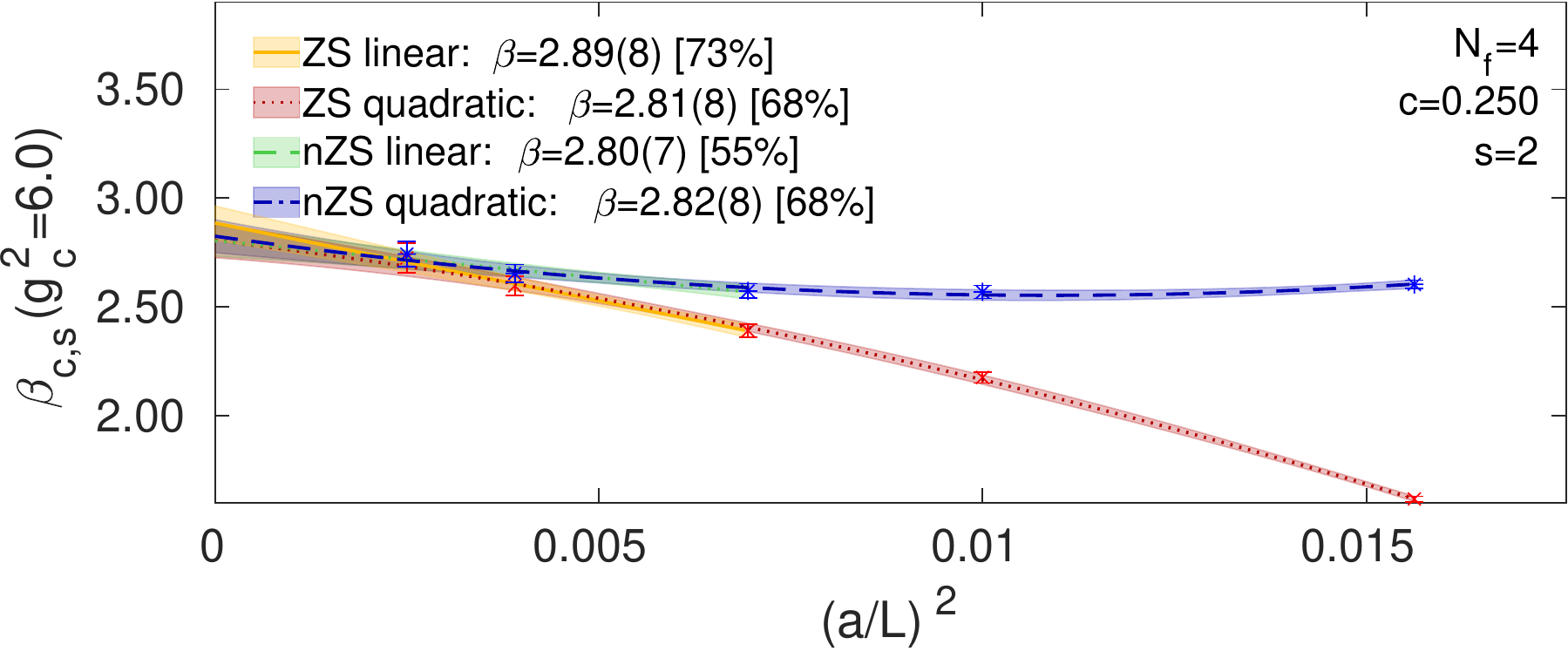}
  \end{minipage}
  \caption{Discrete step-scaling $\beta$-function for $N_f=4$ in the $c=0.250$ gradient flow scheme for our preferred nZS (left) and ZS (right) data sets. The symbols in the top row show our results for the finite volume discrete $\beta$ function with scale change $s=2$. The dashed lines with shaded error bands in the same color of the data points show the interpolating fits. We consider two continuum limits: a linear fit (black line with gray error band) in $a^2/L^2$ to the three largest volume pairs and a quadratic fit to all volume pairs (black dash-dotted line). The $p$-values of the continuum extrapolation fits are shown in the plots in the second row. Further details of the continuum extrapolation at selected $g_c^2$ values are presented in the small panels at the bottom where the legend lists the extrapolated values in the continuum limit with $p$-values in brackets. Only statistical errors are shown.}  
  \label{Fig.Nf4_beta_c250}
\end{figure*}

%%%%%%%%%%%%%%%%%%%%%%%%%%%%%% BIBLIOGRAPY %%%%%%%%%%%%%%%%%%%%%%%%%
%\clearpage
\bibliography{../General/BSM}
\end{document}

%% file: gcSq_Nf6_nZS_ZS.tex
8 & 8.50 & 631 & 1.0638(19)  & 1.1180(20) & 0.53(8) & 1.0584(15)  & 1.1376(17) & 0.52(8) & 1.0524(12)  & 1.1713(13) & 0.50(8)\\ 
 8 & 7.00 & 631 & 1.4641(29)  & 1.5388(30) & 0.61(10) & 1.4542(23)  & 1.5630(25) & 0.58(9) & 1.4429(18)  & 1.6058(20) & 0.56(8)\\ 
 8 & 6.00 & 631 & 1.9748(38)  & 2.0754(40) & 0.58(9) & 1.9540(29)  & 2.1001(31) & 0.50(6) & 1.9308(23)  & 2.1489(25) & 0.47(5)\\ 
 8 & 5.20 & 631 & 2.7586(54)  & 2.8991(57) & 0.51(6) & 2.7161(43)  & 2.9192(46) & 0.48(5) & 2.6693(33)  & 2.9709(37) & 0.45(5)\\ 
 8 & 4.80 & 631 & 3.4835(73)  & 3.6610(77) & 0.57(8) & 3.4154(62)  & 3.6708(67) & 0.61(10) & 3.3403(51)  & 3.7176(56) & 0.63(10)\\ 
 8 & 4.50 & 631 & 4.403(10)  & 4.627(11) & 0.58(9) & 4.2933(86)  & 4.6144(92) & 0.57(9) & 4.1731(69)  & 4.6444(76) & 0.57(9)\\ 
 8 & 4.30 & 631 & 5.473(14)  & 5.752(14) & 0.56(8) & 5.302(11)  & 5.698(12) & 0.57(8) & 5.1135(92)  & 5.691(10) & 0.57(8)\\ 
 8 & 4.20 & 631 & 6.285(18)  & 6.605(19) & 0.6(1) & 6.066(14)  & 6.520(15) & 0.6(1) & 5.824(12)  & 6.481(14) & 0.7(1)\\ 
 8 & 4.15 & 631 & 6.913(25)  & 7.265(27) & 0.8(1) & 6.641(21)  & 7.138(23) & 0.8(1) & 6.339(17)  & 7.055(19) & 0.8(2)\\ 
 8 & 4.10 & 631 & 7.764(31)  & 8.159(32) & 0.8(1) & 7.418(26)  & 7.973(28) & 0.8(1) & 7.031(20)  & 7.825(23) & 0.7(1)\\ 
 8 & 4.05 & 631 & 9.012(75)  & 9.471(79) & 2.5(7) & 8.552(61)  & 9.191(65) & 2.4(7) & 8.027(48)  & 8.933(54) & 2.4(7)\\ 
 \hline 
10 & 8.50 & 631 & 1.0836(22)  & 1.1050(23) & 0.63(10) & 1.0769(19)  & 1.1073(19) & 0.62(10) & 1.0696(15)  & 1.1149(16) & 0.62(10)\\ 
 10 & 7.00 & 619 & 1.5039(30)  & 1.5336(31) & 0.6(1) & 1.4905(24)  & 1.5326(25) & 0.58(9) & 1.4763(19)  & 1.5388(19) & 0.53(9)\\ 
 10 & 6.00 & 614 & 2.0336(39)  & 2.0738(40) & 0.51(6) & 2.0105(32)  & 2.0673(33) & 0.48(5) & 1.9858(26)  & 2.0699(28) & 0.50(4)\\ 
 10 & 5.20 & 625 & 2.8857(61)  & 2.9427(63) & 0.57(8) & 2.8373(49)  & 2.9175(50) & 0.53(7) & 2.7863(38)  & 2.9042(40) & 0.50(6)\\ 
 10 & 4.80 & 631 & 3.6804(92)  & 3.7530(93) & 0.7(1) & 3.6037(73)  & 3.7055(75) & 0.7(1) & 3.5228(57)  & 3.6719(60) & 0.63(10)\\ 
 10 & 4.50 & 631 & 4.762(14)  & 4.856(14) & 0.9(2) & 4.628(11)  & 4.759(12) & 0.8(2) & 4.4888(86)  & 4.6788(90) & 0.8(1)\\ 
 10 & 4.30 & 631 & 5.955(18)  & 6.073(18) & 0.9(2) & 5.763(14)  & 5.926(15) & 0.8(2) & 5.563(11)  & 5.799(12) & 0.8(1)\\ 
 10 & 4.20 & 631 & 6.973(28)  & 7.110(29) & 1.3(3) & 6.716(22)  & 6.906(23) & 1.1(2) & 6.450(17)  & 6.723(18) & 1.0(2)\\ 
 10 & 4.15 & 601 & 7.715(25)  & 7.867(25) & 0.8(2) & 7.419(21)  & 7.629(21) & 0.7(1) & 7.107(17)  & 7.408(18) & 0.7(1)\\ 
 10 & 4.10 & 601 & 8.694(31)  & 8.865(32) & 0.8(2) & 8.337(26)  & 8.572(26) & 0.8(1) & 7.956(21)  & 8.293(22) & 0.7(1)\\ 
 \hline 
12 & 8.50 & 629 & 1.1018(22)  & 1.1123(22) & 0.63(10) & 1.0942(18)  & 1.1089(18) & 0.59(9) & 1.0861(14)  & 1.1074(14) & 0.57(9)\\ 
 12 & 7.00 & 606 & 1.5344(37)  & 1.5490(37) & 0.8(2) & 1.5202(30)  & 1.5406(30) & 0.8(1) & 1.5049(24)  & 1.5345(24) & 0.7(1)\\ 
 12 & 6.00 & 601 & 2.0886(43)  & 2.1085(43) & 0.6(1) & 2.0650(35)  & 2.0927(35) & 0.60(10) & 2.0389(28)  & 2.0789(29) & 0.58(9)\\ 
 12 & 5.20 & 601 & 3.0271(90)  & 3.0559(91) & 1.1(2) & 2.9689(69)  & 3.0087(70) & 0.9(2) & 2.9077(53)  & 2.9648(54) & 0.9(2)\\ 
 12 & 4.80 & 601 & 3.883(12)  & 3.920(12) & 1.1(2) & 3.7946(97)  & 3.8455(98) & 1.0(2) & 3.7013(75)  & 3.7739(76) & 0.9(2)\\ 
 12 & 4.50 & 603 & 5.077(20)  & 5.126(20) & 1.4(3) & 4.924(15)  & 4.990(15) & 1.3(3) & 4.766(11)  & 4.859(12) & 1.1(3)\\ 
 12 & 4.30 & 601 & 6.410(27)  & 6.471(27) & 1.6(4) & 6.188(21)  & 6.271(21) & 1.5(4) & 5.959(15)  & 6.076(16) & 1.3(3)\\ 
 12 & 4.20 & 607 & 7.608(26)  & 7.680(26) & 1.0(2) & 7.304(21)  & 7.402(21) & 0.9(2) & 6.998(16)  & 7.135(17) & 0.9(2)\\ 
 12 & 4.15 & 620 & 8.419(38)  & 8.499(38) & 1.6(4) & 8.072(30)  & 8.181(30) & 1.5(4) & 7.722(23)  & 7.873(24) & 1.4(3)\\ 
 \hline 
16 & 8.50 & 300 & 1.1307(42)  & 1.1341(42) & 1.0(3) & 1.1216(34)  & 1.1264(34) & 1.0(3) & 1.1120(28)  & 1.1189(28) & 0.9(3)\\ 
 16 & 7.00 & 291 & 1.5822(54)  & 1.5871(54) & 0.9(2) & 1.5689(41)  & 1.5756(41) & 0.7(2) & 1.5537(31)  & 1.5634(31) & 0.6(1)\\ 
 16 & 6.00 & 291 & 2.197(11)  & 2.204(11) & 1.6(5) & 2.1684(82)  & 2.1777(83) & 1.3(4) & 2.1368(57)  & 2.1501(58) & 0.9(3)\\ 
 16 & 5.20 & 281 & 3.231(16)  & 3.241(16) & 1.6(5) & 3.166(13)  & 3.180(13) & 1.4(4) & 3.0979(94)  & 3.1172(95) & 1.2(4)\\ 
 16 & 4.80 & 281 & 4.236(26)  & 4.249(26) & 1.9(7) & 4.132(20)  & 4.150(20) & 1.7(6) & 4.022(16)  & 4.047(16) & 1.7(6)\\ 
 16 & 4.50 & 281 & 5.625(29)  & 5.643(29) & 1.5(5) & 5.443(22)  & 5.466(22) & 1.3(4) & 5.256(16)  & 5.289(16) & 1.0(3)\\ 
 16 & 4.30 & 281 & 7.341(48)  & 7.363(48) & 2.1(8) & 7.040(38)  & 7.071(38) & 1.9(7) & 6.742(29)  & 6.784(29) & 1.8(6)\\ 
 16 & 4.20 & 283 & 8.775(57)  & 8.801(57) & 1.7(6) & 8.383(44)  & 8.419(44) & 1.5(5) & 7.995(32)  & 8.045(32) & 1.3(4)\\ 
 16 & 4.15 & 272 & 9.736(72)  & 9.765(72) & 2.3(9) & 9.281(54)  & 9.321(55) & 2.0(7) & 8.835(42)  & 8.890(42) & 1.9(7)\\ 
 16 & 4.10 & 282 & 11.093(80)  & 11.127(80) & 2.2(8) & 10.550(58)  & 10.596(58) & 1.8(6) & 10.023(42)  & 10.086(43) & 1.5(5)\\ 
 16 & 4.05 & 273 & 13.51(12)  & 13.55(12) & 3(1) & 12.801(85)  & 12.856(85) & 2.3(9) & 12.128(62)  & 12.203(62) & 1.8(7)\\ 
 \hline 
20 & 8.50 & 219 & 1.1408(61)  & 1.1422(61) & 1.5(6) & 1.1348(48)  & 1.1368(48) & 1.4(5) & 1.1276(36)  & 1.1306(36) & 1.1(4)\\ 
 20 & 7.00 & 224 & 1.6263(72)  & 1.6284(73) & 1.2(4) & 1.6130(49)  & 1.6159(50) & 0.8(2) & 1.5972(38)  & 1.6014(38) & 0.7(2)\\ 
 20 & 6.00 & 197 & 2.282(15)  & 2.285(15) & 1.4(5) & 2.251(12)  & 2.255(12) & 1.2(4) & 2.2162(87)  & 2.2219(88) & 1.1(3)\\ 
 20 & 5.20 & 197 & 3.401(21)  & 3.406(21) & 1.7(6) & 3.331(17)  & 3.337(17) & 1.7(6) & 3.257(13)  & 3.265(13) & 1.5(5)\\ 
 20 & 4.80 & 203 & 4.546(31)  & 4.552(31) & 1.7(7) & 4.422(25)  & 4.430(25) & 1.7(6) & 4.295(19)  & 4.306(19) & 1.6(6)\\ 
 20 & 4.50 & 186 & 6.205(57)  & 6.213(57) & 3(1) & 5.974(44)  & 5.985(44) & 3(1) & 5.743(34)  & 5.757(34) & 2(1)\\ 
 20 & 4.30 & 192 & 8.15(12)  & 8.16(12) & 6(3) & 7.796(95)  & 7.810(95) & 5(3) & 7.445(71)  & 7.464(71) & 5(2)\\ 
 20 & 4.20 & 227 & 9.786(79)  & 9.798(79) & 2.4(10) & 9.318(55)  & 9.334(55) & 1.8(7) & 8.858(40)  & 8.881(40) & 1.6(5)\\ 
 20 & 4.15 & 224 & 11.04(12)  & 11.06(12) & 4(2) & 10.476(83)  & 10.494(83) & 3(1) & 9.919(55)  & 9.944(55) & 2.5(10)\\ 
 20 & 4.10 & 231 & 12.59(17)  & 12.61(17) & 6(3) & 11.93(13)  & 11.95(13) & 5(3) & 11.285(99)  & 11.314(99) & 5(2)\\ 
 \hline 
24 & 8.50 & 210 & 1.1629(64)  & 1.1637(64) & 1.6(6) & 1.1565(49)  & 1.1575(49) & 1.3(5) & 1.1486(33)  & 1.1501(33) & 0.9(3)\\ 
 24 & 7.00 & 206 & 1.6812(81)  & 1.6823(81) & 1.1(4) & 1.6606(66)  & 1.6620(66) & 1.1(3) & 1.6391(52)  & 1.6412(52) & 1.0(3)\\ 
 24 & 6.00 & 208 & 2.389(16)  & 2.390(16) & 2.1(9) & 2.349(11)  & 2.351(11) & 1.4(5) & 2.3071(78)  & 2.3100(78) & 1.1(4)\\ 
 24 & 5.20 & 207 & 3.587(42)  & 3.590(42) & 5(3) & 3.503(33)  & 3.506(33) & 5(2) & 3.415(26)  & 3.419(26) & 4(2)\\ 
 24 & 4.80 & 202 & 4.865(32)  & 4.868(32) & 2.1(8) & 4.713(25)  & 4.717(25) & 2.0(8) & 4.561(20)  & 4.566(20) & 1.9(7)\\ 
 24 & 4.50 & 220 & 6.616(92)  & 6.620(92) & 6(3) & 6.371(67)  & 6.376(68) & 6(3) & 6.123(46)  & 6.131(46) & 4(2)\\ 
 24 & 4.30 & 215 & 8.99(13)  & 9.00(13) & 6(3) & 8.563(93)  & 8.570(93) & 5(3) & 8.141(67)  & 8.151(67) & 4(2)\\ 
 24 & 4.20 & 209 & 11.09(16)  & 11.09(16) & 7(3) & 10.47(12)  & 10.48(12) & 6(3) & 9.871(85)  & 9.883(85) & 5(3)\\ 
 24 & 4.15 & 196 & 12.51(17)  & 12.52(17) & 6(3) & 11.77(13)  & 11.78(13) & 6(3) & 11.06(10)  & 11.07(10) & 6(3)\\ 
 \hline 
32 & 8.50 & 190 & 1.2100(88)  & 1.2102(88) & 2.3(10) & 1.1982(69)  & 1.1985(69) & 2.1(9) & 1.1858(51)  & 1.1863(51) & 1.8(7)\\ 
 32 & 7.00 & 182 & 1.749(11)  & 1.749(11) & 1.9(8) & 1.7278(86)  & 1.7283(86) & 1.7(7) & 1.7049(67)  & 1.7056(67) & 1.5(5)\\ 
 32 & 6.00 & 178 & 2.513(26)  & 2.514(26) & 3(2) & 2.469(20)  & 2.469(20) & 3(1) & 2.422(15)  & 2.423(15) & 3(1)\\ 
 32 & 5.20 & 181 & 3.866(57)  & 3.867(57) & 6(3) & 3.777(44)  & 3.778(44) & 6(3) & 3.681(33)  & 3.683(33) & 5(2)\\ 
 32 & 4.80 & 201 & 5.370(36)  & 5.371(36) & 2.0(8) & 5.204(29)  & 5.205(29) & 1.9(7) & 5.030(23)  & 5.032(23) & 1.7(7)\\ 
 32 & 4.50 & 198 & 7.63(13)  & 7.64(13) & 11(6) & 7.308(88)  & 7.310(88) & 8(4) & 6.982(54)  & 6.985(54) & 5(3)\\ 
 32 & 4.30 & 173 & 10.65(11)  & 10.65(11) & 3(2) & 10.092(74)  & 10.095(74) & 3(1) & 9.537(52)  & 9.541(52) & 2.1(9)\\ 
 32 & 4.20 & 181 & 13.06(13)  & 13.06(13) & 4(2) & 12.31(10)  & 12.32(10) & 4(2) & 11.582(80)  & 11.587(80) & 4(2)\\ 
 \hline 
40 & 8.50 & 100 & 1.2209(87)  & 1.2210(87) & 1.9(9) & 1.2143(66)  & 1.2144(66) & 1.5(7) & 1.2054(50)  & 1.2056(50) & 1.2(5)\\ 
 40 & 7.00 & 98 & 1.763(18)  & 1.763(18) & 3(2) & 1.755(15)  & 1.755(15) & 3(1) & 1.741(12)  & 1.741(12) & 2(1)\\ 
 40 & 6.00 & 101 & 2.682(71)  & 2.682(71) & 8(4) & 2.625(55)  & 2.625(55) & 8(4) & 2.565(42)  & 2.565(42) & 7(4)\\ 
 40 & 5.20 & 102 & 4.167(96)  & 4.167(96) & 7(4) & 4.061(73)  & 4.062(73) & 6(4) & 3.949(53)  & 3.950(53) & 5(3)\\ 
 40 & 4.80 & 102 & 5.797(58)  & 5.798(59) & 3(2) & 5.614(47)  & 5.615(47) & 3(2) & 5.420(38)  & 5.421(38) & 3(1)\\ 
 40 & 4.50 & 121 & 8.77(16)  & 8.77(16) & 6(3) & 8.36(13)  & 8.36(13) & 5(3) & 7.93(10)  & 7.93(10) & 5(3)\\ 
 40 & 4.30 & 161 & 12.14(24)  & 12.15(24) & 11(6) & 11.48(18)  & 11.48(18) & 10(6) & 10.82(13)  & 10.82(13) & 9(5)

%% file: gcSq_Nf4_nZS_ZS.tex
8 & 8.50 & 591 & 1.0726(19)  & 1.1273(20) & 0.50(6) & 1.0652(16)  & 1.1448(17) & 0.49(6) & 1.0574(13)  & 1.1769(14) & 0.49(6)\\ 
 8 & 7.00 & 591 & 1.4792(27)  & 1.5546(29) & 0.48(7) & 1.4650(23)  & 1.5746(25) & 0.50(6) & 1.4501(18)  & 1.6139(20) & 0.49(6)\\ 
 8 & 6.00 & 591 & 1.9934(37)  & 2.0950(39) & 0.50(4) & 1.9673(30)  & 2.1144(32) & 0.48(4) & 1.9397(24)  & 2.1588(26) & 0.46(4)\\ 
 8 & 5.20 & 591 & 2.8199(62)  & 2.9636(65) & 0.57(8) & 2.7642(51)  & 2.9709(55) & 0.56(8) & 2.7054(41)  & 3.0110(46) & 0.57(8)\\ 
 8 & 4.80 & 591 & 3.5794(86)  & 3.7618(91) & 0.7(1) & 3.4881(67)  & 3.7489(72) & 0.61(10) & 3.3927(50)  & 3.7760(55) & 0.54(8)\\ 
 8 & 4.50 & 591 & 4.568(11)  & 4.801(12) & 0.62(10) & 4.4175(86)  & 4.7478(92) & 0.56(8) & 4.2618(66)  & 4.7432(73) & 0.53(8)\\ 
 8 & 4.30 & 591 & 5.739(16)  & 6.032(17) & 0.6(1) & 5.501(13)  & 5.912(14) & 0.61(10) & 5.2550(100)  & 5.849(11) & 0.61(10)\\ 
 8 & 4.25 & 591 & 6.157(26)  & 6.471(27) & 1.1(2) & 5.884(20)  & 6.323(22) & 1.0(2) & 5.602(15)  & 6.235(17) & 0.9(2)\\ 
 8 & 4.20 & 591 & 6.675(24)  & 7.015(26) & 0.8(2) & 6.357(19)  & 6.832(21) & 0.8(1) & 6.029(14)  & 6.710(16) & 0.7(1)\\ 
 \hline 
10 & 8.50 & 591 & 1.0948(20)  & 1.1164(20) & 0.48(4) & 1.0860(16)  & 1.1167(17) & 0.46(4) & 1.0769(13)  & 1.1224(14) & 0.45(4)\\ 
 10 & 7.00 & 591 & 1.5273(33)  & 1.5575(33) & 0.7(1) & 1.5099(27)  & 1.5526(27) & 0.6(1) & 1.4919(21)  & 1.5550(22) & 0.62(10)\\ 
 10 & 6.00 & 591 & 2.0844(41)  & 2.1256(42) & 0.51(7) & 2.0516(33)  & 2.1096(34) & 0.50(7) & 2.0178(27)  & 2.1033(28) & 0.50(7)\\ 
 10 & 5.20 & 591 & 2.9850(77)  & 3.0440(79) & 0.7(1) & 2.9202(59)  & 3.0027(61) & 0.7(1) & 2.8528(43)  & 2.9736(45) & 0.56(8)\\ 
 10 & 4.80 & 591 & 3.886(11)  & 3.962(11) & 0.8(1) & 3.7681(85)  & 3.8746(88) & 0.7(1) & 3.6493(66)  & 3.8038(69) & 0.7(1)\\ 
 10 & 4.50 & 591 & 5.068(17)  & 5.168(18) & 0.9(2) & 4.872(13)  & 5.009(13) & 0.8(1) & 4.6764(87)  & 4.8744(90) & 0.58(8)\\ 
 10 & 4.30 & 591 & 6.523(36)  & 6.652(37) & 2.0(5) & 6.213(26)  & 6.388(27) & 1.6(4) & 5.906(18)  & 6.155(19) & 1.3(3)\\ 
 10 & 4.25 & 591 & 7.029(25)  & 7.168(26) & 0.9(2) & 6.676(20)  & 6.864(21) & 0.9(2) & 6.327(15)  & 6.594(16) & 0.8(2)\\ 
 \hline 
12 & 8.50 & 623 & 1.1213(22)  & 1.1320(22) & 0.62(10) & 1.1106(18)  & 1.1255(18) & 0.59(9) & 1.0995(14)  & 1.1211(14) & 0.51(6)\\ 
 12 & 7.00 & 606 & 1.5684(39)  & 1.5833(39) & 0.9(2) & 1.5492(31)  & 1.5699(32) & 0.8(2) & 1.5289(24)  & 1.5589(24) & 0.8(1)\\ 
 12 & 6.00 & 620 & 2.1595(54)  & 2.1800(54) & 0.9(2) & 2.1243(43)  & 2.1527(44) & 0.8(2) & 2.0870(34)  & 2.1280(34) & 0.8(2)\\ 
 12 & 5.20 & 609 & 3.1609(73)  & 3.1909(74) & 0.6(1) & 3.0814(61)  & 3.1227(61) & 0.6(1) & 2.9992(47)  & 3.0581(48) & 0.59(9)\\ 
 12 & 4.80 & 861 & 4.158(11)  & 4.198(11) & 1.0(2) & 4.0240(84)  & 4.0779(85) & 0.9(1) & 3.8867(62)  & 3.9630(63) & 0.8(1)\\ 
 12 & 4.50 & 626 & 5.544(18)  & 5.597(18) & 1.1(2) & 5.309(14)  & 5.380(14) & 1.0(2) & 5.073(10)  & 5.173(11) & 0.9(2)\\ 
 12 & 4.30 & 600 & 7.300(27)  & 7.370(27) & 1.1(2) & 6.913(20)  & 7.006(20) & 1.0(2) & 6.533(15)  & 6.661(15) & 0.9(2)\\ 
 \hline 
16 & 8.50 & 499 & 1.1556(26)  & 1.1592(27) & 0.6(1) & 1.1443(21)  & 1.1492(21) & 0.55(8) & 1.1323(17)  & 1.1393(17) & 0.51(8)\\ 
 16 & 7.00 & 481 & 1.6538(46)  & 1.6588(46) & 0.9(2) & 1.6289(33)  & 1.6359(33) & 0.7(1) & 1.6031(25)  & 1.6130(25) & 0.6(1)\\ 
 16 & 6.00 & 483 & 2.3103(71)  & 2.3173(71) & 0.9(2) & 2.2650(56)  & 2.2747(57) & 0.8(2) & 2.2181(45)  & 2.2319(45) & 0.8(2)\\ 
 16 & 5.50 & 495 & 2.9172(97)  & 2.9261(97) & 1.1(2) & 2.8465(76)  & 2.8587(77) & 1.0(2) & 2.7736(61)  & 2.7909(62) & 1.0(2)\\ 
 16 & 5.20 & 482 & 3.477(13)  & 3.488(13) & 1.3(3) & 3.377(10)  & 3.392(10) & 1.2(3) & 3.2748(78)  & 3.2952(78) & 1.1(2)\\ 
 16 & 4.80 & 483 & 4.710(25)  & 4.724(25) & 2.0(6) & 4.534(19)  & 4.553(19) & 1.7(4) & 4.356(14)  & 4.383(14) & 1.4(4)\\ 
 16 & 4.60 & 489 & 5.741(32)  & 5.759(32) & 2.4(7) & 5.486(24)  & 5.510(24) & 2.1(6) & 5.233(18)  & 5.266(18) & 1.8(5)\\ 
 16 & 4.50 & 492 & 6.537(49)  & 6.557(50) & 3(1) & 6.211(32)  & 6.238(32) & 2.3(7) & 5.889(21)  & 5.926(21) & 1.6(4)\\ 
 16 & 4.30 & 493 & 9.070(55)  & 9.098(55) & 2.5(8) & 8.473(40)  & 8.509(40) & 2.2(7) & 7.901(29)  & 7.950(29) & 1.9(5)\\ 
 16 & 4.25 & 482 & 10.019(73)  & 10.049(73) & 3(1) & 9.319(53)  & 9.359(53) & 3.0(10) & 8.653(38)  & 8.707(38) & 2.5(8)\\ 
 16 & 4.20 & 482 & 11.350(85)  & 11.385(86) & 3(1) & 10.489(56)  & 10.534(56) & 2.4(7) & 9.677(38)  & 9.737(38) & 1.8(5)\\ 
 \hline 
20 & 8.50 & 303 & 1.1827(50)  & 1.1842(50) & 1.2(4) & 1.1706(38)  & 1.1726(39) & 1.1(3) & 1.1578(29)  & 1.1608(29) & 0.9(2)\\ 
 20 & 7.00 & 303 & 1.6976(68)  & 1.6998(68) & 1.1(3) & 1.6740(57)  & 1.6769(57) & 1.1(3) & 1.6491(46)  & 1.6533(46) & 1.1(3)\\ 
 20 & 6.00 & 303 & 2.4230(80)  & 2.4260(80) & 0.8(2) & 2.3765(61)  & 2.3807(61) & 0.7(2) & 2.3272(49)  & 2.3332(49) & 0.7(1)\\ 
 20 & 5.20 & 301 & 3.751(26)  & 3.756(26) & 2.5(9) & 3.635(19)  & 3.642(19) & 1.9(7) & 3.517(13)  & 3.526(13) & 1.6(5)\\ 
 20 & 4.80 & 306 & 5.218(51)  & 5.225(51) & 5(2) & 5.007(38)  & 5.015(38) & 4(2) & 4.792(28)  & 4.805(28) & 3(1)\\ 
 20 & 4.65 & 303 & 6.119(37)  & 6.126(37) & 2.2(8) & 5.841(29)  & 5.851(29) & 2.0(7) & 5.560(22)  & 5.574(22) & 1.8(6)\\ 
 20 & 4.60 & 303 & 6.634(68)  & 6.643(68) & 4(1) & 6.286(51)  & 6.297(51) & 3(1) & 5.945(37)  & 5.960(37) & 3(1)\\ 
 20 & 4.50 & 305 & 7.705(89)  & 7.715(90) & 5(2) & 7.237(66)  & 7.250(67) & 4(2) & 6.785(47)  & 6.803(47) & 4(2)\\ 
 20 & 4.30 & 303 & 10.882(98)  & 10.896(98) & 4(2) & 10.074(73)  & 10.092(73) & 4(2) & 9.305(54)  & 9.329(55) & 3(1)\\ 
 20 & 4.25 & 302 & 12.20(12)  & 12.21(12) & 4(2) & 11.232(95)  & 11.252(95) & 4(2) & 10.324(76)  & 10.350(76) & 5(2)\\ 
 \hline 
24 & 8.50 & 201 & 1.2069(70)  & 1.2077(70) & 1.9(7) & 1.1949(56)  & 1.1960(56) & 1.8(7) & 1.1818(44)  & 1.1833(44) & 1.6(6)\\ 
 24 & 7.00 & 201 & 1.781(14)  & 1.782(14) & 3(1) & 1.749(11)  & 1.751(11) & 3(1) & 1.7172(86)  & 1.7194(86) & 3(1)\\ 
 24 & 6.00 & 201 & 2.543(23)  & 2.545(24) & 3(1) & 2.489(18)  & 2.491(18) & 2(1) & 2.433(14)  & 2.436(14) & 2.3(9)\\ 
 24 & 5.20 & 351 & 4.073(43)  & 4.075(43) & 6(2) & 3.931(33)  & 3.935(33) & 5(2) & 3.787(24)  & 3.792(24) & 4(2)\\ 
 24 & 4.80 & 577 & 5.812(57)  & 5.816(57) & 8(3) & 5.534(40)  & 5.539(40) & 6(2) & 5.261(29)  & 5.267(29) & 5(2)\\ 
 24 & 4.50 & 408 & 8.726(75)  & 8.731(75) & 5(2) & 8.165(55)  & 8.173(55) & 4(2) & 7.621(37)  & 7.631(38) & 3(1)\\ 
 24 & 4.30 & 321 & 12.71(14)  & 12.72(14) & 6(3) & 11.70(11)  & 11.71(11) & 6(3) & 10.737(80)  & 10.750(80) & 5(2)\\ 
 \hline 
32 & 8.50 & 121 & 1.251(10)  & 1.251(10) & 2(1) & 1.2374(84)  & 1.2377(84) & 2.0(9) & 1.2228(64)  & 1.2233(64) & 1.7(8)\\ 
 32 & 7.00 & 105 & 1.900(12)  & 1.900(12) & 1.2(5) & 1.861(10)  & 1.862(10) & 1.1(5) & 1.8214(89)  & 1.8222(89) & 1.1(5)\\ 
 32 & 6.00 & 140 & 2.732(27)  & 2.733(27) & 3(1) & 2.674(21)  & 2.674(21) & 2(1) & 2.611(16)  & 2.612(16) & 1.9(9)\\ 
 32 & 5.50 & 152 & 3.680(63)  & 3.681(63) & 5(3) & 3.565(49)  & 3.566(49) & 5(3) & 3.445(37)  & 3.446(37) & 5(2)\\ 
 32 & 5.20 & 161 & 4.595(65)  & 4.596(65) & 4(2) & 4.419(52)  & 4.420(52) & 4(2) & 4.240(41)  & 4.241(41) & 4(2)\\ 
 32 & 4.80 & 123 & 6.910(99)  & 6.912(99) & 5(3) & 6.545(73)  & 6.547(73) & 4(2) & 6.182(50)  & 6.184(50) & 3(2)\\ 
 32 & 4.60 & 171 & 9.301(94)  & 9.303(94) & 4(2) & 8.648(76)  & 8.650(76) & 4(2) & 8.027(59)  & 8.030(59) & 4(2)\\ 
 32 & 4.50 & 121 & 10.930(75)  & 10.932(75) & 3(2) & 10.133(65)  & 10.136(65) & 3(2) & 9.372(57)  & 9.376(57) & 3(2)\\ 
 \hline 
40 & 8.50 & 91 & 1.291(11)  & 1.291(11) & 2(1) & 1.2753(98)  & 1.2754(98) & 3(1) & 1.2593(78)  & 1.2595(78) & 2(1)\\ 
 40 & 7.00 & 111 & 1.949(33)  & 1.949(33) & 5(3) & 1.914(25)  & 1.914(25) & 4(2) & 1.877(18)  & 1.877(18) & 4(2)\\ 
 40 & 6.00 & 121 & 3.015(35)  & 3.015(35) & 3(2) & 2.922(27)  & 2.922(27) & 3(2) & 2.828(21)  & 2.829(21) & 3(1)\\ 
 40 & 5.20 & 103 & 5.014(93)  & 5.015(93) & 7(4) & 4.824(70)  & 4.825(70) & 6(3) & 4.628(52)  & 4.629(52) & 5(3)\\ 
 40 & 4.80 & 111 & 8.175(49)  & 8.176(49) & 2(1) & 7.649(38)  & 7.650(38) & 1.8(8) & 7.150(29)  & 7.151(29) & 1.6(7)\\ 
 40 & 4.65 & 143 & 10.18(15)  & 10.18(15) & 5(3) & 9.49(11)  & 9.49(11) & 5(2) & 8.816(78)  & 8.818(78) & 4(2)\\ 
 40 & 4.60 & 135 & 11.18(20)  & 11.18(20) & 9(5) & 10.35(16)  & 10.35(16) & 9(5) & 9.55(13)  & 9.55(13) & 9(5)